\documentclass{JHEP3}
\usepackage{amsfonts,amsbsy,latexsym,amssymb,amscd,amstext}
\usepackage{epsfig}
\usepackage{graphicx}
\usepackage{amsmath}
\usepackage{bm}
\usepackage{placeins}
\usepackage{verbatim}   % useful for program listings
\usepackage{subfigure}  % use for side-by-side figures

\newcommand{\HG}{\hat{\Gamma}}
\newcommand{\be}{\begin{equation}}
\newcommand{\ee}{\end{equation}}
\newcommand{\ba}{\begin{array}}
\newcommand{\ea}{\end{array}}
\newcommand{\baa}{\begin{array}}
\newcommand{\eaa}{\end{array}}
\newcommand{\bea}{\begin{eqnarray}}
\newcommand{\eea}{\end{eqnarray}}
\newcommand{\half}{{1\over2}}

\newcommand{\Tr} {{\rm Tr}}
\newcommand{\bb}{\bar{b}}
\newcommand{\p}{p}
\newcommand{\q}{q}
\newcommand{\m}{k} % esto es el flujo magnetico. 
\newcommand{\kb}{\bar{k}} % el dual
\newcommand{\pc}{p^{(c)}} %momento de color
\newcommand{\ps}{p^{(s)}} % espacial
\newcommand{\pt}{p}  % total
\newcommand{\vpc}{\vec{p}^{\,(c)}} % color con flecha
\newcommand{\vps}{\vec{p}^{\,(s)}}
\newcommand{\vpt}{\vec{p}}
\newcommand{\vqt}{\vec{q}}
\newcommand{\vqpt}{\vec{q}\,'}
\newcommand{\qt}{q}
\newcommand{\ET}{{\cal E}}
\newcommand{\FS}{F(\vpt,\vqt,-\vpt-\vqt)}
\newcommand{\FSS}{F^2(\vpt,\vqt,-\vpt-\vqt)}
\newcommand{\ev}{\hat{e}}
\newcommand{\llat}{\lambda_L}
\newcommand{\HT}{{\cal H}}

\title{Spatial volume dependence for  2+1 dimensional SU(N) Yang-Mills theory}

\author{ Margarita Garc\'{\i}a P\'erez$^{a}$, Antonio Gonz\'alez-Arroyo
$^{a,b}$ and Masanori Okawa$^{c}$ \\
  $^a$ Instituto de F\'{\i}sica Te\'orica UAM/CSIC, Nicol\'as Cabrera 13-15, \\
  Universidad Aut\'onoma de Madrid, E-28049--Madrid, Spain \\
  $^b$ Departamento de F\'{\i}sica Te\'orica, C-XI \\
       Universidad Aut\'onoma de Madrid, E-28049--Madrid, Spain \\
  $^c$ Graduate School of Science, Hiroshima University,\\
Higashi-Hiroshima, Hiroshima 739-8526, Japan \\    

E-mail: \email{margarita.garcia@uam.es, antonio.gonzalez-arroyo@uam.es,okawa@sci.hiroshima-u.ac.jp}
 }

\abstract{
We study the 2+1 dimensional SU(N) Yang-Mills theory on a finite
two-torus with  twisted boundary conditions. 
Our goal is to study the interplay between the rank of the group $N$,
the length  of the torus $L$  and the $Z_N$ magnetic flux.
After presenting the classical and quantum formalism, we analyze the  
spectrum of the theory using perturbation theory to one-loop and using 
Monte Carlo techniques on the lattice. In perturbation theory, 
results to all orders depend on the  combination $x=\lambda N L$ and an angle
$\tilde{\theta}$ defined in terms of the magnetic flux ($\lambda$ is `t 
Hooft coupling). Thus, fixing the angle, the system exhibits a form of 
volume independence ($NL$ dependence).  The numerical  results interpolate 
between our  perturbative calculations and the confinement regime.
They are consistent with x-scaling and provide  interesting information
about the k-string spectrum and effective string theories. The occurrence  
of tachyonic instabilities is also analysed. They seem to be avoidable 
in the large N limit with a suitable scaling of the magnetic flux. 
}

\preprint{
IFT-UAM/CSIC-13-066\\ FTUAM-13-12 \\HUPD-1306\\ \\
{\it Dedicated to the Memory of Misha Polikarpov}}

\date{\today}

\begin{document}

{\vskip 1cm}

\section{Introduction}
\label{s.intro}

Yang-Mills theories are a fantastic laboratory to test our
understanding of quantum field theory: very simple to formulate,
but containing a extremely rich phenomenology. Despite the enormous
progress made, we are not yet at the level of claiming a full
understanding of many of the entailing phenomena, such as confinement. 
Although most of the phenomenological interest lies in 3+1 space-time
dimensions, the 2+1 case has also been subject of interest. Proving 
confinement is not so much of a challenge in that case,
because even the abelian theories with monopoles enjoy this property in 2+1
dimensions~\cite{polyakov}. However, we still face the challenge of
having a successful computational method to provide the full spectrum
of the theory. No doubt that the lattice approach~\cite{wilson} provides 
numerical values for the string tension and spectrum~\cite{teper}. 
Nevertheless, it would be very desirable to have semi-analytical approaches
that could provide approximate results and a deep understanding of the
relevant dynamics involved. From that point of view, the 2+1 case,
being simpler, is also a perfect laboratory for ideas, methods and
techniques applicable to the higher dimensional case and to other
theories. There is an extensive literature on the subject where these 
ideas have been put forward.  For example, we signal out the work done
by Nair, who  proposed an approach which can lead to the desired analytical 
control~\cite{nair}. 

The present  work started with the concrete goal of  analyzing  the interplay
between the dependence of the theory  on the rank of the
gauge group $N$ and on the volume of space-time. Both dependencies are
presumably strongly linked to each other. The intriguing connection among both 
quantities (rank and volume) was initially put forward
by Eguchi and Kawai~\cite{EK}. In particular, it was suggested that
when the rank of the group goes to infinity, finite volume effects
vanish.  However, the validity of this result, often called large N
reduction,  rests upon  assumptions about whether  certain symmetries are
unbroken. As a matter of fact, the single lattice point {\em reduced model}
proposed in Ref.~\cite{EK} was soon disproved~\cite{QEK} due to breaking of the
necessary $Z(N)^4$ symmetry. Two possible ways out were soon proposed; 
one in Ref.~\cite{QEK} and the other in Refs.~\cite{TEK1,TEK2}.

Here we will not be directly concerned with the reduced models. We will 
adopt a more general viewpoint, and simply consider the dynamics of
SU(N) Yang-Mills fields living  on a  finite spatial manifold but in 
infinite (euclidean) time. This is an appropriate setting for a
Hamiltonian description in which the spectrum of the system can be
studied.  In particular, we will choose the spatial manifold to be a two-torus
with euclidean metric and size $L_1\times L_2$. The choice of the torus 
allows the introduction of a certain topology in the space of SU(N)
gauge fields through the selection of the boundary conditions.
This was first realized by `t Hooft~\cite{tbc1}, and
following his nomenclature, they are called {\em twisted boundary
conditions}. In this particular 2-dimensional situation, they are associated
to the introduction of a discrete modulo N magnetic flux through space. 
Boundary conditions have a strong impact on the finite volume dynamics 
at weak coupling. This is the reason why in Ref.~\cite{TEK1,TEK2} it
was proposed that using twisted boundary conditions the reduction idea 
could be rescued. The corresponding reduced model is hence known as 
the {\em twisted Eguchi-Kawai} model (TEK). Here, we will not attempt a full
description of the model and a review of its properties, since as
mentioned previously, our approach is more general. It is nevertheless
necessary to comment some aspects of the history of reduced models 
in as much as it affects the selection of our framework. 

Already in Ref.~\cite{TEK2} and  in Ref.~\cite{EguchiNakayama} an
alternative derivation of reduction was given based on perturbation theory
within the twisted reduction scheme. Despite the absence of spatial
degrees of freedom,  a suitable description of the Lie algebra mimics
the Fourier expansion on a finite lattice. The Feynman rules remember 
the non-commutative nature of the group through the presence of momentum
dependent phases  in place of the structure constants of the group.
These phases have an important effect in selecting planar diagrams
over the rest. Indeed, only for planar diagrams the overall phases of 
a diagram cancel, as proven in the aforementioned  references and in
Ref.~\cite{APOGA}. The peculiar Feynman rules were encountered, perhaps 
not unsurprisingly, many  years later in the context of quantum field
theories on  non-commutative space (for a review see \cite{douglasnekrasov}).
Indeed, the presence of momentum dependent phases has little to do
with the lattice nature of the TEK model. A continuum version of the rules
was given very early~\cite{GAKA}, having an obscure interpretation until its
reappearance in the context of non-commutative space. The quick
development of the field following its appearance in the string theory
literature led to the identification of problems within the
perturbative regime. These manifest themselves in that the negative
character of the self-energy makes  certain low momentum
modes become tachyonic, signalling the instability of the
perturbative vacuum~\cite{Hayakawa:1999yt}-\cite{Armoni:2003va}. The fate of the model has to 
be addressed 
non-perturbatively~\cite{Bietenholz:2006cz} and in this 
respect the TEK model played a role as a regulated version of the non-commutative theory~\cite{ambjorn}. 

Apart from the afore-mentioned problems, numerical studies~\cite{IO,TV,AHHI}
showed the appearance of symmetry breaking  for the TEK model at large values 
of $N$. This was interpreted as a first order transition to a symmetry
breaking vacuum configuration which at small coupling appears as a
metastable state.  The connection between these problems and the ones 
mentioned previously in relation with tachyonic instability remains
unclear. In any case, the previously mentioned  interpretation of the 
origin  of the problem in terms of metastable states, allowed to find 
a solution. It turns out~\cite{TEK3} that,
rather than keeping the integer magnetic flux associated to twist fixed as the 
rank of the group $N$ goes to infinity, one has to scale this flux in 
a concrete way with $N$. This prescription has led to results which
are not only free of symmetry breaking, but also reproduce the results
obtained by direct extrapolation to large N in a rather spectacular
way~\cite{AGAMOI}. Despite the numerical success, it is clear that a
deeper understanding is welcome. The present paper, which addresses 
the 2+1 dimensional case, is a step in that direction. The lesson to 
be learned is that particular care has to be paid to the dependence 
on the total magnetic flux and its scaling properties with N.

Up to now our motivation has been centered upon large N physics,
however, the interest of volume dependence at finite N is also relevant 
for other purposes. Certainly, from a practical point of view, 
since non-perturbative results obtained on the lattice 
are subject to these corrections. From a more fundamental standpoint 
the use of the space-time volume was advocated as a good parameter 
to monitor the transition from the small box perturbative region to the
large volume confinement regime. This program had the goal of
providing new calculational techniques or new insights into the nature 
of non-perturbative phenomena~\cite{luscher,vanbaal:2000zc}. Here, of
course,  boundary conditions do matter, and the use of twisted boundary
conditions was argued to produce a smoother transition in several 
perturbative~\cite{zahed,korthals,daniel}  and non-perturbative
works~\cite{rtn,martinez}. It has also been advocated as a
tool to determine the electric-flux spectrum and employed to
analyze the confinement/deconfinement transition~\cite{Kovacs:2000sy}-\cite{Hasenfratz:1989tp}. 
To conclude this lengthy motivation we should
mention that recently a fundamental line of work has also advocated 
the use of spatial  size  as a monitoring parameter with a similar
goal of developing new calculational techniques and/or insights 
into the nature of some non-perturbative phenomena~\cite{unsal,Armoni:2011dw}. 
In the majority of these new works only one direction is kept finite 
and, in the absence of a phase transition, it allows to continuously 
connect to a dimensionally reduced model, benefiting from the 
use of the  powerful  low-dimensional techniques available. 

After having put our work in perspective and motivated its interest,
we address the description of the
lay-out of the paper. As mentioned earlier we will focus upon SU(N) 
Yang-Mills theory living on a $T_2\times R$, space-time. We will
restrict ourselves to the case in which the fields satisfy
twisted boundary conditions with non-zero magnetic flux $m$ 
on the spatial two-torus of  size $L_1\times L_2$. The case of
periodic boundary conditions demands very different techniques and has
been studied more extensively earlier both in the papers mentioned 
before~\cite{luscher,vanbaal:2000zc}, as in the context of large N
reduction~\cite{narayananneuberger}. Our paper is neatly divided into
three parts. The first part develops the formalism. Since few
researchers are acquainted with this formalism, we have made an effort 
to make a self-contained presentation that could serve as background
material for this and future papers. The 2+1 dimensional case, being
simpler, provides a good starting point before facing the more involved 3+1
case.  The presentation begins with classical aspects exhibiting the two
most useful types of partial gauge fixing. One of the formulations is
particularly well-suited for perturbative calculations, while the other
is useful for semiclassical contributions around non-trivial minima of 
the action.  The connection of these two formalisms is a new result,
whose derivation is given in Appendix~\ref{appendixa}. The quantization of the
system is done in the Hamiltonian formalism in the $A_0=0$ gauge, where 
the Gauss constraint condition is implemented as a condition on
physical states. 

The second part proceeds by setting up the perturbative calculation of
the spectra up to order $g^2$. The whole set of rules depends on the 
size of the torus $L$, the rank of the group $N$, the gauge coupling,
better expressed as `t Hooft coupling $\lambda=g^2N$, and an angle 
$\tilde{\theta}$ which depends on the magnetic flux through space. 
We observe, that for fixed value of this angle all the size dependence 
appears in the combination $LN$ (for $L\equiv L_1=L_2$). This can 
be considered a strong form of reduction, which is valid at  finite
$N$. If, as expected, the zero-electric flux sector of the theory at large
$NL$ becomes independent of this angle, this would imply the standard 
large $N$  reduction result. In this paper we focus specifically on the 
calculation of the lowest energy states in each electric flux sector. 
The most  crucial part of the calculation is the one loop gluon self-energy
contribution. In the Hamiltonian formalism this result appears as 
a sum of three individually divergent contributions. The divergence 
is however $\tilde \theta$ independent, so that the $\tilde \theta$ dependence of the
self-energy can be easily derived. To extract a full finite result one
must take care to regularize the expressions in a gauge invariant way. 
For that purpose we decided to calculate this self-energy following  two
additional gauge invariant procedures. The first being the calculation 
of the vacuum polarization in euclidean space, which later is
dimensionally regularised. The result turns out to be finite and
gives a $\tilde \theta$ dependence which coincides with the one computed earlier 
in the Hamiltonian formulation. Finally, we also use a lattice
regularization. The result extrapolated to vanishing lattice spacing
is again finite and coincides with the previous calculation.
The whole result is summarised in an elegant 
self-energy formula which allows us to study the occurrence or not of
tachyonic instabilities in the theory. Indeed, the formula always
predicts negative energies for sufficiently large values of the
coupling $\lambda$. The problem is that for certain choices of 
the magnetic flux, as the rank of the group gets larger, the coupling 
at which the instability occurs is small enough to validate the
perturbative calculation. On the contrary, if we scale the magnetic 
flux with $N$, following the
prescription of Ref.~\cite{TEK3} generalized to 2+1 dimensions, the 
instabilities occur always at finite values of $\lambda$ at which the 
perturbative calculation is not necessarily trustworthy. This result 
is by itself one of the most important new results contained in our
paper. 

The previous considerations bring us naturally to the third part of
the paper which makes use of the numerical study of the model
discretized on the lattice. Our numerical results are far from being 
a complete test of the model, but were specifically designed to test 
the transition of the ground state energies from the region where 
the perturbative formulas should apply to that in which 
these energies enter the confinement regime giving rise to the 
linearly rising $k$-string spectrum, which will be analyzed. 
Our results provide non-perturbative support to the strong reduction 
conjecture indicating that the effective 
size controlling the linear growth of the 
k-string energies is indeed $LN$ or, more specifically, the dimensionless 
variable $x=\lambda L N /4\pi$. 
The dependence on $x$ and $\tilde \theta$ of the numerical results is fairly 
well described by a 
parameterization that encodes both the perturbative and the large volume asymptotic 
behaviour, including subleading terms in the effective string description of
the electric-flux energies. 
This formula allows to extend the analysis of the appearance of tachyonic instabilities
outside the realm of perturbation theory, and to argue that the prescription of Ref.~\cite{TEK3}
gives rise to non-tachyonic results in all the range of values of $x$.

The paper closes with a summary of our results and a description of
open problems and future prospects.

\section{Yang-Mills fields on a twisted box}
\label{method}

This section will review the basic formalism for describing  gauge fields
living  on a spatial  torus. Here, we will be specific
to the two-dimensional case in a Hamiltonian framework and will derive
several formulas specific for this case. Most of the 
original papers developing the formalism have been cited in the
introduction. Additional  references  and a more complete presentation 
to the field can be found in Ref.~\cite{TONYREV}.

The first step is to define the configuration space. This is given 
by gauge fields on the torus. Namely, we have to define a bundle with
base space the 2-torus and a connection on this bundle. The torus has
euclidean metric and periods given by $L_i \ev_i$, where $\ev_i$ is the
unit vector in the $i$th direction and  $i=1,2$. As is
customary in the Physics literature, we will work with a trivialization of the bundle 
entailed by having a single rectangular patch of size $L_1\times L_2$
and transition functions $\Omega_i(x)$. We will also take the bundle
to be an SU(N) associated bundle in the fundamental representation.
Hence, the transition matrices are $N\times N$ matrices. The
consistency conditions imply:
\be
\Omega_1(x+L_2 \ev_2)\Omega_2(x)= e^{2 \pi i \m/N} \Omega_2(x+L_1
\ev_1)\Omega_1(x)\quad ,
\ee
where $\m$ is an integer modulo  $N$, known as {\em magnetic flux}.
Once the bundle is trivialized the connection is given by the $N\times
N$ traceless hermitian matrix $A_i(x)$. The periodicity condition for
this connection is 
\be
\label{A_per_cond}
A_i(x+L_j \ev_j)= \Omega_j(x) A_i(x)\Omega^\dagger_j(x)+i \Omega_j(x)
\partial_i\Omega^\dagger_j(x)\quad .
\ee

The starting field-space of the system is given by the pair $\{\Omega_i(x), A_i(x) \}$.
However, the physical configuration space is the space of trajectories
under local gauge transformations. A gauge transformation acts on the
starting space as follows:
\begin{eqnarray}
A_i(x) &\longrightarrow& \Omega(x)A_i(x) \Omega^\dagger(x)+i
\Omega(x)\partial_i \Omega^\dagger(x)\quad , \\
\Omega_i(x) &\longrightarrow& \Omega(x+L_i \ev_i) \Omega_i(x) \Omega^\dagger(x)\quad .
\end{eqnarray}
Wilson loops and Polyakov lines are gauge invariant observables, as
well as the eigenvalues of the magnetic field. 

We might partially constrain  gauge transformations  by fixing the value of
$\Omega_i(x)$. Thus, gauge transformations should satisfy the following 
periodicity condition 
$$ \Omega(x+L_i \ev_i)=\Omega_i(x) \Omega(x) \Omega^\dagger_i(x) \quad .$$

There is a symmetry of the action corresponding to a transformation
which multiplies the transition matrices by an element of $\mathbf{Z}_N$
without affecting the gauge field $A_i(x)$. This is not a gauge
transformation since the Polyakov loops are multiplied by an element of the
center. `t Hooft called them {\em singular gauge transformations}.
Once we fix the gauge to specific transition matrices $\Omega_i(x)$,
the symmetry transformation looks as an ordinary  gauge transformation
satisfying the following generalized 
periodicity condition
\be
\label{per_cond}
\Omega(x+L_i \ev_i)=e^{2 \pi i k_i/N} \Omega_i(x) \Omega(x) \Omega^\dagger_i(x) \quad .
\ee
The space of x-dependent SU(N) matrices satisfying the previous
equation will be labelled ${\cal G}(\vec{k})$. The quotient group 
$$ \left(\cup_{\vec{k}} \, {\cal G}(\vec{k})\right)/{\cal G}(\vec{0})
\sim Z_N^2 $$
is a very important symmetry group of our problem. Its
representations are labelled by the electric flux vector $\vec{e}$. 

We might generalize the periodicity condition  for
$\Omega(x)$ (Eq.~\ref{per_cond}) to  
arbitrary $N\times N$ matrices. This defines the vector spaces of
matrix-valued fields ${\cal E}(\vec{k})$. 
These are interesting spaces satisfying
\be
\forall U \in {\cal E}(\vec{k}) \quad {\rm and} \quad V \in {\cal
E}(\vec{k}') \quad {\rm then } \quad UV \in  {\cal E}(\vec{k}+
\vec{k}') \quad .
\ee
These spaces admit ${\cal G}(\vec{k})$ as subspaces. Even more, one
can take a non-singular  matrix $U(x)$ in ${\cal E}(\vec{k})$ and decompose
it uniquely as a product 
\be
U=\Omega H\quad ,
\ee
where $\Omega \in {\cal G}(\vec{k})$ and $H$ is hermitian and positive
definite. Furthermore, $H$ belongs to ${\cal H}_0$, which is the
subspace of ${\cal E}(\vec{0})$  corresponding to hermitian matrices.  

A very important property of the space of connections is that it is an
affine space, in which $ {\cal H}_0$ is the associated vector space.
This means that a generic  connection $A_i(x)$ satisfying Eq.~\ref{A_per_cond}
can be written as 
\be
\label{eq.qi}
A_i(x) =  A_i^{(0)}(x) + g\, Q_i(x) \quad ,
\ee
where 
$ A_i^{(0)}(x)$ is a particular representative of the space, and
$Q_i(x) $ runs over the space of  traceless elements of ${\cal H}_0$. 

We conclude this general presentation with two additional comments.
The first is about the form of infinitesimal gauge transformations. 
These are elements of ${\cal G}(\vec{0})$ 
of the form $\mathbf{I}+ i \omega + \ldots$, where $\omega$ is an
infinitesimal  traceless element of ${\cal H}_0$. The second comment refers to
operators acting on these spaces. In particular, we point out that the 
covariant derivative operator with respect to any compatible gauge field 
transforms an element  of  ${\cal E}(\vec{k})$ into a new element of
the same space.

All the previous results are valid for all choices of twist matrices
$\Omega_i(x)$. In practice, there are two main choices which have been
used in the literature and which have relative advantages. The first choice
is given by the constant non-commuting twist matrices $\Omega_i(x)=
\Gamma_i$. This is particularly well suited for perturbation theory,
since $A_i=0$ is a possible connection in this case (even for
non-vanishing magnetic flux $\m$). The second choice  is given by
x-dependent commuting (often called abelian) matrices:
\be
\Omega_i(x)=\exp\{i \bar{B} \epsilon_{i j} L_i x_j/2  \}\quad ,
\ee
where $\bar{B}$ is a traceless diagonal matrix whose elements satisfy
\be
\bar{B}^a L_1 L_2= \frac{2 \pi \m}{N} +2 \pi q^a\quad ,
\ee
where the $q^a$ are integers. 

In the following two subsections we will develop the two formulations
in turn. The connection among the two descriptions is accounted for by a
non-trivial gauge transformation. Its form is derived in
Appendix~\ref{appendixa}.

\subsection{First Formalism}

In this subsection we will develop the expression of the gauge fields
in the formalism with constant twist matrices $\Omega_i(x)=\Gamma_i$.
They satisfy
\begin{equation}
\label{MCOM}
\Gamma_1 \Gamma_2 = e^{2 \pi i \m/N} \Gamma_2 \Gamma_1\quad ,
\end{equation}
where  $\m$ is the magnetic flux. In the case that $\m$ and $N$ are
co-prime, this equation defines the matrices $\Gamma_i$ uniquely 
modulo global gauge transformations (similarity transformations).
The $\Gamma_i$ matrices belong to SU(N) and verify
the following conditions:
\be 
 \Gamma_i^N= \pm \mathbf{I} \quad ,
\ee
 for N odd or even respectively (For simplicity we will assume $N$ is
 odd and coprime with $\m$ in the following). 

In this formalism, the gauge fields $A_i(x)$ have to satisfy the
following periodicity formulas:
\begin{equation}
\label{tbc}
A_i(x+ L_j \ev_j)=\Gamma_j A_i(x) \Gamma^\dagger_j\quad .
\end{equation}
Notice,  that $A_i=0$  satisfies this condition and, hence, is 
 an admissible connection. Thus, we can choose it as our particular  connection 
$A_i^{(0)}$ of the previous section. Hence,
the space of connections is the space of traceless elements of ${\cal
H}_0$. Thus, in this formalism,
it is convenient to rename the hermitian element $Q_i$, in Eq.~\ref{eq.qi}, as $A_i$.
This is equivalent to reabsorbing a factor $1/g$ in the definition
of $A_i(x)$, so that the potential energy to leading order has no $g$ dependence.

To solve the constraint Eq.~\ref{tbc},  we introduce a basis 
of the space of $N\times N$ matrices $\HG(\vpc)$ satisfying:
\begin{equation}
\Gamma_i\HG(\vpc)\Gamma_i^\dagger = e^{i L_i\pc_i}
\HG(\vpc) \quad .
\end{equation}
The index $\vpc$ varies over vectors $(\frac{2 \pi n_1}{L_1 N},
\frac{2 \pi n_2}{L_2 N})$
with $n_i$  integers  defined modulo N. Thus, there are
$N^2$ such matrices. Indeed, the solution is unique modulo a
multiplicative factor. Using a suitable normalization condition 
this solution is given by 
\be
\label{defHG}
\HG(\vpc) = \frac{1}{\sqrt{2N}}\, e^{i \alpha(\vpc)} \Gamma_1^{-\kb n_2} \Gamma_2^{\kb n_1}\quad ,
\ee
where $\kb$ is an integer satisfying $ \m \kb = 1 \bmod N$,
and $\alpha(\vpc)$ an arbitrary phase factor, which will  be fixed later.

Now one  can expand our gauge fields in this basis
\be
A_i(x)= \sum'_{\vpc} \hat{A}_i(x,\vpc)\, e^{i
\vpc\vec{x}}\, \HG(\vpc)\quad .
\ee
The prime means that we exclude $\vpc=0$ from the sum, since
$A_i(x)$ is traceless (for SU(N) group).
The boundary conditions imply that $\hat{A}_i(x,\vpc)$ is
periodic, so it can be expanded in the normal Fourier series. 
Introducing  the momenta $\ps_i=2 \pi m_i/L_i$ with $m_i$ an integer, 
we obtain 
\be
A_i(x)=  {\cal N}\sum'_{\vec{P}} \hat{A}_i(\vpt)\,  e^{i
\vpt\vec{x}}\,  \HG(\vpc) \quad ,
\ee
where ${\cal N}=1/\sqrt{L_1 L_2}$ and  $\vpt= \vps+ \vpc$. 
This can be interpreted by saying that the total momenta $\vpt$ 
is composed of two pieces: a colour-momentum part $\vpc$ and a 
spatial-momentum part $\vps$. 
Notice that (if we neglect the prime in the sum) the range of values of 
$\vpt$ is just that of a theory defined on a box of size $(N\cdot
L_1) \times (N \cdot L_2)$. Furthermore, the Fourier coefficients
$\hat{A}_i(\vpt)$ are simple complex numbers and not vectors, so
that the formalism looks as if the theory had no colour, but was
defined on a bigger spatial box. 

To make the formulas more elegant we recall that  $\vpt$ can be
decomposed uniquely into its two components, so that we might write 
$\HG(\vpt)$ instead of $\HG(\vpc)$ in the previous formulas.
We might even take advantage of this modification to make the phase
$\alpha$ appearing in Eq.~\ref{defHG} dependent on $\vpt$ rather
than on its colour-momentum part only. 

The hermiticity
properties of $A_i$ imply:
\be
\hat{A}^*_i(-\vpt)= e^{i(\alpha(-\vpt)+\alpha(\vpt))}\, e^{-2 \pi i n_1 n_2 \kb/N}
\hat{A}_i(\vpt) \quad .
\ee
To make the resemblance with ordinary Fourier decomposition even more
apparent, we may choose the phases $\alpha(\vpt)$ such as to
impose the matrix condition 
\be
\HG(-\vpt)=\HG^\dagger(\vpt) \quad ,
\ee
implying that the coefficients satisfy 
\be
\hat{A}_i(-\vpt)= \hat{A}_i^*(\vpt) \quad .
\ee
A particularly natural and elegant  choice is 
\be
\label{alphachoice}
\alpha(\vpt)= \frac{\theta}{2}\pt_1 \pt_2\quad ,
\ee
where $\theta$ is given by 
\be
\label{thetadef}
\theta=\frac{\kb N L_1 L_2}{2 \pi}\quad .
\ee
To conclude this discussion, we give the final decomposition of our
vector potential 
\be
\label{decomposition}
A_i(x)=  {\cal N}\sum'_{\vpt} \hat{A}_i(\vpt)\,  e^{i
\vpt\vec{x}}\,  \HG(\vpt)\quad .
\ee
The inverse formula will also be needed in what follows, and is given by 
\be
\label{inverse}
\hat{A}_i(\pt)= 2 {\cal N}\int dx  \,
\mathrm{Tr}(\HG(-\vpt)A_i(x))\, e^{-i \vpt\vec{x}}\quad ,
\ee
where the spatial integral extends over the 2-torus of size $L_1L_2$.

Let us now compute the magnetic field of the theory: 
\begin{equation}
B(x)=\partial_1 A_2-  \partial_2 A_1  -i g [A_1, A_2]\quad .
\end{equation}
It satisfies the same boundary conditions as the vector potential, and
can, henceforth, be decomposed similarly in terms of the complex
coefficients  $\hat{B}(\vec{P})$. Using the previous formulas one can
obtain the connection among the coefficients as follows:
\begin{equation}
\hat{B}(\vpt)= i \pt_1 \hat{A}_2(\vpt
) -i \pt_2 \hat{A}_1(\vpt) 
+ g{\cal N} \sum'_{\vqt} \sum'_{\vqpt} \delta(\vqt+\vqpt-\vpt)
\hat{A}_1(\vqt)\hat{A}_2(\vqpt) F(-\vpt, \vqt, \vqpt)\quad ,
\end{equation}
where 
\be
F(-\vpt, \vqt, \vqpt)= -2i {\rm
Tr}(\HG(-\vpt)\, [\HG(\vqt), \HG(\vqpt)])\quad .
\ee

The coefficients $F(-\vpt, \vqt, \vqpt)$ are basically the
structure constants of the SU(N) Lie algebra for our particular basis. 
From its definition one concludes that they are cyclic-symmetric and
anti-symmetric under the exchange of any two indices. Under  complex
conjugation 
\be
F^*(-\vpt, \vqt, \vqpt)= F(\vpt, -\vqt, -\vqpt)\quad . 
\ee

Finally, we can give the explicit expression for $F(-\vpt, \vqt, \vqpt)$
for the particular choice of the phases
$\alpha(\vpt)$ given earlier:
\be
\label{eqdeF}
F(\vpt,\vqt,-\vpt-\vqt)= -\sqrt{\frac{2}{N}}  \sin\left(\frac{\theta}{2}
(\vpt\times \vqt)\right)\quad ,
\ee
where $\vpt\times \vqt=\pt_1\qt_2-\pt_2\qt_1$ and $\theta$ was defined 
in Eq.~\ref{thetadef}. We point out that the three momenta which are
arguments of $F$ sum up to zero. Thus, taken in a given order they
define and oriented triangle. In this geometrical description the
argument of the sine in Eq.~\ref{eqdeF} is $\theta$ times the area
(taken with sign) of the triangle.

The classical dynamics of the Yang-Mills fields on the 2-torus can be 
formulated in terms of the Fourier coefficients $\hat A(\vpt)$ and its 
corresponding velocities and/or  conjugate momenta. The potential
energy of the Yang-Mills fields, for example, becomes
\be V= \int dx\,  {\rm Tr}( B(x) B(x))= \frac{1}{2} \sum_{\vpt}
|\hat{B}(\vpt)|^2\quad .
\ee

Before describing the quantization of the system in these coordinates,
let us comment upon a class of alternative descriptions of the space of
gauge fields which turns out to be useful in computing certain
non-perturbative effects.

\subsection{Second Formalism}

As mentioned previously, another possible choice of the transition
matrices is given by the abelian ones:
$\Omega_i(x)=\exp\{i \bar{B} \sum_j \epsilon_{i j} L_i x_j/2  \} $
where $\bar{B}$ is a traceless diagonal matrix whose elements satisfy
$ \bar{B}^a L_1 L_2= \frac{2 \pi \m}{N} +2 \pi q^a$
where the $q^a$ are integers. Indeed, any possible value of the
integers $q^a$ provides  a different choice.

The goal is once more that of parameterizing the space of gauge fields
satisfying the boundary conditions. One must first identify one particular
connection $A_i^{(0)}(x)$ belonging to this space. A particularly
simple one is given by
\be
A_i^{(0)}(x) = -\frac{1}{2} \bar{B} \epsilon_{ i j} x_j \quad ,
\ee
which corresponds to a uniform magnetic field of magnitude $\bar{B}$. A
general gauge field is then given by
\be
A_i(x)= A_i^{(0)}(x) + g Q_i(x) 
\quad ,
\ee
where the $Q_i$ transform homogeneously:
\be
Q_i(x+L_j \ev_j)= \Omega_j(x) Q_i(x) \Omega^\dagger_j(x) 
\quad .
\ee
These conditions naturally split the gauge field degrees of freedom 
into those associated to diagonal components, which are periodic, and
off-diagonal components which satisfy properties related  to those of 
Jacobi theta functions. The whole treatment is similar to that
appearing in the abelian projection of non-abelian gauge theories. 
The matrix $\bar{B}$ defines a direction in colour space and a
corresponding subgroup which leaves this direction  invariant. For
coinciding eigenvalues ($q^a=q^b$ for some $a$ and $b$) the subgroup is
still non-abelian.

Notice, that this formalism, although completely general, signals out 
the gauge configuration corresponding to a uniform  magnetic field.
Obviously, this is an extremum of the potential energy with a value
\be
\label{Vlowest}
V= \frac{L_1 L_2}{g^2} {\rm Tr}( \bar{B}^2) = \frac{4 \pi^2}{L_1
L_2 g^2} \left(\sum_a
(q^a)^2 - \frac{\m^2}{N}\right)
\quad .
\ee
This is not the absolute minimum, which we know has vanishing energy. 
The value of $Q_i(x)$ at which this absolute minimum is achieved is non-trivial 
and can be found in appendix~\ref{appendixa}. This makes ordinary perturbation
theory very impractical in this formalism. On the contrary, the
formalism is  particularly well-suited to study fluctuations around the
uniform magnetic field configurations associated to $Q_i(x)=0$. 
Being extrema of the  potential energy,  they are  the equivalent of
{\em sphalerons} for our 2+1 theory. They may play an important
dynamical role in the transition from small to large torus sizes. 
The first configurations expected to play a role in this context are those
having minimal energy within each magnetic flux sector sector $\m$. 
This is achieved if we take $q^a=-1$ for $a=1,{\ldots} , \m$, and $q^a=0$ for
$a=\m+1,{\ldots}, N$. Thus, the minimal energy associated to a constant field strength
solution for non-vanishing magnetic flux $\m$ is
\be
V_{\rm min} = \frac{4 \pi^2 \m}{L_1 L_2 g^2} (1-\frac{\m}{N})
\quad,
\ee
which goes to zero when the area goes to infinity. 
If we take $\m$ of order $N$ ($\m=\alpha N$), then the minimal energy
becomes
\be
V_{\rm min} =  \frac{4 \pi^2 N}{L_1 L_2 g^2} \alpha (1-\alpha)
\quad .
\ee
Different situations arise depending on whether one takes the large N
or the infinite volume limit first.

As mentioned previously the configurations corresponding to a constant 
magnetic field $A_i(x)=A_i^{(0)}(x)$ are  extremals of the potential 
energy. However, as we will see, they are not local minima, but rather 
saddle points. To study this aspect one has to perturb around the
solution, which is equivalent to taking $Q_i$  non-zero and small. 
Our goal would be to expand  the potential up to order quadratic 
in the $Q_i$ fields. The sign of the eigenvalues of the quadratic form 
determines  the stability or not of this solution. 

Our first step would be to compute the magnetic field 
\be
\label{magnetic2}
B(x)= \bar{B} + g (\bar{D}_1 Q_2(x)- \bar{D}_2 Q_1(x)) -i g^2 [Q_1(x), Q_2(x)]
\quad ,
\ee
where the  symbol $\bar{D}$ stands for the covariant derivative in the 
background field of $A_i^{(0)}(x)$. Now we plug this expression into
the formula for the potential energy 
\be
V=\frac{1}{ g^2} \int dx \,{\rm Tr} (B^2(x))=  \frac{1}{g^2} \int dx \, (\sum_a (B^{a a})^2 +\sum_{a\ne b} |B^{a b}|^2)
\quad ,
\ee
and keep terms up to order $Q^2$. The lowest order is given by
Eq.~\ref{Vlowest}, the linear term vanishes, and the  quadratic term
takes the form
\be
 \int dx \, \Tr\left(  (\bar{D}_1 Q_2(x)- \bar{D}_2
Q_1(x))^2 -2 i \bar{B} [Q_1(x), Q_2(x)] \right)
\quad .
\ee
This expression has two parts. The  first  is given by the square of
the parenthesis which is  positive semidefinite, but the last part
could give rise to instabilities since the sign is not determined.
By a simple inspection it is easy to see that the potentially negative 
term only involves off-diagonal fluctuations $Q_{a b}$, with $a\ne b$.
It is clear that, to this quadratic order, all off-diagonal components
are decoupled from others except its transpose $Q^{b a}=Q^{a b\, *}$.
Thus, it is enough to  concentrate in one particular pair of
indices $a,b$, and simplify the notation by writing $Q(x)\equiv Q^{a b}(x)$. Its
contribution to the potential density becomes:
\be
2 \bb \, {\rm Im}(Q_2 Q_1^*)+  |(\bar{D}_1 Q_2 -\bar{D}_2 Q_1)|^2
\quad ,
\ee
where $\bb\equiv(\bar{B}^a-\bar{B}^b)=\frac{2 \pi q}{L_1 L_2}$, 
and $q=q^a-q^b$ is an integer. We will take $q$ to be positive, which
can always be done by exchanging the colour indices. It is convenient 
to spell out the form of the covariant derivative
\be
(\bar{D}_i Q) = \partial_i Q +
\frac{i}{2}\bb\, \epsilon_{i k} x_k Q
\quad .
\ee
It is also interesting (and necessary) to take into account the 
boundary conditions that must be satisfied by the fluctuations.
They are given by 
\be
\label{BC2}
Q(x+L_j \ev_j)= \exp\{i \bb \epsilon_{j k} L_j x_k/2\}\, Q(x)
\quad .
\ee

Now we can  express the covariant derivatives  as follows
\bea
\bar{D}_1= \sqrt{\frac{\bb}{2}} (\mathbf{a}+ \mathbf{a}^\dagger)\quad ,\\
\bar{D}_2= -i \sqrt{\frac{\bb}{2}} (\mathbf{a} -\mathbf{a}^\dagger)
\quad ,
\eea
where the operators $\mathbf{a}^\dagger$ and $\mathbf{a}$ satisfy 
the algebra of creation and annihilation operators. Using these
operators one could construct a basis of the space of fields 
satisfying the boundary conditions Eq.~\ref{BC2} by using the 
eigenstates of the harmonic oscillator 
\be
\Psi_n(x)= \frac{(\mathbf{a}^\dagger)^n}{\sqrt{n!}} \Psi_0(x)
\quad ,
\ee
where the field $\Psi_0(x)$ behaves  like the ground state of the 
harmonic oscillator and is annihilated by the operator 
$\mathbf{a}$. The existence and properties of this state will be 
clarified below.  The way the operators act on the $\Psi_n$ states 
replicates the formulas of the harmonic oscillator. 

Equipped with this technology we go back to the expression of the 
fluctuation potential. It is convenient to define the combinations 
$Q_{\pm}\equiv Q_1 \pm i Q_2$. In terms of them, the covariant
derivative part of the magnetic field becomes 
\be
i\sqrt{\frac{\bb}{2}}(\mathbf{a}Q_--\mathbf{a}^\dagger Q_+)
\quad ,
\ee
while the potentially dangerous contribution to the potential energy
density  becomes 
\be
\frac{\bb}{2}(|Q_+|^2-|Q_-|^2)
\quad .
\ee
As anticipated it can become negative. This could happen, for example, 
by taking $Q_+=0$. This negative value cannot be compensated by a positive
contribution from the square of the covariant derivative term if we 
take $Q_-\propto \Psi_0$, since in that case the other contribution
vanishes (choosing $\Psi_1$ would have exactly compensated the
negative term). Hence, we end up concluding that the constant field 
strength solution  is unstable under deformations with $Q_+=0$
and $Q_-\propto \Psi_0$.

Let us now verify the existence of the $\Psi_0$ configuration and
write out its form. For that purpose we introduce the 
complex coordinate $z=(x_1+i x_2)/L_1$. It is interesting
to write the operator $\mathbf{a}$ in terms of
derivatives of the complex variables. One has:
\be
\mathbf{a}=\frac{1}{\sqrt{2\bb}}\Big (\frac{2}{L_1}\frac{\partial }{\partial
\bar{z}}+ \frac{\bb L_1}{2}z\Big )
\quad ,
\ee
where $\bar{z}$ is the complex conjugate of $z$. 
Indeed, if we parameterize the fields as 
\be
\label{parameterization}
Q(x)=\exp\{-\frac{L_1^2 \bb}{4} (z\bar{z}-z^2)\}\, \chi(x) 
\quad ,
\ee
one sees that the destruction operator acts on 
$\chi(x)$ just as a derivative with respect to the complex conjugate 
coordinate $\bar{z}$. The conclusion is that $\Psi_0$ is given 
by Eq.~\ref{parameterization} with  $\chi(x)$  being a holomorphic function
$\chi(z)$.

The final step is to study the form of the boundary conditions expressed
in terms of $\chi(z)$. We leave the calculation to the reader. The
result  is that for $q=1$ the boundary conditions coincide with those 
satisfied by the Jacobi theta functions $\theta_3$ (see for example
Ref.~\cite{Tata}). Indeed, this is the unique holomorphic function 
satisfying the boundary conditions up to multiplication by a constant. 
For $q>1$ there are  indeed $q$-linearly independent solutions, which
can be obtained by multiplication of theta functions. In that case 
there there are $q$ linearly-independent $\Psi_n$ basis vectors for
each $n$.

Considering now all possible off-diagonal elements of $Q$,
the total number of unstable modes is 
\be
\sum_{a<b} |q_a-q_b| >0
\quad .
\ee
We recall that $\sum_a q_a=-\m$, which makes the previous number
strictly positive for non-zero flux. Indeed, the minimum number of
unstable modes turns out to be $(N-\m)\m$, and is achieved precisely 
for the same configurations studied before having least potential 
energy $V_{\rm min}$. It is perhaps not surprising that both criteria 
lead to the same configuration.

The study of the dynamical role played by these  spatial
configurations should follow a similar pattern to the one 
played by sphalerons in 4d theories. We address the reader to the 
literature on the subject~\cite{sphalerons}. In any case, this study 
demands considerable effort  and will not be included in this paper.

\subsection{Hamiltonian formulation in 2+1 dimensions}
The previous description has been essentially classical. The goal was
to find a parameterization of the gauge potentials encoding the
boundary conditions. In this section we will briefly describe the 
operator formalism approach to its quantum mechanical formulation. 
The neatest  way to quantize the system in a Hamiltonian context is 
to choose the $A_0=0$ gauge.  In this gauge, the electric field
$E_i(x)=E_i^a(x) \lambda_a$ is canonically conjugate to the vector
potential $A_i^a(x)\lambda_a$:
\be 
[E^a_i(x), A^b_j(y)]= \frac{1}{i} \delta_{i j} \delta_{a b} \delta(x-y)
\quad ,
\ee
with $\lambda_a$ a generator of the SU(N) Lie algebra in the
fundamental representation, and normalized as
$\mathrm{Tr}(\lambda_a \lambda_b)=\frac{1}{2}\delta_{a b}$.
The Hamiltonian of the system is given by
\be
H=\int dx \; \left[ {\rm Tr} (E_i(x) E_i(x)) + {\rm Tr} (B^2(x)) \right]
\quad .
\ee

However, this is not quite the end of the story because the
physical Hilbert space does not coincide with the naive Hilbert
space of functionals of the spatial vector potential, due to gauge 
invariance. At the classical level, the equations of motion derived
from this Hamiltonian system only give three of the non-abelian Maxwell equations. 
The additional equation is the, so-called, Gauss-law or Gauss-constraint
\be
D_i E_i(x) =0
\quad ,
\ee
which acts as an initial condition preserved by the remaining equations.
At the quantum level, the Gauss-constraint is imposed as a restriction
on the physical states of the system. It is not hard to see that this 
condition amounts to demanding that physical states are invariant under gauge
transformations continuously connected with the identity.

The previous results apply both on the plane and for our  torus
case.  One can verify that the boundary conditions for the electric field 
are similar to those satisfied by the magnetic field $B(x)$. Hence,
for the first formalism, having constant twist matrices, a similar Fourier
decomposition Eq.~\ref{decomposition} applies as well. All expressions can be 
now expressed in terms of the Fourier coefficients, now transformed
into operators. The canonical commutation relations become
\be
[E_i^\dagger(\vpt), \hat{A}_j(\vqt)]= \frac{1}{i} \delta_{i j}
\delta(\vpt-\vqt)
\quad ,
\ee
and the Hamiltonian is given by 
\be
H = \frac{1}{2} \sum_{\vpt} (E_i(\vpt)E_i^\dagger(\vpt)+
B(\vpt)B^\dagger(\vec{p})) 
\quad .
\ee
Notice that the hermiticity properties of the classical field  implies that 
$\hat{A}^\dagger_j(\vqt)=\hat{A}_j(-\vqt)$ and the same relation
holds for the electric field. Once we are working in Fourier space, it is much more
convenient also to work in the  transverse and longitudinal gluon basis. For 
that purpose we define the vectors
$u_L(\vpt)=\vpt/|\vpt|$ and $u_T(\vpt)=\epsilon\vpt/|\vpt|$.
The matrix $\epsilon$ is the completely antisymmetric tensor with two
indices and $\epsilon_{1 2}=1$. Now we can decompose
$$ \hat A_i(\vpt)= i (u_L(\vpt))_i A_L(\vpt) +  i (u_T(\vpt))_i
A_T(\vpt) \quad , $$
with the same decomposition applying for $E_i(\vpt)$. The presence
of the complex factor $i$ is justified to preserve the property 
$A^\dagger_{L, T}(\vqt)=A_{L, T}(-\vqt)$. On the other hand, the
commutation relations for the longitudinal and transverse gluons still 
maintain the canonical form
\be
[E_T^\dagger(\vpt), A_T(\vqt)] = \frac{1}{i} \delta(\vpt-\vqt) \quad , 
\ee
\be
[E_L^\dagger(\vpt), A_L(\vqt)] = \frac{1}{i} \delta(\vpt-\vqt) \quad .
\ee

Now one can study the spectrum of the theory by means of perturbation theory in
the coupling constant $g$. The standard methodology has to be
supplemented with the Gauss constraint condition, which also depends on
$g$. This will be explicitly carried out in the next section, but here
we will anticipate the result to lowest order. 

To leading order the Gauss constraint amounts to the condition that
the physical states do not depend on the longitudinal vector field. On
this subspace the lowest order Hamiltonian is given by
\be
H_0 = \frac{1}{2} \sum_{\vpt} ( E_T(\vpt)  E^\dagger_T(\vpt)+
|\vpt|^2  A_T(\vpt)  A^\dagger_T(\vpt))
\quad .
\ee
This has the typical spectrum of a collection of free transverse
gluons. To show this, one introduces creation-annihilation operators in 
the standard way  
\bea
A_T(\vpt) &=&\frac{1}{\sqrt{2 |\vpt|}} (
a^\dagger(\vpt)+a(-\vpt))
\quad , \\
E_T(\vpt)&=&i \sqrt{\frac{|\vpt|}{2}}
(a^\dagger(\vpt)-a(-\vpt)) 
\quad ,
\eea
and rewrite the Hamiltonian as:
\be
H_0 = \frac{1}{2} \sum_{\vpt}  |\vpt| (a^\dagger(\vpt) a(\vpt)+
a(\vpt) a^\dagger(\vpt)) 
\quad .
\ee
The energies of the gluons are given by $|\vpt|>0$, so that the
theory has a gap.  

Before describing the lowest lying spectrum at $g=0$, let us re-examine 
the appearance, in this context, of the  $Z_N^2$ symmetry, whose dual is the
mod N electric flux defined by `t Hooft. Specializing the general
formalism constructed in the preamble to the case of constant twist
matrices, we see that gauge transformations belonging to ${\cal
G}(\vec{l})$ have to satisfy:
\be
\Omega(x+ L_i \ev_i) = e^{ 2 \pi i l_i/N} \Gamma_i
\Omega(x) \Gamma_i^\dagger 
\quad .
\ee
We can choose as representative of this space a constant gauge
transformation:
\be
\Omega_{\vec{l}}(x) =\Gamma_1^{\kb l_2} \Gamma_2^{-\kb l_1}
\quad .
\ee
Any other element of ${\cal
G}(\vec{l})$ is obtained by combining this gauge transformation with
an element of  ${\cal G}(\vec{0})$.

An element of the Hilbert space  that transforms under the
operator which implements  these gauge transformations in the
following way:
\be
\mathbf{U}(\Omega_{\vec{l}}) \, |\Psi\rangle = e^{i 2 \pi
\vec{e}\cdot\vec{l}/N} |\Psi\rangle  
\quad ,
\ee
is said to carry electric flux $\vec{e}$. In the same way, one can
assign electric flux quantum numbers to operators acting on Hilbert
space. It is quite obvious with this definition that the vector potential
operators $\hat{A}(\vpt)$ carry electric flux given by its colour 
momentum as follows:
\be
e_i= \kb \epsilon_{i j} n_j = \kb N L_j \epsilon_{i j}
\pc_j/(2 \pi) \bmod N
\quad ,
\ee
or the inverse
\be
\label{electric}
\pc_i= -\frac{2 \pi \m }{N L_i} \epsilon_{i j} (e_j \bmod N)
\quad .
 \ee
Since  electric flux is a conserved quantum number, the Hilbert space
can be split into the direct sum of the $N^2$ subspaces associated to different
values of the electric flux. Notice, that the conservation of electric
flux follows here from momentum  conservation on the vertices.

Focusing on gauge invariant operators, the vector potentials act
simply as representatives of the Polyakov lines. To see this, we
recall the expression of a Polyakov line operator:
\be
\label{eq.pol}
{\cal P}(\gamma)\equiv \mathrm{Tr}\left( T\exp\{ -i g \int_\gamma dx_i A_i(x)\}
\Gamma_2^{\omega_2} \Gamma_1^{\omega_1}\}\right) 
\quad ,
\ee
where $\gamma$ is closed curve on the 2-torus and $\vec{\omega}$ its
corresponding winding number. The symbol $T\exp$ stands for the
path-ordered exponential, where the order of matrix multiplication
follows left-to-right the order of the path. 
If we parameterize the curve in terms of the 
parameter $\tau$ ranging from 0 to 1, we have  a function $x_i(\tau)$
satisfying
\be
x_i(0)\equiv  x^{(ini)}_i  \quad x_i(1)\equiv  x^{(fin)}_i =
x^{(ini)}_i + L_i \omega_i 
\quad ,
\ee
where $\vec{x}^{(ini)}$ and  $\vec{x}^{(fin)}$ represent the initial and final points
of the path. The result does depend on the actual path $\gamma$, but
is reparametrization invariant. 

Now, if  we expand the T-exponential, the leading non-vanishing term for
non-trivial winding is linear in the vector potential. To compute this
term, we use the Fourier  expansion (Eq.~\ref{decomposition}) and decompose
the Fourier coefficients into longitudinal and transverse parts. The result is
\be
 g {\cal N} \sum_{\vpt}  \frac{1}{|\vpt|} \int_0^1 d
 \tau \, (A_L(\vpt) \frac{d u}{d \tau} +  A_T(\vpt) \frac{d v}{d
 \tau}) e^{iu} \,\mathrm{Tr}(\HG(\vpt)\Gamma_2^{\omega_2}
 \Gamma_1^{\omega_1})
\quad ,
\ee
where $u(\tau)=\vpt\cdot\vec{x}(\tau)$ and
$v(\tau)=\vec{x}(\tau)\times\vpt$. The trace can we evaluated and it 
fixes the colour momentum to be the one given by the formula
Eq.~\ref{electric} with $\vec{e}=\vec{\omega}$. With this constraint
the variable $(u(1)-u(0))/(2 \pi)$ becomes an integer. As a consequence, the
term proportional to the longitudinal field becomes the integral of a
total derivative of a periodic function, hence, it vanishes. 

The final result is that the Polyakov loop is, to leading order in
perturbation theory, a linear combination of transverse gluon fields 
with different momenta but with a common $\vpc$ corresponding to
electric flux $\vec{\omega}$. The coefficients depend on the particular
momentum and on the path $\gamma$. A  simple example is  
given by a straight line path 
\be
x_i(\tau)=  x^{(ini)}_i + \tau L_i \omega_i
\quad .
\ee
In this case, the only non-vanishing coefficient corresponds to momentum
$\vqt= -\frac{2 \pi \m }{N L_i} \epsilon_{i j} \omega_j$. For this
momentum value $u(\tau)=0$. Hence, the computation is quite simple, and
the result becomes 
\be
{\cal P}(\gamma)= \sqrt{\frac{\lambda}{2}} {\cal N} l(\gamma) e^{i \vqt\vec{x}^{(ini)}}
e^{i \alpha(\vqt)}\, A_T(\vqt) 
\quad ,
\ee
where $l(\gamma)=\sqrt{L_1^2 \omega_1^2+ L_2^2\omega_2^2}$ is the
length of the straight line path. Notice that for $L_1=L_2$ the prefactor
multiplying $A_T$ becomes $\sqrt{\lambda}||\vec{\omega}||/\sqrt{2}$.

From these considerations   it is easy to extract and interpret the spectrum to 
zeroth order in perturbation theory. The first excited states over the
vacuum correspond to single gluon states with momenta
$\vpt=(\pm\frac{ 2 \pi }{L_1 N},0)$ and $\vpt=(0,\pm\frac{ 2 \pi
}{L_2 N})$. These states carry electric flux and are, therefore, the
states with minimal energy within their corresponding sector. 
Since, electric flux is a good quantum number there is no mixing among these
states. In the next section we will compute the contribution to these
energies to the next order in perturbation theory, adopting the form
of a gluon self-energy graph. 

In general, in a given electric flux sector there will be single gluon
state having the minimal energy within each sector. However, it is
easy to see that there are always multiple gluon states which are
degenerate in energy, for example those made up of an appropriate
number of minimal energy gluons. Notice, that at large volumes the
energies of the electric flux sectors should grow linearly with the
torus length giving rise to the string tension and the k-string
spectrum.  

The most important sector is that with vanishing electric flux, since 
it remains light in the infinite volume limit. The vacuum belongs to
this space. In perturbation theory,
the first excited state  in this sector  above the vacuum is a two-gluon state, 
in which the gluons have minimal and opposite momenta. For $L_1=L_2$, there 
is a two-fold degeneracy of levels, which is broken at higher orders producing states that 
can be classified according to the representations of the
cubic group in 2-d ($Z^4$). This being abelian, all representations are one
dimensional. For large volumes these give rise to the glueball spectrum.

\section{Perturbative calculations of the mass spectra}

\newcommand{\E}{\mathcal{E}}
\subsection{Perturbative calculation in the Hamiltonian approach} 
\label{s:hamiltonian}
Here we will give our derivation of the corrections to the energy
levels of the 2+1 Yang-Mills field theory in a finite 2-torus with twist
within the Hamiltonian framework. Our starting point is the quantized
version of the theory in the $A_0=0$ gauge, explained in the previous
section. Eventually, however, our calculation will involve only
transverse gluons, so that it can be considered to reside in the
Coulomb gauge. Thus, the paragraphs  that will follow can be regarded as 
a derivation for the Hamiltonian formulas  in the Coulomb gauge. Expressions
coincide with  those appearing in the literature\cite{lee}. 

Let us first describe  the perturbative construction in an schematic way.
For that purpose, we write the Hamiltonian as 
\be
H= \frac{1}{2} E_L^2+ H_0+ g H_1+ g^2 H_2 \quad ,
\ee
where $E_L$ denotes the longitudinal electric field and $H_0$ is 
the lowest order Hamiltonian given in the previous section and
depending on transverse gluons only. 
On the other hand, we can expand the  Gauss constraint operator $G$ in
powers of $g$:
\be
G=E_L+ g \delta G \quad .
\ee

For a given eigenstate, we can expand the energy and wave function in powers
of $g$:
\be
\E=\E_0+ g^2 \delta \E + \ldots \quad , 
\ee
\be
\Psi= \Psi_0+ g \Psi_1+g^2 \Psi_2+\ldots  \quad . 
\ee
Now we look for eigenstates of $H$ which are simultaneously annihilated 
by the $G$. To lowest order we have states which only depend on
transverse gluons and verify
\be
H_0 \Psi_0= \E_0 \Psi_0 
\quad .
\ee
These were studied in the previous section.

To the next order, the Gauss constraint imposes
\be
E_L \Psi_1= -\delta G \Psi_0 
\quad .
\ee
Obviously, this formula implies that  $\Psi_1$ depends on 
longitudinal gluon fields as well. 
Applying $E_L$ to both sides one gets 
\be
E_L^2 \Psi_1= -[E_L,\delta G] \Psi_0 
\quad ,
\ee
where we have used the fact that $\Psi_0$ only depends on transverse
gluons and is therefore annihilated by $E_L$. 
Now we are ready to look at the eigenvalue equation to order 
$g$. We get 
\be
(\E_0-H_0)\Psi_1= \left( H_1- \frac{1}{2} [E_L, \delta
G]\right) \Psi_0
\quad .
\ee
Both sides of the equation are polynomials in the longitudinal 
vector potentials $A_L$. The equality must be satisfied for the
coefficients. In particular, the purely transverse part of both sides  
have to match. We might introduce for that purpose the projector 
${\cal P}_T$ onto the purely transverse part (equivalent to setting
$A_L=0$). Then, if we define $\Psi_1^T\equiv {\cal P}_T\Psi_1$.
and take it to be  orthogonal to $\Psi_0$, we may write 
\be
\Psi_1^T= (\E_0-H_0)^{-1}{\cal P}_T \left( H_1- \frac{1}{2} [E_L, \delta
G]\right) \Psi_0 
\quad .
\ee

To second order, the Gauss constraint equation gives
\be
E_L \Psi_2=-\delta G \Psi_1
\quad ,
\ee
and the eigenvalue equation 
\be
\delta \E \Psi_0+ \E_0 \Psi_2= H_0 \Psi_2+ \frac{1}{2} E_L^2
\Psi_2+ H_2\Psi_0+H_1 \Psi_1 
\quad .
\ee
Again the equation should hold  at all values of $A_L$. In particular,
it also holds when restricting both sides to the transverse gluon space.
Our goal is to obtain $\delta \E$, and this can be done by projecting 
both sides of the restricted equation onto the $\Psi_0$ state:
\be
\label{deltaE}
\delta \E= \Psi_0^*  {\cal P}_T H_2\Psi_0 + \Psi_0^* {\cal P}_T H_1'
(E_0-H_0)^{-1} H_1' \Psi_0 + \frac{1}{2}\Psi_0^*
{\cal P}_T (\delta G)^2 \Psi_0
\quad ,
\ee
with $H_1'=  {\cal P}_T (H_1 - \frac{1}{2} [E_L, \delta G])$. We have used the symbolic 
notation $\Psi_0^* A \Psi_0$ to mean the 
matrix element of the operator $A$ on the lowest order wave function. 
The derivation has been very schematic, but it properly reflects the fact
that the energy is a sum of three terms, each corresponding  to a
possible diagram. 

Notice  that the final calculation only involves
transverse gluons, and that we have avoided defining longitudinal
gluons (as particles) at all. We have also avoided using any scalar
product in the space of functionals containing longitudinal vector
potentials: only the transverse gluon space is a Hilbert space. 
The longitudinal field $A_L$ appears explicitly in the wave-function 
and $E_L$ acts as a derivative with respect to this field. Once the
appropriate derivatives are taken, the projection ${\cal P}_T$ amounts
to setting the remaining $A_L$ to zero. Nevertheless, the perturbative
formulas that we have obtained with our procedure   differ
from those obtained by setting the  longitudinal vector potential
directly to zero in the original Hamiltonian. The last term, for
example, would not have appeared. Indeed, this modified Hamiltonian coincides 
with the one obtained for the Coulomb gauge formulation in the
literature~\cite{lee}.

The next step in the calculation will be to write down explicitly the form of the 
operators $H_1$, $H_2$ and $\delta G$ in terms of the components 
of the vector potentials. For example $\delta G$ becomes
\begin{eqnarray}
\nonumber 
\delta G(\vec{P})&=&  {\cal N} \sum_{\vqt}
\frac{F(-\vpt,\vqt,\vpt-\vqt)}{|\vpt||\vqt||\vpt-
\vqt|}\times
\\
\label{deltaGform}
&\{&\vqt\cdot(\vpt-\vqt)\left(A_L(\vqt)E_L(\vpt-\vqt)+
A_T(\vqt)E_T(\vpt-\vqt)\right)
\\
\nonumber
&-&(\vqt\times\vpt)\,\left(A_T(\vqt)E_L(\vpt-\vqt)-
A_L(\vqt)E_T(\vpt-\vqt)\right)  \} \quad .
\end{eqnarray}
Notice that the factor $F$ vanishes if two of its arguments are
collinear. This brings in an important simplification of the previous
formulas, since in implies that the commutator  $ [E_L,\delta G]$
appearing in the expression of $H_1'$ vanishes. 
This is so because one has to differentiate the previous formula with
respect to $A_L(\vpt)$, giving a  $\delta(\vqt-\vpt)$ factor
multiplying $F$.

It is clear that expressions like Eq.~\ref{deltaGform} are rather
lengthy and hard to work with. For that reason we will introduce some
new notation that will simplify the manipulation and presentation of
the remaining formulas. In particular the symmetry properties under
exchange of momenta plays an important role. Thus, we will rename 
the three momenta that are arguments of $F$ as $\vpt_{(i)}$, 
$\vpt_{(j)}$ and $\vpt_{(k)}$ instead of $-\vpt$, $\vqt$ and
$\vpt-\vqt$ respectively. To label the terms associated to these 
three  discrete momenta arguments we will simply use $i$, $j$ and $k$.
Furthermore, the term corresponding to momentum $-\vpt_{(i)}$ will be labelled
$\bar{i}$. With this notation we can combine the $F$ factor with others 
introducing the three index tensor 
\be
\lambda_{i j k} =  {\cal N}\,
\frac{F(\vpt_{(i)},
\vpt_{(j)},\vpt_{(k)})}{|\vpt_{(i)}||\vpt_{(j)}||\vpt_{(k)}|}\,
\delta(\vpt_{(i)}+\vpt_{(j)}+\vpt_{(k)})
\quad ,
\ee
which is totally antisymmetric with respect to the exchange of its
indices. In addition we introduce the following  two-index tensors
\be
 S_{j k}= \vpt_{(j)}\cdot \vpt_{(k)} \quad \quad A_{j
k}=\vpt_{(j)}\times \vpt_{(k)}  
\quad ,
\ee
which are symmetric and antisymmetric respectively. Finally,  we introduce
an index $\alpha$ taking two values $T$ and $L$ and two
2$\times$2 matrices  $\delta$ (the $2\times 2$ identity matrix) 
and $\epsilon=i\sigma_2$. 

With the help of this notation we can rewrite Eq.~\ref{deltaGform}
as follows
\be
 \delta G_{\bar{i}} = \lambda_{i j k} \,(S_{j k} A_j^\alpha
E_k^\alpha- A_{j k} \epsilon_{\alpha \beta} A_{j}^\alpha E_k^\beta) 
\quad ,
\ee
where the vector potential and electric field components have been
renamed in an obvious way (for example $A_j^1\equiv
\hat{A}^T(\vpt_{(j)})$). The ability of the notation to condense the
formulas is obvious.

We can proceed in the same way with the magnetic field. Using the 
natural symbol $B_{\bar{i}} \equiv \hat{B}(-\vpt_{(i)})$ we obtain
\be
B_{\bar{i}} = |\vpt_{(i)}| \,( A_{\bar{i}}^T- \frac{g}{2}\, \lambda_{i j
k}\, ( A_{j k} A_j^\alpha A_k^\alpha + S_{j k} \epsilon_{\alpha \beta}
A_j^\alpha A_k^\beta))
\quad .
\ee
Indeed, according to our previous derivations we only need to use the
transverse part of $B_{\bar{i}}$  to compute the transverse part 
of $H_1$ and $H_2$. Notice that the last term (involving $S_{j k}$)
would not contribute to this  transverse part.

The last step would be to express  the transverse vector potential 
and electric field in terms of creation and annihilation operators
using the formulas given before. As an example we show below the 
expression of  the $O(g)$ part of the magnetic field  $B_{\bar{i}}$
\be
-\frac{g}{4} \sum_{j k}   \frac{\lambda_{i j k}}{
\sqrt{|\vpt_{(j)}||\vpt_{(k)}|}}
\, A_{j k}\, |\vpt_{(i)}| \,
\left( a_k^\dagger a_j^\dagger + 2 a_k^\dagger a_{\bar{j}}+
a_{\bar{k}} a_{\bar{j}} \right)
\quad ,
\ee
where $a_k^\dagger\equiv a^\dagger(\vpt_{(k)})$ and so on.  

A little more care is required to deal with the operator ${\cal P}_T (\delta G)^2
{\cal P}_T$. One cannot directly set the longitudinal components to
zero,   because the longitudinal electric field acts on the longitudinal vector
potential to produce a term of the form
\be
i \sum_{i,j,k} \lambda^2_{i j k} A^2_{j k}  A_{\bar{k}}^T E_k^T
\quad .
\ee
However, the matrix element of this operator between one state and
itself gives only a constant shift of all the levels, including the
vacuum energy. Hence, finally, it is also enough to keep only the 
purely transverse part of $\delta G_{\bar{i}}$ given by:
\be
\frac{i}{4} \sum_{j k}   \frac{\lambda_{i j
k}}{\sqrt{|\vpt_{(j)}||\vpt_{(k)}|}}
\, S_{j k} \, \left( 2 (|\vpt_{(k)}|+|\vpt_{(j)}|) a_k^\dagger
a_{\bar{j}}+
(|\vpt_{(k)}|-|\vpt_{(j)}|) (a^\dagger_k a^\dagger_j+
a_{\bar{j}} a_{\bar{k}})\right)
\quad .
\ee

Now we have all the ingredients to compute the three operators
that enter the  $O(g^2)$ contribution to the spectrum. In the following 
sections we will use these formulas to compute the corrections to 
the lowest order energies.

\subsubsection{Self-energy}
We can apply the previous formulas to the computation of the
self-energy of the gluons, which provides the leading correction to the
electric-flux energies. The  result  is the sum of three terms
corresponding to those appearing in Eq.~\ref{deltaE}. With our symbolic notation
the calculation is simple. 

The last term in  Eq.~\ref{deltaE} is associated to the additional
term in the Coulomb gauge Hamiltonian. For the self-energy we need only 
to express the part of that operator having one creation and one
annihilation operator. Making use of our previous expression 
for $\delta G_i$ we get 
\be
 \frac{1}{2} \sum_i \delta G_i \delta G_{\bar{i}} = \sum_k a^\dagger_k
 a_k \left(\frac{1}{4}\sum_{i j} \lambda^2_{i j k} S^2_{j k}
 \frac{(|\pt_{(j)}|^2+|\pt_{(k)}|^2}{\pt_{(j)} \pt_{(k)}}\right)+\ldots
\quad .
 \ee
From this formula and our previous definitions one can read out the
expression of the contribution to the self-energy from this term
(labelled  $\Sigma^{(3)}(\vpt)$):
\be
\Sigma^{(3)}(\vpt)= \frac{g^2 {\cal N}^2}{4} \sum_{\vqt}'
\frac{F^2(\vpt,\vqt, -\vpt-\vqt)\,
(\vpt\cdot\vqt)^2\, (|\vpt|^2+ |\vqt|^2)}
{|\vpt|^3|\vqt|^3|\vpt+\vqt|^2}
\quad .
 \ee
 
 The contribution of  $H_2$ to the self-energy  can be obtained by 
 selecting the part containing one creation and one annihilation
 operator given by
 \be
 H_2= \sum_k a^\dagger_k  a_k \left(\frac{1}{4}\sum_{i j} \lambda^2_{i
 j k} A^2_{j k} \frac{|\vpt_{(i)}|^2}{|\vpt_{(j)}| |\vpt_{(k)}|}\right)+\ldots
\quad .
 \ee
This leads to the $\Sigma^{(1)}(\vpt)$ contribution :
\be
\Sigma^{(1)}(\vpt)=\frac{g^2 {\cal N}^2}{4} \sum_{\vqt}'
\frac{F^2(\pt,\vqt, -\vpt-\vqt)\,
(\vpt\times\vqt)^2}{|\vpt|^3|\vqt|^3}
\quad .
\ee

Finally we have  the part involving $H_1$. The transverse part of
$H_1$ can be expressed in terms of creation-annihilation operators as
follows:
\be
H_1^T= -\frac{|\vpt_{(i)}|^2}{2} \lambda_{i j k} A_{j k} A_i^T A_j^T A_k^T=
-\frac{ \lambda_{i j k} A_{j k}|\vpt_{(i)}|^2}{4 \sqrt{2|\vpt_{(i)}|
|\vpt_{(j)}| |\vpt_{(k)}|}} \,
(a^\dagger_i+a_{\bar{i}})(a^\dagger_j+a_{\bar{j}})(a^\dagger_k+a_{\bar{k}})
\quad .
\ee
Plugging this expression on the second term of Eq.~\ref{deltaE} one
gets 
\bea
\Sigma^{(2)}(\vpt)&=&\frac{ \lambda^2_{i j k} A^2_{j k}}{16
|\vpt_{(i)}| 
|\vpt_{(j)}| |\vpt_{(k)}|} (|\vpt_{(i)}|^2 +
|\vpt_{(j)}|^2 + |\vpt_{(k)}|^2)^2 \times \\
&\Big (&\frac{1}{|\vpt_{(i)}| -
|\vpt_{(j)}| - |\vpt_{(k)}|}-\frac{1}{|\vpt_{(i)}| + 
|\vpt_{(j)}|+ |\vpt_{(k)}|} \Big )\nonumber
\quad ,
\eea
which after undoing our notation becomes
\bea
\nonumber
\Sigma^{(2)}(\vpt)= -\frac{g^2 {\cal N}^2}{8} \sum_{\vqt}'
\frac{F^2(\vpt,\vqt, -\vpt-\vqt)\,
(\vpt\times\vqt)^2 }{|\vpt|^3|\vqt|^3}\times \\
\frac{(|\vpt|^2+|\vqt|^2+|\vpt+\vqt|^2)^2(|\vqt|+|\vpt+\vqt|)}{
|\vpt+\vqt|^3((|\vqt|+|\vpt+\vqt|)^2-|\vpt|^2)}
\quad .
\eea

The final result is then the sum of the three contributions
$\Sigma^{(1)}(\vpt)+\Sigma^{(2)}(\vpt)+\Sigma^{(3)}(\vpt)$ which are
all individually divergent. We can combine them to give
\be 
g^2 \delta \ET(\vpt) =
\frac{g^2 {\cal N}^2}{4} \sum_{\vqt} F^2(\vpt,\vqt,-\vpt-\vqt) (A_1
(\vec p \cdot \vec q)^2 +
A_2 |\vec p \times \vec q|^2 )
\quad ,
\ee
with
\be
A_1 = {|\vpt|^2 + |\vqt|^2 \over |\vpt+\vqt|^2|\vpt|^3|\vqt|^3}
\quad ,
\ee
and
\be
A_2=  {1 \over 2 |\vpt|^3|\vqt|^3} \Big ( 2 - { (|\vpt|^2 + |\vqt|^2 +
|\vpt+\vqt|^2)^2 (|\vqt|+|\vpt+\vqt|) \over |\vpt+\vqt|^3
((|\vqt|+|\vpt+\vqt|)^2-|\vpt|^2)}\Big )
\quad .
\ee
Although, the sums over momenta are divergent, we might perform a
subtraction at a given value of $\theta$ and evaluate numerically
these sums with a sharp momentum cut-off. Performing the subtraction 
at a non-zero value of $\theta$ and then taking increasing values for
the cut-off, we obtain a smooth function of $\theta$
vanishing at the subtraction point. 

Forgetting about the divergent character of the sums involved, we can simplify 
the final expression considerably using the following
identities:
\be
{2 |\vec p \times \vec q|^2 \over (|\vec q|+|\vec p+\vec q|)^2-|\vec p|^2} = 
{|\vec q| |\vec p +\vec q| \sin^2 \phi \over  (1+\cos \phi)} = 
|\vec q| |\vec p +\vec q| -  \vec q \cdot (\vec p +\vec q) 
\quad ,
\ee
\be
(|\vec q|+|\vec p+\vec q|) (|\vec q| |\vec p +\vec q| - \vec q \cdot (\vec p +\vec q) ) 
= |\vec q| \ (\vec p \cdot (\vec p +\vec q) )- |\vec p+\vec q| \ (\vec p \cdot \vec q)
\quad ,
\ee
and
\be
(|\vec p|^2 + |\vec q|^2 + |\vec p+\vec q|^2)^2 = 4 (|\vec p+\vec q|^2 
- \vec p \cdot \vec q)^2
\quad .
\ee
Making use of them one obtains 
\be 
|\vec p \times \vec q|^2 \ A_2=  {|\vec p \times \vec q|^2 \over
|\vpt|^3|\vec q|^3 } - {(|\vec p+\vec q|^2 - \vec p \cdot \vec q)^2
\over |\vpt|^3 |\vec q|^2 |\vec p+\vec q|^2 } \Big ( {\vec p \cdot (\vec p +\vec q) \over 
|\vec p+\vec q| }- {\vec p \cdot \vec q \over
|\vec q| }
\Big )
\quad .
\ee
We can now do the following change of variables to the second term on
the right hand side of the previous equation
\be
\vqt \longrightarrow  -\vpt-\vqt
\quad .
\ee
obtaining:
\be 
|\vec p \times \vec q|^2 \ A_2= {1 \over |\vec p|^3} \Big \{ 
{|\vec p \times \vec q|^2 \over |\vec q|^3 } +
{2 (\vec p \cdot \vec q) (|\vec p+\vec q|^2 - \vec p \cdot \vec q)^2
\over |\vec q|^3 |\vec p+\vec q|^2 }  
 \Big \}
\quad .
\ee

Plugging this in the expression of the energy and after some trivial manipulation one arrives at: 
\be
g^2 \delta \ET(\vpt) =\frac{g^2 {\cal N}^2}{4} \sum_{\vqt}' F^2(\vpt,\vqt,-\vpt-\vqt)
\Big \{
{1 \over |\vec p| |\vec q|} + 2 (\vec p \cdot \vec q){ |\vec p|^2 + |\vec q|^2 \over 
|\vec p|^3 |\vec q|^3 }
\Big\}
\quad .
\ee
The second term is odd over $\vqt$ and should vanish when summed over. 
Hence, our final result becomes 
\be
g^2 \delta \ET(\vpt) = \frac{g^2 {\cal N}^2}{4|\vec p|} \sum_{\vqt}' 
\frac{F^2(\vpt,\vqt,-\vpt-\vqt)}{|\vec q|} 
\quad .
\ee
Although the expression is also divergent, we have verified that
performing the same  subtraction as before, at some value of $\theta$,
one gets a finite result which coincides the previous one. This justifies the
validity of our manipulations, at least for the finite subtracted
piece. 

The question now is: Does gauge invariance dictate what is the correct 
subtraction to do? Rather than trying to resolve the puzzle within
the Hamiltonian formulation, we have followed  the   standard procedure to 
introduce a gauge invariant regularization which respects the Ward
identities. This has driven us to the derivation of  the gluon
self-energies within an euclidean context. This will be addressed  
in  the following two subsections.

\subsubsection{Energy levels in the zero electric flux sector}

The Hamiltonian formalism presented previously can be used also to
study the sector of vanishing electric flux. The vacuum belongs to 
this sector. In the large volume limit, the  spectrum of excited states 
should have  a well defined limit, corresponding to the spectrum of glueballs.
To lowest order in perturbation theory the first
excited state over the vacuum corresponds to a pair of minimum momentum 
gluons of opposite sign $|\vpt,-\vpt\rangle$ with energy $2|\vpt|$. The
order $g^2$ corrections to these energies can be computed using the
Coulomb gauge Hamiltonian  derived in the previous subsection. For the
sake of the calculation it is convenient to arrange this Hamiltonian
into a sum of normal ordered operators. The constant piece is
irrelevant, while the $a^\dagger a$ term adds the  self-energy
contribution of each of the two gluons. Thus, up to this level, the 
energy of the two-gluon state is twice the energy of one gluon, and
there is no interaction energy. To get an interaction energy, one needs to 
consider the part of the Hamiltonian having 2 creation and two
annihilation operators. Such a term could come from  $H_2$, from the 
$(\delta G)^2$ term or from the piece quadratic in $H_1$. However, 
it is easy to see that the matrix element of these operators between 
two identical two-gluon states vanishes. This is due to the presence of 
the $F$ symbol which vanishes if two of its arguments are collinear 
vectors. This is unavoidable for a state in which all the external
momenta are collinear. 

There is one situation in which the Hamiltonian formalism produces 
an interaction term at order $g^2$. This occurs in case of degeneracy
of levels. For example, if $L_1=L_2$ there are two minimum momenta, 
$\vpt_{(1)}\equiv\frac{2 \pi}{N L}(1,0)$ and  $\vpt_{(2)}\equiv\frac{2 \pi}{N
L}(0,1)$, with equal energy. The corresponding zero-momentum 2-gluon states 
will be labelled $|1\rangle$ and $|2\rangle$. The relevant calculation will
then be 
\be
\ET_I= \langle  1| H_{\tiny \rm eff} |2\rangle
\quad ,
\ee
where $\ET_I$ is the interaction energy of  the two-gluons and
$H_{\tiny \rm eff}$ is the effective Hamiltonian at this order (sum of
three contributions). Using
the rotational invariance of the Hamiltonian, one sees that the new 
eigenstates of the Hamiltonian are
$|\pm\rangle\equiv\frac{1}{\sqrt{2}}(|1\rangle\pm|2\rangle)$ with
interaction energies $\pm \ET_I$ respectively.

One can use the general formulas given before to compute the
contribution  to the interaction energy coming from the three terms that 
make up the Hamiltonian at this order. The $\delta G$
term gives a vanishing contribution.
The contribution to $\ET_I$ coming from the $H_2$ part becomes
\be 
\frac{\lambda}{4 \pi^2} \sin^2(\tilde{\theta}/2) \quad , 
\ee
and that from the term quadratic in $H_1$ is
\be
-\frac{\lambda}{ \pi^2} \sin^2(\tilde{\theta}/2) \quad ,
\ee
where we have introduced a new angle
\be
\label{defthetatilde}
\tilde{\theta}\equiv \frac{2 \pi \kb}{N}
\quad ,
\ee
which depends on the magnetic flux $\m$ of the box through the coprime 
integer $\kb$.
The sum of these two contributions is negative, and this implies that the state of
minimum energy is the rotationally  invariant state. 

We will not proceed any further. Although, the  vanishing electric
flux sector is very interesting, the numerical study to be presented 
in the following section has focused on the  non-zero sector. The
latter is simpler to study, and looks like the obvious first step 
in understanding the dynamics of the system, and the transition from 
small to large volumes. Nevertheless, the formulas given in this paper 
would enable an straightforward application to other states at this 
and higher orders. 
\subsection{Perturbative calculation in the Euclidean approach}
\label{s:euclidean}

In the previous subsection we developed the perturbative expansion in
the Hamiltonian formulation. The self-energy contribution of the gluon 
is a sum of individually divergent diagrams. Applying a sharp cut-off 
regularization breaks gauge invariance, hence our result could only be
fixed up to an additive constant. To fix this constant one needs a
gauge invariant regularization. This will be done in this section
using the more conventional Euclidean approach. Indeed, we will use two 
different well-known gauge invariant regularization procedures. In the
first subsection we will use a continuum formulation and dimensional
regularization. In the following one we will use a lattice
regularization. We will show that the continuum limit is well defined
and gives identical results to the ones obtained by dimensional
regularization. The agreement of our three procedures in what respects 
the electric flux dependence of the self-energy provides a very strong 
verification of the validity of our calculation. 

In the following paragraphs we will explain the general procedure for 
extracting the self-energy within the Euclidean approach. Then, in the 
next two subsections we will address the computation of the self-energy 
in the continuum and on the lattice respectively. We opted by 
focusing on the calculation itself  within
the text, but for completeness we added the necessary background
material in  appendices~\ref{appendixb} and \ref{appendixc}.  The readers are also invited to
consult previous references which address similar perturbative
calculations with twisted boundary conditions~\cite{korthals,daniel},
\cite{Snippe:1996bk} - \cite{Luscher:1985wf}. 

A gauge invariant definition of the gluon self-energies follows by
analyzing the exponential decay at large times of Polyakov-loop
correlators. As explained earlier the winding number of the loop
coincides modulo $N$ with the electric flux. We will sacrifice
generality to clarity and consider only straight loops winding $e_1$ times 
around the $x$ direction. 
Their expression is a particular case of the
general formula given before and reads
\be 
\label{eq.polx}
{\cal P}_1 (t,y; e_1) = {\rm Tr} \Big [\,\prod_1^{e_1} {\rm T \exp}
\Big (-ig\int_0^{L_1} A_1(x) dx \Big )  \Gamma_1  \Big ] \quad .
\ee
We recall that these operators carry electric flux $\vec e=(e_1,0)$.
Notice that they do not depend on $x$, but only on $y$ (and $t$).
Furthermore, it follows from the general formalism explained in
section~\ref{method} that,  in the twisted box, they are not periodic under
translations by one period in the $y$ direction:   
\be   
{\cal P}_1(t,y+ L_2; e_1) = e^{i {2 \pi \m e_1\over N}}  {\cal P}_1 (t,y; e_1) \quad,
\ee
where $\m$ is the magnetic flux. Hence, they can be Fourier expanded  
as a sum of momenta of the form $p_2L_2N/(2\pi)=\m e_1 \bmod N$. This
is just a particular case of the relation between momenta and electric
flux of gluon fields explained in the previous section.

Now when computing the correlation of two Polyakov loops in
perturbation theory we have to expand the ordered exponential as we did 
in the previous section after Eq.~\ref{eq.pol}. To the order we are
working this turns out to be proportional to the correlator of two
transverse gluon fields. This is the Fourier transform of the
transverse gluon propagator written in momentum space. The exponent 
of the exponential decay in time --the energy of the state-- is given
by the poles of the Euclidean propagator: $p_0=i\ET$. To determine 
the pole, one uses the formula for the inverse propagator obtained by 
resuming the Lippmann-Schwinger series
\be
D_{\mu \nu }^{\, (-1)}  (\p)= P_{\mu \nu }^{\, (-1)}  (\p) - \Pi_{\mu \nu } (\p)\quad ,
\ee
where $P_{\mu \nu }$ is the tree-level propagator given in 
Eq.~(\ref{eq:prop}), and $\Pi_{\mu \nu }$ is the vacuum polarization. 
Thus, the energy is determined imposing that the inverse propagator, 
projected to transverse components, vanishes for Euclidean momentum
$\p=(i \ET(\vpt), \vpt)$.
The resulting energy  $\ET(\vpt)$ satisfies the following
dispersion relation:
\be
\ET^2(\vpt)\equiv  \vpt^2 + g^2 \delta \ET^2(\vpt)  = \vpt^2- \sum_{\mu \nu} \varepsilon_\mu \Pi_{\mu \nu }^{\rm on-shell}(\vpt) 
\varepsilon_\nu^*
= \vpt^2- \sum_{\mu} \Pi_{\mu \mu }^{\rm on-shell}(\vpt)\quad ,
\label{eq:deltaE2}
\ee
where the last expression on the right hand side holds if the Ward identity 
($\p_\mu \Pi_{\mu \nu }(\p) = 0$) is preserved by the regularization.

The minimum energy within each electric flux sector is obtained by 
taking the minimal value of Eq.~\ref{eq:deltaE2} as $\vpt$ runs over all
of the allowed momenta. For the particular case that we were studying 
$\vec e =(e_1,0)$, we have $p_1=0$ and $p_2L_2N/(2\pi)=\m e_1 \bmod N$. 

Notice, that in this construction one naturally obtains a formula for 
the square of the energy. To the order we are working the 
expression for the energy itself becomes 
\be
\ET(\vpt)= |\vpt| + {1 \over 2 |\vpt|} \, g^2\delta \ET^2(\vpt)
\quad .
\ee
which can be directly compared with the result of the Hamiltonian
formulation.

\subsubsection{The Euclidean self-energy correction in the continuum}

In this subsection we will apply the previous prescription to
determine the corrections to the energies of the gluons. For simplicity 
we will take the box to be symmetric $L_1=L_2\equiv L$. The starting
point is the expression of the vacuum polarization $\Pi_{\mu\nu}$ to
one-loop order. The derivation is given in Appendix \ref{appendixb}. 
It follows the standard procedure with minor modifications associated
to the boundary conditions. The final result in the Feynman gauge reads:
\bea
\Pi_{\mu \nu} (\p)&=&
{ g^2 {\cal N}^2 \over 2} \int {d \q_0 \over 2 \pi} \sum_{\vqt}
{F^2(\p,\q,-\p-\q)  \over q^2 (\p+\q)^2} \times
\\
&&\Big((2 \q_\nu - 3 \p_\nu) (\p_\mu + 2 \q_\mu)
- 2  \delta_{\mu \nu} \q^2
+ 8 (\q_\mu \p_\nu - \delta_{\mu \nu} \ \p\cdot \q)  \Big )\quad .
\nonumber
\eea
It is important to take into account the  Ward identity $\p_\mu
\Pi_{\mu\nu}(\p)=0$ to guarantee gauge invariance. Using the
expression of the vacuum polarization we obtain 
\be
\p_\mu \Pi_{\mu\nu}(\p) = {g^2 {\cal N}^2 \over 2} \! \int {d \q_0 \over
2 \pi} \sum_{\vqt} F^2(\p,\q,-\p-\q)
\Big \{  {\q_\nu \over \q^2} - {\p_\nu+\q_\nu \over (\p+\q)^2}
\!+\! \frac{3\p_\nu}{2} \Big( {1 \over \q^2} - {1 \over (\p+\q)^2}\Big ) \Big  \}
\ee
We have written it in a way in which it is obvious that it vanishes by
making use of the translation invariance of the measure
$\q\longrightarrow \q+\p$. Thus, the regularization procedure must
respect this property. A sharp cut-off in spatial momenta is not
valid, for example.

After performing  the integration  over $\q_0$, we obtain:
\be
g^ 2\delta \ET^2(\vpt) \equiv \ET^2(\vpt) - |\vpt|^2 =  {g^2 {\cal N}^2
\over 2 |\vpt|^2}\sum_{\vqt} F^2(\p,\q,-\p-\q) 
   (2|\vpt|^2+ \vpt\cdot \vqt ) \Big ( {1\over |\vqt|} - 
{1 \over |\vpt + \vqt|} \Big ).  
\ee
Now using  shift symmetry, which is necessary to preserve the Ward
identity, we arrive at: 
\be
g^ 2\delta \ET^2(\vpt) =  {g^2 {\cal N}^2  \over 2 }\sum_{\vqt}
{F^2(\p,\q,-\p-\q)  
\over |\vqt|}\quad ,
\ee
As mentioned previously, this simple formula coincides with the one 
that can be obtained by manipulating the result of the Hamiltonian
computation.

To evaluate the previous expression we first  use the relation:
\be
\int_0^\infty {d t \over \sqrt{t}} e^{-t A \pi} = {1 \over \sqrt{A}}
\ee
to cast the self-energy in the form:
\be
g^ 2\delta \ET^2(\vpt) =  
{ \lambda   \over 2 \pi N L}\sum_{\vec k} \sin^2\Big({\pi \, \vec k \cdot \vec e \over N}
\Big ) \int_0^\infty {d t \over \sqrt{t}} e^{-t  \pi |\vec k|^2}
\, , 
\ee
where we have set the  loop-momentum $\vqt$ to:
$ \vqt = 2 \pi  \vec k/NL$, and we have used the relation between the external 
momentum and the electric flux to rewrite 
$F\propto \sin (\pi\,  \vec k \cdot \vec e /N)$.
Recalling the definition of the Jacobi $\theta_3$ function~\cite{Tata}:
\be
\theta_3(z,it) = \sum_{k \in {\bf Z}} \exp\{-t \pi k^2 + 2 \pi i k z\}\quad ,
\ee
and using the trigonometric relation $2 \sin^2(a) = (1-\cos(2a))$, we decompose
the self energy calculation in two parts
$\delta \ET^2 = \delta \ET_a^2 + \delta \ET_b^2$ with:
\bea
g^ 2\delta \ET_a^2 (\vpt)&=& {\lambda \over 4 \pi N L } \int_0^{\infty} {dt \over \sqrt{ t}
}
\Big(\theta_3^2(0,i t) - 1\Big) \quad ,\\
g^ 2\delta \ET_b^2 (\vpt)&=& -{\lambda \over 4 \pi  N L } \int_0^{\infty} {dt \over \sqrt{ t}
}
\Big( \theta_3(z_1,i t) \, \theta_3(z_2,i t) - 1\Big)\quad ,
\eea
and  $z_i = e_i / N $.
The first term is independent of the electric flux and ultraviolet (UV) divergent. 
The second one carries all the electric flux dependence and  is UV finite for 
$\vec z \ne \vec 0$. 

To analyze the singularity structure of $\delta \ET_a^2$ one can make use of
the duality  relations of the $\theta_3 $ function:
\be
\theta_3 (z,it) = {1 \over \sqrt{t}} \, e^{-{\pi z^2 \over t}} \sum_{k
\in {\bf Z}} \exp\{-{\pi k^2 \over 
t} + {2 \pi z k \over t}\}\quad.
\ee
Using this, we can rewrite $\delta \ET_a$ as:
\be
g^ 2\delta \ET_a^2 (\vpt)= {\lambda \over 4 \pi N L } \int_0^1 {dt \over \sqrt{ t}} \Big(\sum_{\vec k} 
 \frac{1}{t} \, e^{- {\pi \vec k^2 \over t}}    -1 \Big )
+  {\lambda \over 4 \pi N L }  \int_1^{\infty} {dt \over \sqrt{ t}} \ \Big(\theta_3^2(0,it) -1 \Big) \quad .
\ee
The second term in the r.h.s is finite, but the first one diverges for $\vec k = \vec 0$
as:
\be
\int_0^1 {dt \over t \sqrt{ t}}\quad .
\ee
This integral can be regularized by extending the result to $d$
dimensions and by analytical continuation to $d=2$.
This gives
\be
\int_0^1 dt  \ t^{-{d+1\over 2 }} = {2 \over 1-d} \longrightarrow -2 \quad . 
\ee
With this, we finally obtain a finite expression for the first term:
\be
g^ 2\delta \ET_a^2 (\vpt) = {\lambda \over 4 \pi N L} \int_0^{\infty} {dt \over \sqrt{ t}}
\Big(\theta_3^2(0,it) - 1 - {1 \over t}\Big)\quad .
\ee
Putting everything together we get 
\be
g^2\delta \ET^2 (\vpt) = {\lambda \over 4 \pi N L} \int_0^{\infty} {dt \over \sqrt{ t}}
\Big(\theta_3^2(0,it) -\theta_3(z_1,i t) \, \theta_3(z_2,i t)- {1 \over t}\Big)\quad .
\label{eq:deltaE}
\ee
where $z_i=e_i/N$.

\FIGURE[ht]{\centerline{
\psfig{file=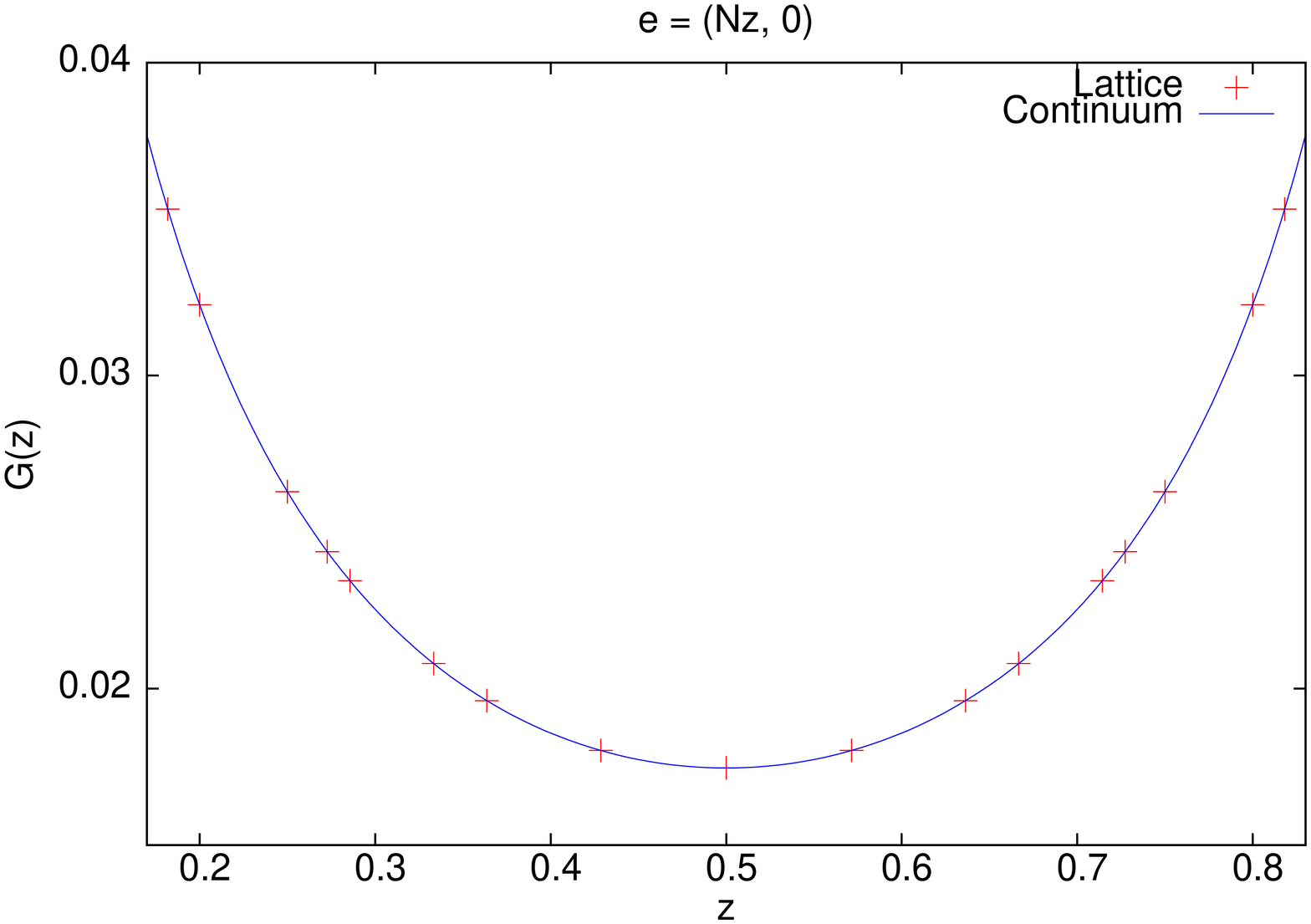,angle=-90,width=7cm}
\psfig{file=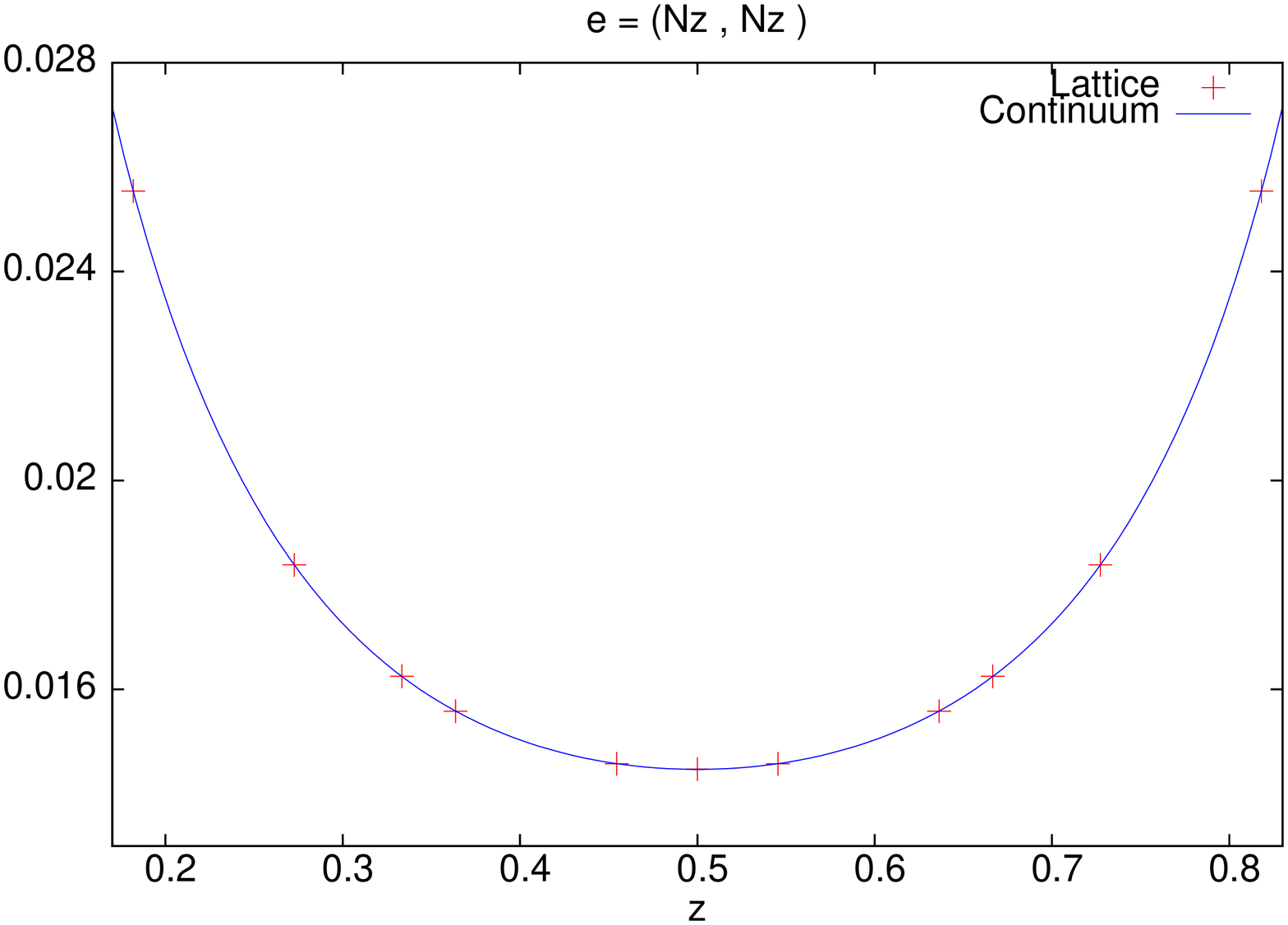,angle=-90,width=7cm}}
\caption{\label{fig.theta}
We show, for electric flux $\vec e =(N z, 0)$ (left),
and $\vec e =(N z,N z)$ (right), the function $G(z)$ that
gives the one-loop correction to the energy of a one-gluon state, 
through Eq.~\ref{elenergform}. The blue line corresponds
to the continuum expression Eq.~\ref{eq:Gtheta}, while the red points are derived 
using a lattice regularization in the calculation of the self-energy.}
}

The final result is quite compact and has many interesting properties
that we want to  comment  about. The first important point is that the
correction does not depend on the value of $\vpt$ but only on the
electric flux. Thus, the minimum energy within each sector corresponds
to the minimum momentum $\vpt_{\rm min}=\frac{2 \pi \vec{n}}{NL}$. We
will write it as follows:
\be
\label{elenergform}
{\ET^2(\vec e)  \over \lambda^2} =  {|\vec n|^2 \over 4 x^2}  
- {1 \over x } \, G \Big ({\vec e \over N}\Big )\quad ,
\ee
where we have introduced the variable $4 \pi x= \lambda N L $, and the function:
\be 
G ({\vec z} ) = -{1 \over 16 \pi^2} \int_0^\infty 
{dt \over \sqrt{ t}} \,
\Big(\theta_3^2(0,it) -\theta_3(z_1,i t) \, \theta_3(z_2,i t)- {1 \over t}\Big)\quad 
.
\label{eq:Gtheta}
\ee
This is a natural way to write it, since in 2+1 dimensions $\lambda$ has
dimensions of energy, and appears as the natural unit. The three terms 
in Eq.~\ref{elenergform} and $x$ are all dimensionless. Since the first
term appearing in the dispersion relation is just the momentum squared,
it is natural to interpret the correction as the mass squared. However, 
it is negative, since $G(\vec{z})$ is positive. As an illustration, we 
display in Fig. \ref{fig.theta} the $\vec z= \vec e /N$ dependence of
$G (\vec z)$ for two different cases: $\vec z= (z,0)$, and $\vec z=  (z,z)$.
Notice that it peaks at $e/N$ close to $0$ and $1$. This feature is
very relevant and will be commented about later. 
The red crosses in the figure result from a calculation of the self-energy 
correction using the lattice regularization that will be presented below. 
It agrees amazingly well with the continuum determination.

\subsubsection{The Euclidean self-energy for the Wilson lattice regularization}

Finally, in this subsection, we will present the calculation of the one-loop Eucliden self-energy 
using a lattice regularization. For the derivation we will make use of the 
results for the four dimensional case, with non-trivial twist on a two-torus, 
that have been obtained previously in \cite{Snippe:1996bk}-\cite{Luscher:1985wf}. 
Without much effort they can be translated into the three dimensional
set up that we are facing here. For the sake of completeness, 
the generalization to 3 dimensions of the vacuum polarization 
derived in \cite{Snippe:1996bk},~\cite{Snippe:1997ru} will be reproduced 
in Appendix~\ref{appendixb}.

The starting point is a 3-dimensional lattice with $N_s$ sites in the 
spatial direction and infinite number of points in time. We will consider a 
discretization of the continuum action based on the Wilson plaquette action:
\be
S_W =   N b \sum_{n \in {\bf Z}^3}  \sum_{\mu \ne \nu} 
\Big ( N - {\rm Tr} P_{\mu \nu } (n) \Big ) \quad ,
\label{eq.wilson}
\ee
where  $b= 1/(\lambda_L a)$, with $a$ the lattice spacing and $\lambda_L$  't Hooft coupling on the lattice. 
$P_{\mu \nu } (n)$ represents the plaquette, written 
in terms of the SU($N$) link matrices $U_\mu(n)$ as:
$$P_{\mu \nu } (n)= U_\mu (n) U_\nu(n + e_\mu) U_\mu^\dagger (n + e_\nu) 
U_\nu^\dagger (n) $$
In the twisted box that we are considering, the spatial links satisfy the boundary conditions
\be
\label{tbclink}
U_i (n + N_s \ev_j) = \Omega_j(n) U_i(n) \Omega_j^\dagger(n+\ev_i) \quad .
\ee
The continuum limit is taken by sending $N_s\rightarrow \infty$ and the lattice spacing $a\rightarrow 0$,
while keeping $L=a N_s$ constant. Note that in the Hamiltonian set up we 
are dealing with, the number of lattice points in the time direction is 
infinite and the momentum is hence cut-off at an UV scale $\pi/a$ for 
both spatial and temporal components.  From now on and for simplicity we will set $a=1$.

For the Wilson lattice action, we can repeat the arguments used in the 
continuum and derive the lattice dispersion relation by imposing that the
inverse lattice propagator, projected over transverse components, vanishes for lattice
momentum $p=(i\ET_L, \vpt)$. This gives: 
\be
\sinh^2( \ET_L /2) = \sum_i \sin^2( \p_i/2) - {1\over 4} \sum_{\mu \nu} \varepsilon_\mu 
\Pi_{\mu \nu }^L(\vpt)
\varepsilon_\nu^* \Big|_{\rm on-shell}
\quad ,
\ee
where, at lowest order in $\lambda$, the lattice on-shell condition amounts to
$p_0 = i \ET^0_L$, with:
\be
\sinh^2(\ET^0_L/2) = \sum_i \sin^2(\p_i/2) \quad .
\ee
The correction to the gluon energy in the continuum can be derived by evaluating  
$\ET_L$ numerically, using the expressions for the lattice vacuum polarization 
presented in Appendix~\ref{appendixc}, and extrapolating the results to the continuum limit 
by taking $ N_s\rightarrow \infty$.

Here we will present a slightly modified version of the lattice 
dispersion relation that differs from the one above in the corrections 
at order $a^2$. The reason to do that is that the corresponding analytic 
formulas simplify considerably.
For that, we will make use of the continuum expression on the right hand-side of
Eq.~\ref{eq:deltaE2} to approximate the lattice dispersion relation by:
\be
\ET_L^2(\vpt) \equiv  
4 \sum_i \sin^2(\p_i/2) - \sum_{\mu} \Pi_{\mu \mu }^L(\vpt)\Big|_{\rm on-shell}\quad .
\ee
Using the formulas for the 
vacuum polarization presented in Appendix~\ref{appendixc}, one can easily 
derive the following expression at leading order in $\lambda_L$:
\bea
&& \delta \ET_L^2(\vpt)  \equiv  -  
\sum_{\mu} \Pi_{\mu \mu }^L\Big|_{\rm on-shell}(\vpt) = \nonumber \\
&=&  {\lambda_L \over 4}   
+   {g^2_L {\cal N}^2 \over  6 } \int_{-{\pi}}^{{\pi }}
{d \q_0 \over 2 \pi} \sum_{\vqt}'\sum_{ \mu} {\FSS\over  \widehat \q^2} 
 \Big ( 
 \widehat \q_\mu^2 
+{ 3\widehat{(\p+2\q)}_\mu^2  - 6 \tilde V^{(3)}_\mu \over 2 \widehat {(\p+\q)}^2}
-2 \tilde V^{(4)}_\mu   \Big ) \Big|_{\rm on-shell} \nonumber \\
&-& {g^2_L {\cal N}^2 \over 4  } \int_{-{\pi}}^{{\pi }}
{d \q_0 \over 2 \pi} \sum_{\vqt}'\sum_{ \mu} \Big ( {1\over 2 N} - {\FSS\over 6}\Big ) 
 \Big ( {\widehat \q_\mu^2 \widehat \p_\mu^2 \over \widehat \q^2}    
\Big ) \Big|_{\rm on-shell} \quad,
\label{eq:deltaEL}
\eea
with
\bea
\tilde V^{(3)}_\mu &=& \cos^2\Big ( { (\p+\q)_\mu \over 2}\Big )  \widehat{(\q - \p)}^2 +  
\cos^2\Big ( { \q_\mu \over 2}\Big ) \widehat{(2\p+\q)}^2 +
\sum_\rho \cos^2\Big ({ \p_\rho \over 2}\Big )\widehat{(2\q+\p)}_\mu^2\nonumber \\
&+&
2 \ \cos\Big ( { (\p+\q)_\mu \over 2}\Big )
\cos\Big ( { \q_\mu \over 2}\Big ) \widehat{(\q-\p)}_\mu \widehat{(2\p+\q)}_\mu \nonumber \\
&-& 2 \ \cos\Big ( { (\p+\q)_\mu \over 2}\Big ) \cos\Big ({ \p_\mu \over 2}\Big ) \widehat{(2\q+\p)}_\mu 
\widehat{(\q-\p)}_\mu\nonumber \\
&-& 2 \ \cos\Big ( { \p_\mu \over 2}\Big ) \cos\Big ({ \q_\mu \over 2}\Big ) \widehat{(2\q+\p)}_\mu
\widehat{(2\p+\q)}_\mu 
\quad , 
\eea
and
\bea
\tilde V^{(4)}_\mu &=& {1 \over 4} \widehat{(\q + \p)}^2 + 3 \cos^2\Big ( { (\q-\p)_\mu\over 2}\Big ) 
-3 \cos( { \q_\mu}) \sum_\rho \cos( { \p_\rho}) \nonumber \\ 
&+& {1 \over 3}  \cos\Big( { \q_\mu \over 2}\Big) \Big \{ \widehat \p_\mu  \widehat{(\q - \p)}_\mu - 
2\  \widehat \p_\mu  \widehat{(\q + \p)}_\mu  - \widehat \q_\mu  \widehat{(2\p)}_\mu\Big \} \nonumber \\
&-& {1 \over 3}  \cos\Big( { \p_\mu \over 2}\Big) \Big \{ \widehat \q_\mu  \widehat{(\q - \p)}_\mu  +
2 \ \widehat \q_\mu  \widehat{(\q + \p)}_\mu  + \widehat \p_\mu  \widehat{(2\q)}_\mu \Big \}\quad .
\eea  
Here, $\widehat \q_\mu = 2 \sin(\q_\mu /2)$, with $\q_i = 2 \pi n_i /N N_s$,
for $n_i =0,\cdots, N_s N-1$, and where the sum over $\vqt$ excludes momenta with
$n_i =0$ (mod $N) \forall i$. 
Note that in the lattice formulation  the linear divergences that arise in
the different contributions to the self-energy cancel away in the sum and the final
expression is UV finite.

To evaluate  Eq.~(\ref{eq:deltaEL}) numerically several steps are in order. 
We first fix the number of colours $N$ and
the value of $\kb$. Then, keeping the number of spatial sites ($N_s$) fixed,
we evaluate the integrals over 
$\q_0$ by discretizing the momenta in units of $2\pi/N_0$.  Following 
Ref.~\cite{Luscher:1985wf}, we improve the convergence of the finite
sum over $\q_0$ by shifting the pole of the propagator through the change of variables:
$$
\q_0 \rightarrow  \q_0 - \half \p_0 - \gamma \sin (\q_0) \quad ,
$$
with $0\le \gamma < 1$, and $\gamma$ tuned close to 1 to improve the convergence of 
the sum. For fixed value of $N_s$, we increase $N_0$ until a stable
result within machine precision is obtained for $ G_L \equiv - x_L \, \delta \ET_L^2 /\lambda_L^2$,
where $4 \pi x_L = \lambda_L N N_s$.
We generate in this way a set of values of $G_L$ for varying $N_s$ and 
extrapolate the results to $N_s \rightarrow \infty$. The extrapolation is obtained from a fit 
of the form:
$$
G_L \Big ({e \over N} \Big )= G \Big ({e \over N} \Big )+ \Big(b_1-b_2 \ln(N N_s)\Big) 
{1\over N^2 N_s^2} + \Big(c_1-c_2 \ln( N N_s) \Big){1\over N^4  N_s^4}  \quad .
$$
As mentioned in the previous section, the continuum-extrapolated results obtained through this procedure 
match perfectly well the ones obtained in the continuum with dimensional regularization.
This is exemplified in Fig.~\ref{fig.theta} for two values of the external momenta. The lines represent the 
continuum results, while the crosses indicate the results derived on the lattice.

\subsection{General comments about the perturbative results and
reduction}
\label{s.remarks}

In the previous subsections we have analysed the perturbative
calculation of the spectrum in our twisted box context. It is
interesting to explore the general properties of this expansion. 
In particular let us  focus on the dependence of our results
on $N$ and $L$. It is clear that in momentum sums the two quantities 
enter in the combination $NL$. There is a slight correction due to the
restriction induced by  SU(N), since there is no component associated 
to vanishing colour momentum. The modification is termed {\em slight} since
it affects only 1 of the $N^2$ degrees of freedom of the U(N) group.
Now let us analyze the presence of  $N$ and $L$ factors in the
vertices. It is not hard to see that all vertices are proportional to 
the factor $g{\cal N}F$. Replacing the expressions given in the text 
we get: $-\frac{\sqrt{2\lambda}}{LN}\sin(\theta{\cal A})$. Notice
that, once more, $L$ and $N$ appear combined as a  product, once we  
express the formula in terms of `t Hooft coupling. The argument of the
sine function is the area of the triangle formed by momenta meeting at a
vertex, multiplied by the parameter $\theta$ appearing in
Eq.~\ref{thetadef}. The parameter can be rewritten (for the square box
case) as 
\be
\theta= \tilde{\theta} \times \Big(\frac{NL}{2 \pi}\Big)^2\quad ,
\ee
with $\tilde \theta$ given by Eq.~\ref{defthetatilde}.

Thus, we conclude that, according to perturbation theory, all physical results 
will depend on $\tilde{\theta}$, $\lambda$ and $LN$. Therefore, if we 
keep $\tilde{\theta}$ fixed, these results will depend jointly on
$LN$. This is a form of {\em volume independence} or {\em reduction},
since finite volume effects would disappear, provided there is a well-defined
large $N$ limit. On the other hand, if we achieve this limit with the
alternative interpretation, L large and N fixed, we expect that the 
spectrum of the sector with vanishing electric flux should become 
independent of $\tilde{\theta}$. The reason is that this angle  only affects
the boundary conditions, and these should become irrelevant in the
large volume limit. This argument does not apply for the non-zero electric
flux sectors, because their masses continue to depend strongly on
the size as we take it to infinity. 

Another important point mentioned earlier, is that all energy scales could 
be expressed in units of $\lambda$. Because of dimensional counting, 
the ratio would appear as a power series expansion in the dimensionless 
quantity $\lambda L$. Combining this with our previous remarks, we
conclude that the relevant expansion parameter would be 
\be
\label{xdef}
x=\frac{\lambda N L}{4 \pi}
\ee
as was used previously in relation with the self-energy expression.

Thus, our final conclusion based on
perturbation theory alone is that, as we take $N$ to infinity keeping 
$\tilde{\theta}$ fixed, we should recover the infinite volume limit. 
There are two provisos in this result. The first is that  non-perturbative
effects might not work the same way. We had indications of that in the study
of the sphaleron solutions. This makes it very important to study the
theory  non-perturbatively to see what the behaviour really is. A
first step in that direction will be presented in the next section.
The other comment is that it becomes rigorously impossible to keep 
$\tilde{\theta}$ fixed as we change $N$. This is so because $\kb/N$ is
an irreducible rational number. To make the statement precise we have
to assume that the final result would depend continuously on
$\tilde{\theta}$. As $N$ grows, one could choose a sequence of values of 
$\kb$ approximating $\tilde{\theta}$. In the next section we will see
some examples. 

Before, let us go back to the analysis of the consequences of our
perturbative calculation of the minimum energy of the electric flux
sectors. The formula  (Eq.~\ref{elenergform}) reproduced below
\be
{\ET^2(\vec e)  \over \lambda^2} =  {|\vec n|^2 \over 4 x^2}
- {1 \over x } \, G \Big ({\vec e \over N}\Big )\quad ,
\ee
complies to the general pattern mentioned earlier. We remind the reader 
that $\vec n$ is given by  the minimum momentum associated to a
particular value of the electric flux 
\be
 -\frac{N}{2}< (n_i=- \m \, \epsilon_{ij} e_j  \, (\bmod   N)) <
\frac{N}{2} \quad.
\ee
As mentioned previously, the correction $\delta \ET^2$ to the energy square
could be  considered the mass-square of the state. This interpretation seems
bizarre since the quantity is negative. A particle with negative mass square
is usually called a {\em tachyon}. A corresponding negative energy
square would signal an instability. The state would be unstable and
decay with a rate proportional to the imaginary part of the energy. The 
phenomenon is, thus, appropriately described as a {\em tachyonic
instability}. Such a situation has been encountered in the context of 
field theories in non-commutative space-times~\cite{Guralnik:2002ru,Bietenholz:2006cz}, 
where it was presented as  signalling spontaneous breaking of $Z_N$ symmetry  and
electric flux condensation. Of course, as mentioned in the
introduction, the question then is whether this tachyonic instability
is present in our theory and, if so, what is its interpretation and
the possible implications.  

Of course,  the negative {\em mass-squared} term does not
necessarily imply any instability. At zeroth-order in perturbation
theory our model has a mass gap, since the minimum momentum cannot
vanish. Thus, for arbitrarily small  coupling, the theory  is stable. 
If we trust the one-loop result as exact, an instability would
necessarily occur at 
\be
x_T(\vec{e})=\frac{|\vec n|^2}{4 G(\frac{\vec{e}}{N})}
\ee
The point is then whether at this value of $x$ the perturbative
formula remains valid or not. Notice that the reliability of this
formula is higher, the smallest the value of $x_T$. Focusing on the 
values of $G$ displayed in Fig.~\ref{fig.theta} it might seem like a fairly
large number. However, notice also that the function grows as $e/N$
approaches $0$ or $1$.  This is not surprising because 
as $e/N$ approaches zero we should recover the UV divergence of the 
original integral.  It is relatively simple to compute the behaviour 
close to the singularity by focusing on the integral producing the
divergence 
\be
\label{Gasymp}
\int_0^1 {dt \over t^{3/2}} \ \exp \Big (-{\pi  |\vec z|^2\over t }\Big  )
=  \, {1 \over \ |\vec z| } + {\rm regular \ terms} \quad ,
\ee
which leads to:
\be
{\ET^2(\vec e)  \over \lambda^2} =  { |\vec n|^2 \over 4 x^2} 
- {N \over  16 \pi^2 x \, |\vec e|  }  \quad ,
\ee
valid for $|\vec e|$ small enough. Plugging this result in the formula 
for the threshold for tachyonic behaviour we get 
\be 
x_T = { 4 \pi^2 |\vec e||\vec n|^2 \over N} \quad .
\ee 
Taking $|\vec e|$ and $|\vec n|$ of order 1 and $N$ going to infinity,
seems to lead unavoidably to a tachyonic instability setting in at
$x=x_T$.

However, $|\vec e|$ and $|\vec n|$ are not unrelated quantities. If we
forget about the modulo $N$ part and simply replace $n_i=-\m\epsilon_{i j} e_j$ 
in the previous formula we get 
\be 
x_T= { 4 \pi^2 |\vec e|^3 \m^2 \over N} \quad .
\ee
where $\m$ is the magnetic flux. If we scale   the magnetic flux
like $N$ (or even $\sqrt{N}$) in the large N limit, we keep 
the tachyonic threshold $x_T$ safely away from the perturbative region. 
On the opposite extreme if we take the lowest momentum value $n=1$,
this corresponds to an electric flux of $\kb$. Hence, to avoid
instabilities we should take the large N limit keeping $\kb/N$ bounded
from below. It is remarkable that our criteria correspond to similar
ones advocated by  Gonz\'alez-Arroyo and Okawa~\cite{TEK3} for large
$N$ reduction to apply for the twisted Eguchi-Kawai model in 4-dimensions.

It is clear that our analysis has opened many questions about the region of 
validity of the perturbative regime and its  consequences that
only a non-perturbative analysis can settle. This is indeed the
purpose of the numerical results that 
will be presented in section~\ref{s.numerical}.

\section{Non-perturbative lattice determination of the mass spectra}
\label{s.numerical}

\subsection{The goals}
In the previous section we have studied, by  perturbative
techniques, the behaviour of 2+1 SU(N) Yang-Mills fields defined on a 
finite spatial torus with twisted boundary conditions. Several 
interesting properties emerged concerning the  $N$ and $L$
dependence of physical  observables. In particular, we observed that
with appropriate choices of the magnetic flux through the box,
physical observables depend jointly on the product $NL$, or rather on
the dimensionless quantity $x=\lambda N L/(4\pi)$. We also observed 
that, in certain circumstances, perturbation theory suggested the
appearance of tachyonic instabilities.  Both phenomena are well 
exemplified by the formula for the ground state energies in the
sectors of non-zero electric flux $\vec{e}$ (Eq.~\ref{elenergform}),
valid for small values of $x$. The purpose of this section is to explore 
the evolution of these quantities for larger values of $x$. This will
allow us to determine the region of validity of the perturbative
formulas and to test if and when tachyonic instabilities will appear. 
Last, but not least, we can check  whether $NL$ scaling continues to
hold in the non-perturbative region. 

In order to justify our particular setting, it is necessary to be more
specific about what quantities will be explored and why. Since we
want to follow the transition from the perturbative to the
non-perturbative region, we have concentrated on those quantities which are
easier to track in both regimes. Thus, we will focus on straight line
Polyakov loops having the minimal and next to minimal momentum values
$\vpt=(\frac{2 n \pi}{NL},0)$ (with $n=1,2$) (we disregard here the
mixed momenta $\vpt=\frac{2 n \pi}{NL}(\pm 1,\pm 1)$ which have lower 
perturbative energy than the $n=2$ states). To simplify notation, in
what follows, we will refer to these quantities as $\ET_1$ and $\ET_2$ 
for the $n=1$ and $2$ cases respectively. The $n=1$ loops are the 
lowest excited states over the vacuum in the perturbative regime. 
As we will see, this will not be the case in general  for the large $x$
region. We should emphasize that, even if we stick to these momenta,
one still covers a wide range of electric flux values by
changing the magnetic flux of the box. We recall that the relation
between the integer  $n$ appearing in the momentum  formula and 
electric flux is $|\vec{e}|= \kb n \bmod N$, where $\kb\m=1 \bmod N$.
Notice that, with twisted boundary conditions, there are no zero-momentum 
states carrying electric flux. 

Concerning the behaviour of these quantities in the perturbative
region it is interesting to make a few observations. While at zero 
order in perturbation the minimum $n=1$ momentum contains a  unique 
state having the lowest energy (indeed  the first excited state over
the vacuum), for the $n=2$ case there are actually two degenerate 
states corresponding to one gluon or two collinear gluons. 
Since these states have the same quantum numbers they can mix. To order 
$\lambda$ the degeneracy is broken by the self-energy term, but  
mixing remains absent. Which  of the two states is lighter, follows 
by comparing $G(2\kb/N)$ with $4G(\kb/N)$, and the answer  depends on 
the value of $\kb$. 

In order to interpret our results it is interesting to revise what is
the expected behaviour of our observables for large torus sizes. 
Confinement dictates that all states carrying non-vanishing electric
flux  should have energies  that grow linearly with the size of
the box $L$. A priori, one expects that the energy per unit length
--the string tension-- should depend only on the value of the electric
flux and neither on the particular momentum chosen $n$, nor on the
magnetic flux through the torus $\m$:
\be
\label{kstrings0}
{\ET_n  \over \lambda}  = {\sigma_{\vec e} \, L \over \lambda} \quad .
\ee
The dependence of the string tension $\sigma_{\vec e}$ on the electric flux 
$\vec{e}=(0,e)$ is a very interesting property, around which considerable
literature has been generated. The expected dependence is as follows:
\be
\label{kstrings}
\sigma_{e}=N\sigma'\,  \phi\Big({e\over N}\Big) \quad ,
\ee
where $\sigma'$ can be fixed in terms of the ordinary string tension, 
which corresponds to one unit of electric flux. The function $\phi(z)$ 
should behave as $z$ for small argument, and by definition should satisfy
$\phi(z)=\phi(1-z)$. There are two forms of $\phi(z)$ which have
appeared repeatedly in the literature when studying various theories,
either exactly or with approximations: 
$$ \phi(z)=z(1-z) $$
and 
$$ \phi(z)=\sin(\pi z)/\pi \quad .$$
We point out that the k-string formula Eq.~\ref{kstrings} is in perfect
agreement  with our $x$ scaling hypothesis, since the linear behaviour 
of the energies can be rewritten as 
\be
{\ET_n  \over \lambda}  =  4 \pi x \frac{\sigma'}{\lambda^2} \, \phi\Big({e\over N}\Big) \quad .
\label{eq.largel}
\ee
We remind the reader that the $x$-dependence holds for fixed values 
of $\kb/N$. 
For $n=1$, $e$ and $\kb$ are just equal, while for $n=2$, $e=2\kb \bmod N$. 

The way in which the linear regime is approached, as $L$ grows from
small to large values, has also been studied intensively for various confining 
theories, including 2+1 dimensional gauge theories. The picture that the 
fluctuating flux tubes can be described by an effective string theory 
has received strong support by recent lattice studies. For example, 
the work of Ref.~\cite{Lucini:2012gg} showed an spectacular agreement 
with the spectrum of the Nambu-Goto string. This goes beyond the
behaviour having an universal character (irrespective of the
particular string theory involved). In this respect, for the
three-dimensional theory, it has been shown~\cite{Luscher:2004ib} that 
the first two coefficients in the expansion of the string energy in
powers of $1/(\sigma L^2)$ are universal.

This predicted $L$ dependence is very important in order to interpret 
our results. However, we might question how these conclusions are
affected by our particular twisted box setting. Previous studies
concerned the case of only one compact direction rather than two.
Furthermore, one might wonder what effect does the presence of the 
twist have upon the string description. As a matter of fact, our work 
might help in giving support to a particular scenario of this type. 
There are remarkable indications that twisted boundary conditions 
could be related  with string models on the background of Kalb-Ramond
$B$-fields. We will explain  below why this is indeed the most natural
choice. The Nambu-Goto prediction for the energy of a closed-string winding
$\vec e$ times around a torus on the background of a Kalb-Ramond $B$-field, is
derived from:
\be
{\ET^2(\vec e) \over \lambda^2} =  \Big ( {\sigma |\vec e|  L \over
\lambda}\Big )^2 - {\pi \sigma \over 3 \lambda^2} + \sum_i \Big ( { \epsilon_{ij}
e_j \, B \over \lambda L }\Big)^2 \quad ,
\ee
The $B$-dependent term in this  expression has an intriguing
analogy with the pertubative expansion for the electric-flux energy.
It suffices to recall that $\epsilon_{ij} e_j = - \kb n_i ({\rm
mod} N)$, with $\vpt= 2 \pi \vec n/NL$ the gluon momentum, to suggest the
identification: $B \equiv -2 \pi \m /N$.
This transforms the $B$-field dependent term into the tree-level
perturbative contribution: $|\vec n|^2  / (4 x^2)$.

On second thoughts, the connection between twisted boundary conditions and strings 
with Kalb-Ramond $B$-fields is also suggested by their common relation 
to non-commutative quantum field theory. In the open string sector, the latter 
appears as a particular low-energy limit of these type of string 
theories~\cite{Seiberg:1999vs}, with  non-commutativity parameter $\theta_{ij} \equiv 
- \epsilon_{ij} L^2/B$. Combining this with the expression
given above for the $B$-field readily reproduces the field-theory $\theta$ 
parameter introduced for the twisted box in Eq.~\ref{thetadef}.

This observation allows to cast the spectrum of the closed Nambu-Goto string 
on the background of a $B$-field into a form fully consistent with the 
conjectured $x$ dependence of all physical observables:
\be
{\ET^2(\vec e)  \over \lambda^2} =
\Big ( {4 \pi \sigma |\vec e| \over  N \lambda^2 }\Big )^2 x^2 - {\pi
\sigma \over 3 \lambda^2} + {|\vec n|^2 \over 4 x^2}
\label{eq.NG}
\quad ,
\ee
connecting, in a very remarkable way, perturbation theory in the twisted box with 
the Nambu-Goto string dynamics. From this formula we deduce that the
first two terms are not affected by the presence of the twisted
boundary conditions, which will allow us to make use of the results
obtained in other settings.

\subsection{The methodology}

Having fixed our goals, we will now describe the way in which we have
set up our methodology to accomplish them. We will be using lattice
gauge theory techniques, which have proved to be a powerful tool to
investigate non-perturbative phenomena in gauge theories. More detailed
aspects of our technique follow.

Twisted boundary conditions can be very easily implemented on the lattice. The starting point is   
the Wilson lattice action given in Eq.~\ref{eq.wilson} with link matrices satisfying the boundary 
conditions specified in Eq.~\ref{tbclink}. It is possible to perform a change of variables  
that allows one to work with periodic links at the price of introducing a 
twist dependent factor in the action. The new action reads
\be
S_W = N b \sum_{n \in {\bf Z}^3}  \sum_{\mu \ne \nu} \Big ( N - z_{\mu \nu}^*(n) 
{\rm Tr} P_{\mu \nu } (n) \Big ) \quad ,
\label{eq.wilson2}
\ee
where $z_{\mu \nu}(n)$ is equal to 1 except for the corner plaquettes in each (1,2) 
plane where it takes the value:
\be
z_{i j}(n) = \exp \Big (i  \epsilon_{ij} {2 \pi \m\over N}\Big ) \quad .
\ee

In order to explore the volume and $N$ dependence as a function of the magnetic flux, 
we have performed a set of Monte Carlo simulations for various values of  
$\m$, the gauge group SU($N$), the lattice size $N_s=L/a$, and $b\equiv 1/(a \llat)$, 
where $a$ is the lattice spacing and $\llat$ is `t Hooft coupling on the lattice. The temporal
extent, with $N_0$ sites, has been chosen sufficiently large to avoid significant finite 
temperature effects. Our simulations allow us to cover a wide range of values of 
$x_L= N N_s /(4 \pi b)$, which is the lattice counterpart of the
dimensionless quantity $x$ introduced earlier. Similarly, multiplying
$b$ by the spectrum as measured on the lattice we get $b\, \ET_L$, the
lattice equivalent to the ratios $\ET/\lambda$. Lattice
counterparts differ from the continuum values by a function of $a\lambda$ 
which goes to zero as $a\lambda\sim 1/b$ goes to zero. Thus, to recover
the continuum results one should take $b$ and $N_s$ to infinity with a
fixed ratio. For the exploratory study, contained in this paper, we will be
moderately ambitious and work at  a few fixed values of $N_s$ and $N$ 
but range over many values of $b$, which will  allow us to scan a wide 
set  of values of $x_L$, ranging from the perturbative regime to the 
domain where confinement of electric fluxes sets in. Notice, however, that 
in this way the lattice corrections would affect mostly the results obtained 
for larger values of $x_L$. 

Our choice of values of $N$ and $N_s$ was designed to test the validity 
of our conjecture that physical results will depend only on the product 
$N L$. With this purpose in mind, we have analyzed four sets of $(N,N_s,N_0)$
values: (5,22,72), (5,14,48),  (7,10,32), and (17,4,32),  with magnetic
fluxes: $\m=1,2$ for SU(5), $\m=1,2,3$ for SU(7), and $\m=1,3,5,8$ for
SU(17). The last three sets have been chosen to have approximately equal values
of $N N_s$. The first, with the same value of $N$ as the second but larger $N_s$, 
was included as a measure of the effects of lattice artifacts.  
The results obtained at the same value of $x_L$ for the first set should be 
closer to the continuum. 

As mentioned previously, our analysis will be focused on the electric-flux states
generated by straight  Polyakov lines that have minimal and next to minimal 
momentum values $\vpt=(\frac{2 n \pi}{NL},0)$ (with $n=1,2$). 
We have considered two different lattice operators projecting onto these 
states. They are represented by the product of link matrices:
\be
{\cal P} (t,x;w)  = {\rm Tr} \Big\{ \prod_{s=0}^{N_s-1} U_2 (t,y+s,x)\Big\}^w\quad ,
\ee
with the two possible choices given by $w=-n\kb \,(\bmod N)$ and 
$w = N-n\kb \,(\bmod N)$.  
The spectrum has been extracted from correlators of these operators 
projected over the appropriate momenta through: 
\be
C_n(t;w) = {1\over N_s}  \sum_{x, \tilde x}  \exp \Big (-i 2 \pi  \, {n \tilde x \over N_s N}\Big ) \, 
\langle {\cal P}( 0, x;w) {\cal P}^\dagger(t, x + \tilde x;w ) \rangle \quad . 
\ee
The ground state mass within each electric-flux sector can be determined as usual 
from the late time decay of the correlator. Of the two operators indicated above, 
we have used in each case the one that shows less contamination of excited states 
and gives a lowest ground-state mass. 

For the type of correlators  that were studied  in this work we could
obtain good results by using smeared Polyakov loops. Hence, it  has
not been   necessary to  implement other state  of the art techniques designed to
improve the projection onto the ground state. This is exemplified in
Fig.~\ref{fig.meff}, where we show several characteristic 
effective mass plots, corresponding to both the perturbative and the confining regimes. 
Before continuing, let us point out  that the smearing procedure has to be
slightly modified in the twisted box by introducing the twist tensor, $z_{ji}$,
in the definition of the staples. With this, smearing proceeds as usual by iteratively 
replacing the spatial link matrices by:
\be
U_i^{(s+1)}(n)  = (1-c) \, U_i^{(s)}(n) + {c \over 2} \sum_{\pm j, j \ne i}
\! z_{ji} \, U_j^{(s)}(n) U_i^{(s)}(n+j) U_j^{(s) \dagger}(n+i)\Big |_{\rm unitarized} \quad .
\ee
The results that will be presented below were obtained with  smearing parameter 
$c = 0.475$ and 20 smearing steps. 

\FIGURE[ht]{\centerline{
\psfig{file=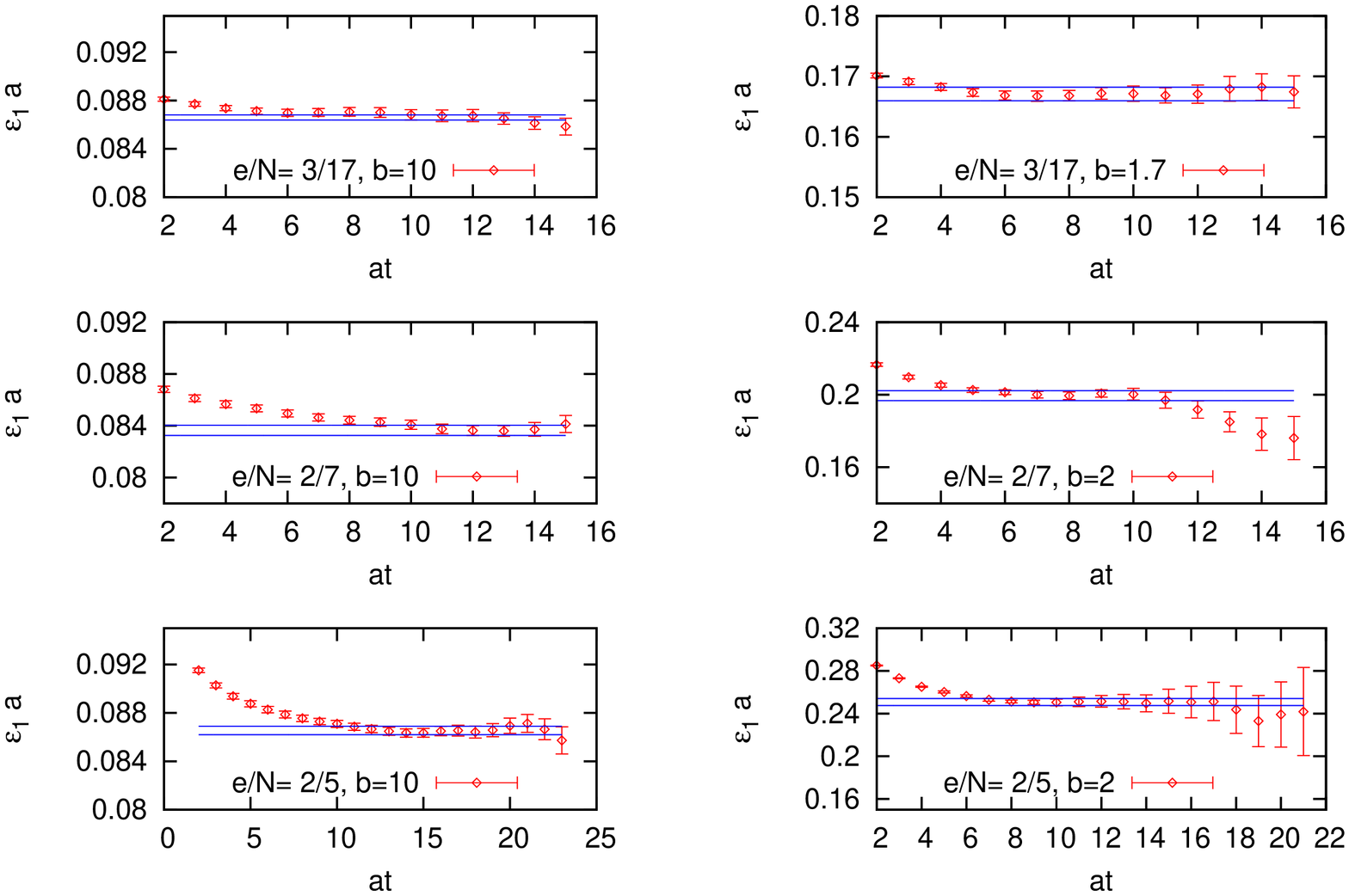,angle=-90,width=15cm}}
\caption{\label{fig.meff}
We show several characteristic effective mass plots for $\ET_1$, the energy of electric flux
$|\vec e|= \kb$ with momentum $|\vpt| = 2 \pi /NL$. They correspond to
the perturbative (Left) and the confining (Right) regimes.
}}

Finally, let us briefly describe the details of the algorithm employed in the Monte Carlo simulations.
It is well know that the standard Creutz's heat bath-algorithm has a very low acceptance rate
both  at weak coupling and for SU(N) gauge theories at large $N$.
To prevent this, we have used instead a heat-bath algorithm based on a proposal
by Fabricius and Haan~\cite{fabricius} that significantly improves the performance 
in these two cases. In  Creutz's method, one sweeps
through all possible SU(2) subgroups of SU(N) with probability given by:
\be
   \, {\rm d}\alpha_0 \, \sqrt{1-\alpha_0^2} \
\exp(4 b N k\alpha_0) \quad .
\ee
For large $b$ and(or) large $N$, the exponential factor nearly
always gives $\alpha_0 \approx 1$, which is most of the times rejected with probability
$\sqrt{1-\alpha_0^2}$.  To circumvent this, Fabricius and Haan introduced the variable
$\alpha_0=1-2\delta^2$, with probability distribution
\be
   \, {\rm d}\delta  \, \delta^2 \,  \sqrt{1-\delta^2} \ {\rm exp}(-8 b N k\delta^2)\quad .
\ee
By introducing a three-dimensional vector  $\vec \delta $, this becomes
\be
  \, {\rm d}^3 \vec \delta   \, \sqrt{1-|\vec\delta |^2} \ {\rm exp}(-8 b N k|\vec\delta |^2) \quad .
\ee
Then the exponential factor nearly always gives $|\vec \delta | \approx 0$ for large
$\beta$ and(or) large $N$,  which is accepted with $\sqrt{1-|\vec\delta |^2}$
probability. With this method at hand, one of our Monte Carlo sweeps consists 
of the heat-bath updating of all possible  SU(2) subgroups of SU(N), followed by 
5 over-relaxation updatings.

A compilation of all our numerical results is presented in the tables of 
Appendix \ref{appendixd}. For each data set, a characteristic ensemble consists 
on $1-2 \times 10^6$ Monte-Carlo sweeps with all the observables
measured every 40 sweeps. 
Both the errors on the correlators as those on the fitted masses have been determined 
using jackknife. Let us finish by mentioning that the characteristic run-times for
the $SU(17)$ lattice, for instance, are 0.15 seconds/Monte-Carlo sweep, running 
on 16 cores of the IFT Hydra cluster (specifications can be found at 
http://www.ift.uam-csic.es/hydra/).

\subsection{Analysis of Numerical Results}
\label{s.results}

In this subsection we will present the numerical results obtained with
the lattice regularization. As explained before, we have restricted our 
analysis to the evaluation of the temporal correlation function of 
straight Polyakov lines with minimal  $\vpt=(\frac{2 \pi }{NL},0)$ and 
next-to-minimal momentum  $\vpt=(\frac{4 \pi }{NL},0)$. The
corresponding energies will be labelled $\ET_1$ and $\ET_2$
respectively. The bare results have been collected in tables in
appendix~\ref{appendixd}. This will allow other researchers to use and 
analyse  the data according to their criteria. In what follows we will 
present our particular analysis of the numerical data.

Let us focus first on the results for the   minimum momentum, with energy
$\ET_1$. A visual representation of the results is given in
Fig.~\ref{fig.deltaE}, in which  the combination $x\ET_1/\lambda$
is displayed as a function of $x$. Each set of points is represented 
in different colours and labelled by the combination $\kb/N$
characteristic of the data sample. They have been grouped into 
subplots which display different ranges of the ratio $\kb/N$,
increasing from left to right. The continuous lines are the results of
a fit to be explained later but, at this level, they are very helpful
to guide the eye.  A simple look at the figure shows the
similarity among the different curves despite the difference in the
value of $N$ and of the magnetic fluxes used. The visual
agreement  is in accordance to our hypothesized universal $x$
behaviour at fixed value of $\tilde \theta=2 \pi \kb/N$. The
differences observed could be explained by the fact that the values of 
$\kb/N$ are not exactly equal for each set of data points. Indeed, the
closest values occur for $N=7$ and $\kb=2$ and $N=7$ and $\kb=5$. The
results of these two cases, appearing in the middle subplot, show a
striking resemblance. 

The general features of the figures also agree with the expectations
described earlier. The lowest order perturbative result, valid for small 
$x$, gives $0.5$, which agrees with the data. The correction comes from the
self-energy term computed in this paper and accounts for the decrease
observed at small but higher  values of $x$. The way in which we have
represented the data allows a very clear recognition of the
perturbative features. If we continue for larger values of $x$, we
observe that the decrease reaches a minimum at around $x\sim 1$, and then
grows very fast for larger $x$.  The fact that the decrease ceases before
the energy becomes negative is precisely what we anticipated. It shows 
that the perturbative formulas cannot be trusted in this case beyond 
values of order 1. Hence, no tachyonic instability occurs in any of
these data sets. We have left out temporarily of the analysis those cases 
in which  perturbation theory indicates the generation of tachyonic 
instabilities at small values of $x$. That cases will be studied in the 
next subsection.

\FIGURE[ht]{\centerline{
\psfig{file=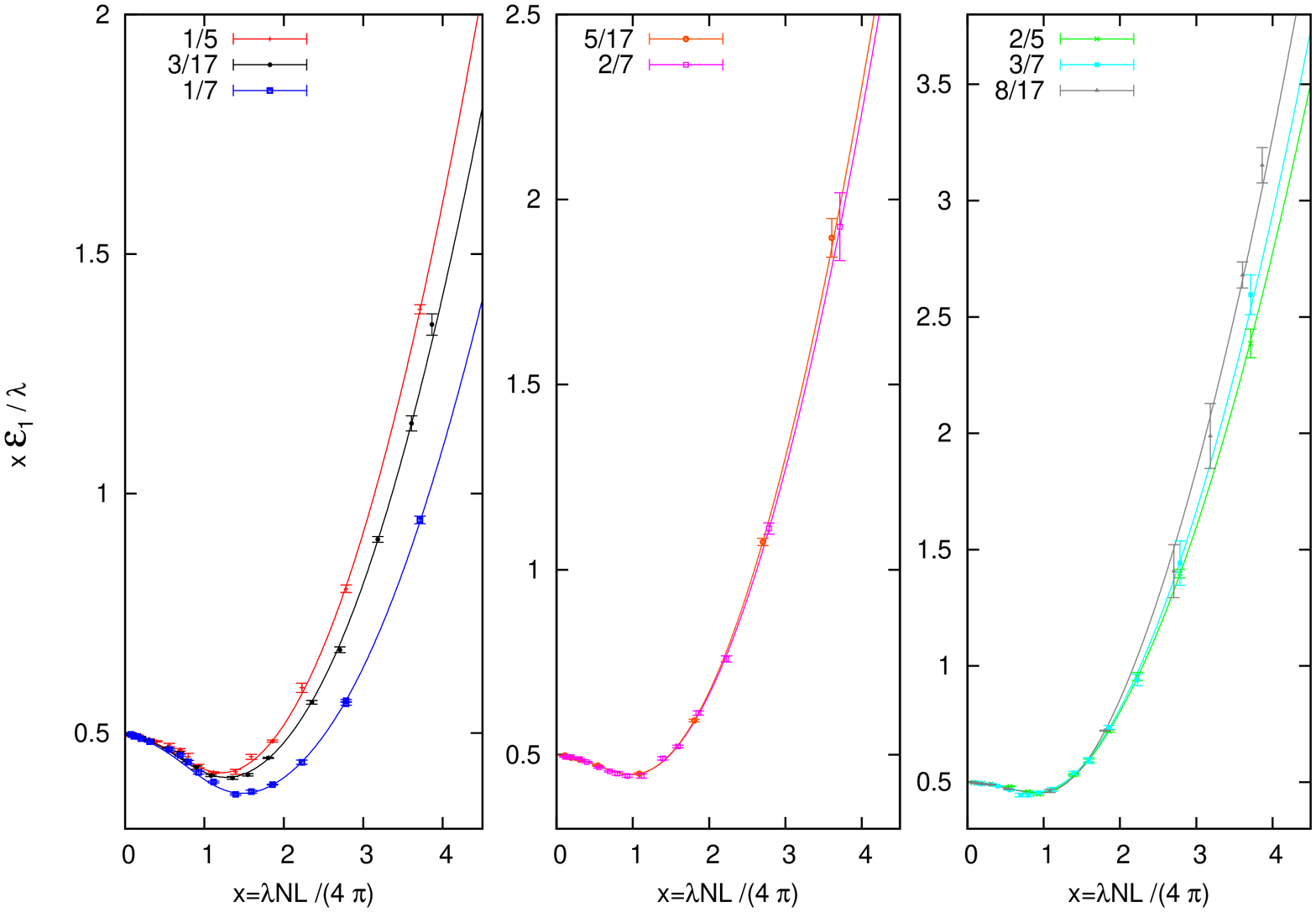,angle=-90,width=15cm}}
\caption{\label{fig.deltaE}
We display $x \ET_1 /\lambda$ as a function  of  $x$.
The lines are fits derived from Eq.~(\ref{eq.fits}) with parameters set
as  discussed in section~\ref{s.results}.
The labels in the plot indicate the value of $\kb /N$.
}}

Finally, we should mention that the rise of the curves at large $x$ is
simply what confinement predicts. To allow a more clear visual
comparison of the results and the expectations, we have displayed in 
Fig.~\ref{fig.deltaE2} the values of $(\ET_1)^2/(\lambda x)^2$ after 
subtracting out $1/(4x^4)$, which is the zero-order
perturbative result for this quantity. The curves look very much
consistent with the expectation that they should tend towards constant
values at infinity with a $1/x^2$ pattern. The constant at infinity is
nothing but the string tension. It is certainly not chance that the
order of the curves corresponds neatly to the order of the values
$e/N=\kb/N$.

\FIGURE[ht]{\centerline{
\psfig{file=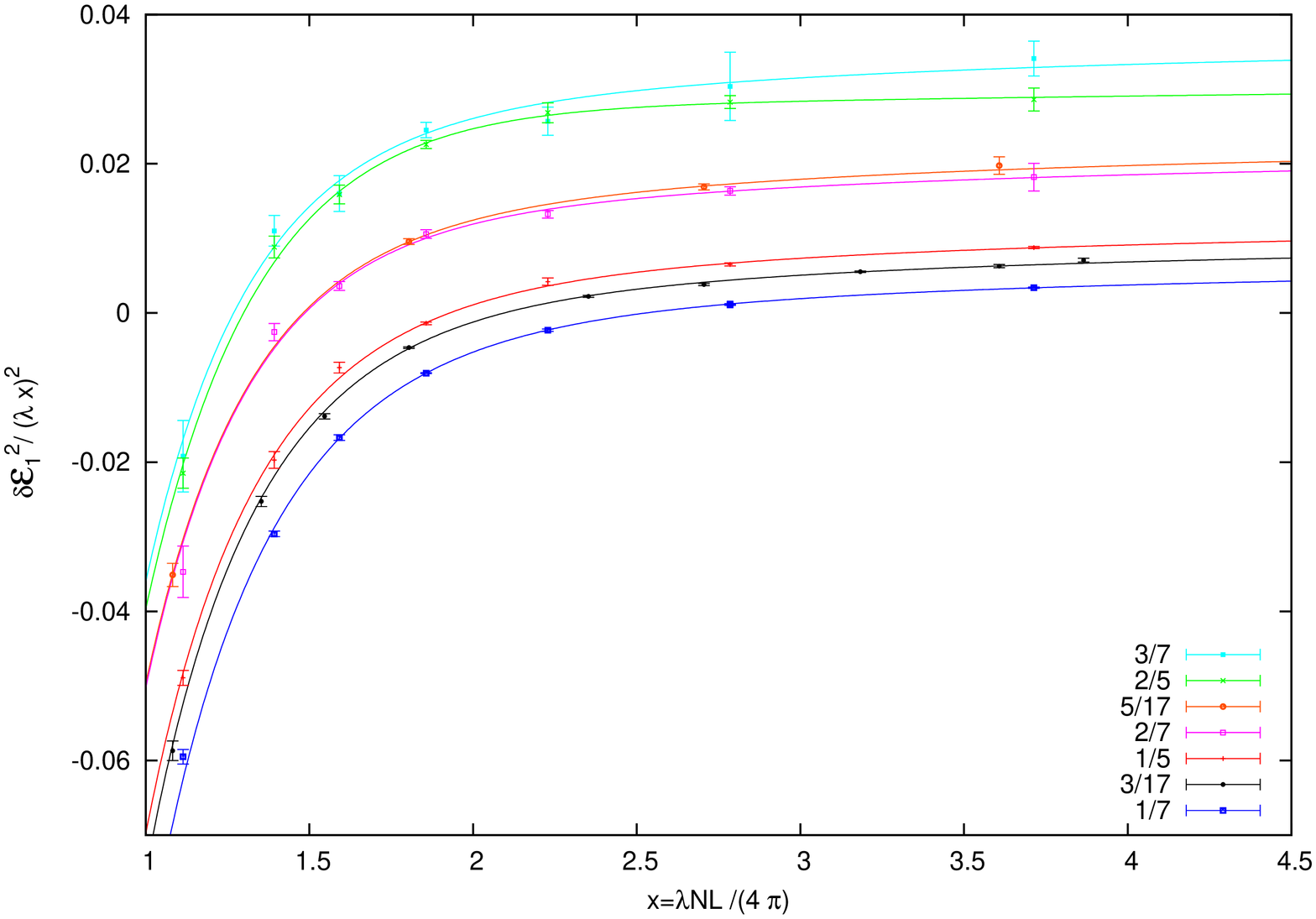,angle=-90,width=15cm}}
\caption{\label{fig.deltaE2}
We display $\ET_1^2 /(\lambda x)^2$ after subtracting out $1/(4 x^4)$, which is the zero-order
perturbative contribution to this quantity.
The lines are fits derived from
Eq.~(\ref{eq.fits}) with parameters determined as discussed in section~\ref{s.results}.
The labels in the plot indicate the value of $\kb /N
$.
}}

\TABLE{
\begin{tabular}{|l|l|l|l|l|l|}
\hline
$\kb$ & N & $G(\kb /N)$& $\gamma$ & ${\cal A}$ & $S_0$ \\
\hline
1&5&0.0322615&0.110(2)&14(3)&7.5(4)$\ \ \ $\\
\hline
2&5&0.0185816&0.177(4)&27(8)&8.0(4)\\
\hline
1&7&0.0446267&0.081(1)&20(2)&8.1(2)\\
\hline
2&7&0.0234449&0.147(5)&12(6)&6.9(7)\\
\hline
3&7&0.0180238&0.190(7)&14(10)&7.2(1)\\
\hline
3&17&0.036346&0.099(2)&14(2)&7.4(3)\\
\hline
5&17&0.0228957&0.152(3)&10(3)&6.9(4)\\
\hline
8&17&0.0175579&0.210(5)&36(76)&10(3)\\
\hline
\end{tabular}
\caption{\label{t.fits}
We present the set of parameters corresponding to the fits used to describe the 
$\ET_1$ data displayed in Figs.~\ref{fig.deltaE} and ~\ref{fig.deltaE2}.
}
}

Thus, visual scrutiny shows that our anticipated results are not
inconsistent with the data. It is necessary to go beyond this stage and
attempt a more quantitative analysis of our results. For that purpose we
have tried to describe the data with a parameterization inspired by the
theoretical considerations presented at the beginning of this section.
Collecting all the different contributions expected from the different
sources, we are led to an expression of the type:  
\be
 {\ET_1^2 \over \lambda^2} =  {1 \over 4 x^ 2}
 + {\alpha \over  x}
+ \beta   + \gamma^2 \,  x^2
\quad ,
\ee
It follows both from the perturbative results and from
the string fluctuation description, that it is advisable to parametrise
$\ET_1^2$ rather than $\ET_1$.
By construction this formula should describe  well the behaviour of our
data in the asymptotically small and large $x$ regimes. This is indeed
the case. This simple ansatz describes also qualitatively well the
transition region between both regimes shown by our data. Not satisfied
with this result, we have aimed at having a quantitatively good
description of the region of intermediate $x$ values. Curiously a
simple additional term of the form 
\be
\frac{{\cal A}}{  x^a}  \, \exp\Big\{-{S_0 \over x}\Big\} \quad ,
\ee
allows us to achieve our goal. Although of a phenomenological nature,
the form of the additional term was inspired by the posible presence of
non-perturbative contributions around additional instanton-like
configurations. The shape is the one arising from a semiclassical
analysis. Indeed, this would not be completely unexpected. Monopole
like contributions have been found to be crucial to achieve confinement 
in abelian gauge theories, and monopoles are just instanton-like in 2+1
dimensions. Furthermore, there are at least some non-perturbative
2-d configurations, studied in section~\ref{method}, which look like
sphalerons and appear as the top of the hill for a corresponding
instanton configuration. Certainly, both the numerical study of the 
transition region, as well as the presence and dynamical contribution 
of semiclassical configurations, deserves further study. For the
time being, we let the reader judge what relevance should be given to
the fact that such a term allows us to describe the transition 
region with rather good precision. 

The additional term adds three parameters, which are however too highly 
correlated to let our data determine them. After exploring 
different choices we found that fixing $a=7/2$ allows us to describe all
the data.  

In summary, the final parameterization used to fit the data in figures \ref{fig.deltaE} - 
\ref{fig.deltaE2} is  of the form:
\be
\label{eq.fits}
 {\ET_1^2 \over \lambda^2} =  {1 \over 4 x^2} + {\alpha \over  x} +{{\cal A} \over x^3 \sqrt{x}} 
e^{-{S_0 \over x}} +  \beta  + \gamma^2 \,  x^2
\quad .
\ee
Before proceeding to present the results of the fits, let us 
discuss a bit more about the origin and  possible values of the parameters.
Our perturbative formulas account for  the first and second terms of our
expression, with  $-\alpha$ replaced by the self-energy  function $G(\vec z)$, 
computed and displayed in section~\ref{s:euclidean}. Therefore, we
fixed this parameter to this value in all our fits, after verifying
that leaving this parameter free does not alter substantially the 
quality of the fit. 
One the other hand, the  $\gamma$ and  $\beta$ terms have the  
form following from a confining linear potential, corrected by 
universal subleading corrections arising in an  effective string 
picture~\cite{Luscher:2004ib}. 
At leading order in $1/(\sigma L^2)$, one expects: $\ET /(\sigma L)=  
1+\pi \rho / (\sigma L^2)$, where the second term is 
the universal L\"uscher correction, with coefficient $\rho = - 1/6$ for the
fundamental closed string in three dimensions.
An interesting question, relevant for the study of the $k$-string dynamics,
is what is the dependence of this coefficient on the
electric flux $e$ (see e.g.~\cite{teper},~\cite{Lucini:2012gg},~\cite{pepe},~\cite{Armoni:2003nz}).
Within our parameterization,  $\rho=2\beta /(N \gamma)$. 
Thus, our results can provide interesting information in this respect.

After these considerations, we present the results of the fits. 
All data points corresponding to particular values of $N$ and $N_s=L/a$,
are well fitted by our formula having four free parameters (${\cal A}$,
$S_0$, $\beta$ and $\gamma$).  From these results one can analyse  the 
behaviour of $\rho$ as a function of the $e/N$ ratio. The results,
displayed  in Fig.~\ref{fig.kstring}, are consistent with a formula of
the type:
\be
\label{eq.rho}
N\rho(z) \equiv {2 \beta \over \gamma} = - {{\cal C} \over   z(1-z)}\quad ,
\ee
with ${\cal C}=1/6$ in this case.  
Fixing this relation allows us to obtain 
equally good fits to the whole sample with less parameter
correlations and, hence, reduced errors. This is just a three
parameter fit which describes the data rather well. The continuum
curves displayed in Figs.~\ref{fig.deltaE} and
~\ref{fig.deltaE2} are the result of these fits. The 
corresponding parameters are presented in
Table~\ref{t.fits}.

\FIGURE[ht]{\centerline{
\psfig{file=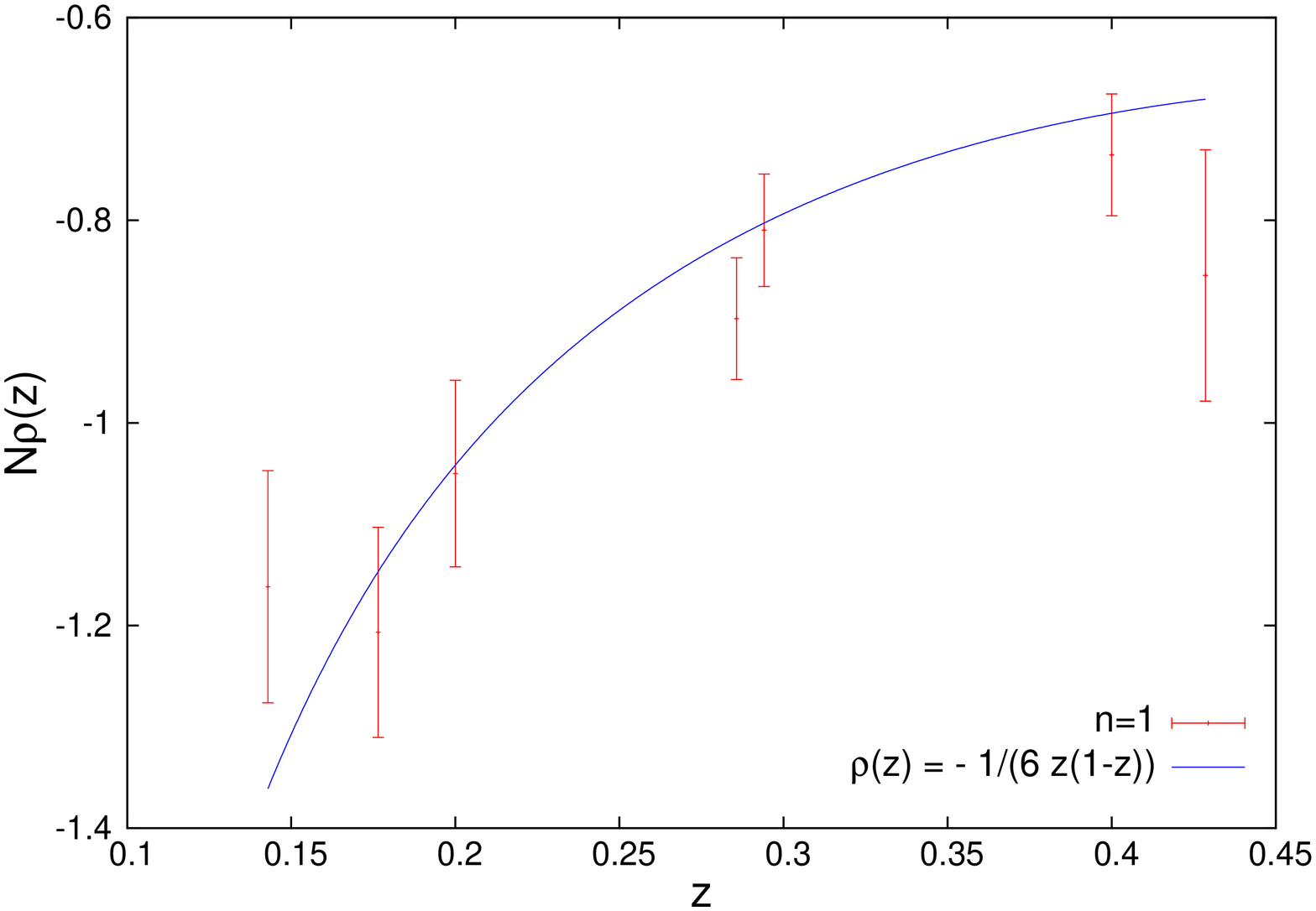,angle=-90,width=7.5cm}
\psfig{file=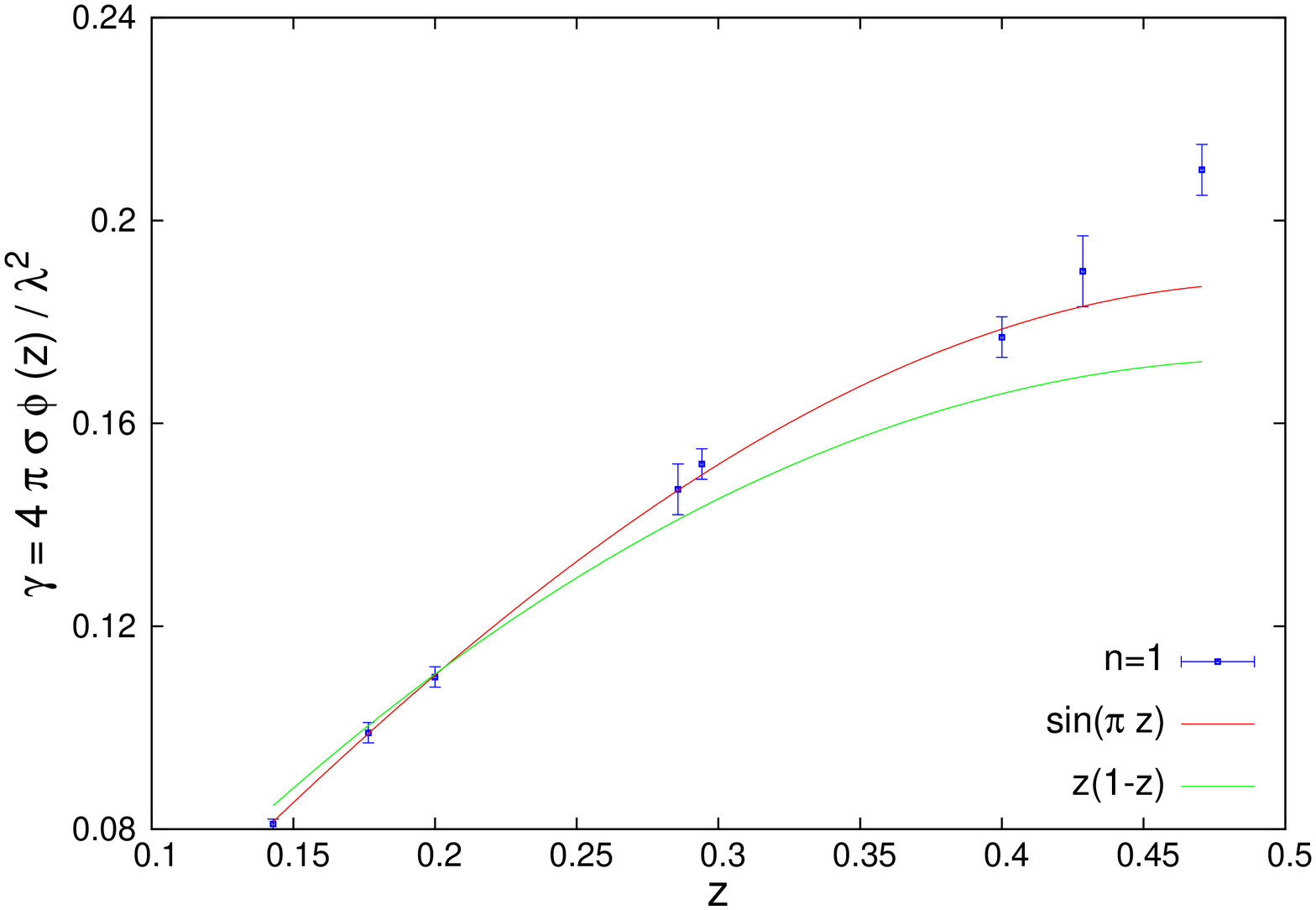,angle=-90,width=7.5cm}
}\caption{\label{fig.kstring}
On the left plot we display the function $N\rho(z)$, given in
Eq.~\ref{eq.rho}, as a function of $z=e/N$. This quantity equals 
($N$ times) the coefficient of
the L\"uscher term in the effective string expansion.
On the right plot we display the function
$\gamma(z) = 4 \pi  \sigma'  \phi(z)/ \lambda^2$, given  in Eq.~\ref{eq.gamma}.
The
red line in the plot is a fit to the Sine scaling formula: $\phi(z) = \sin(\pi z) /\pi$.
The green line corresponds to the prediction from Casimir scaling: $\phi(z) = z(1-z)$.
}}

\FIGURE[ht]{\centerline{
\psfig{file=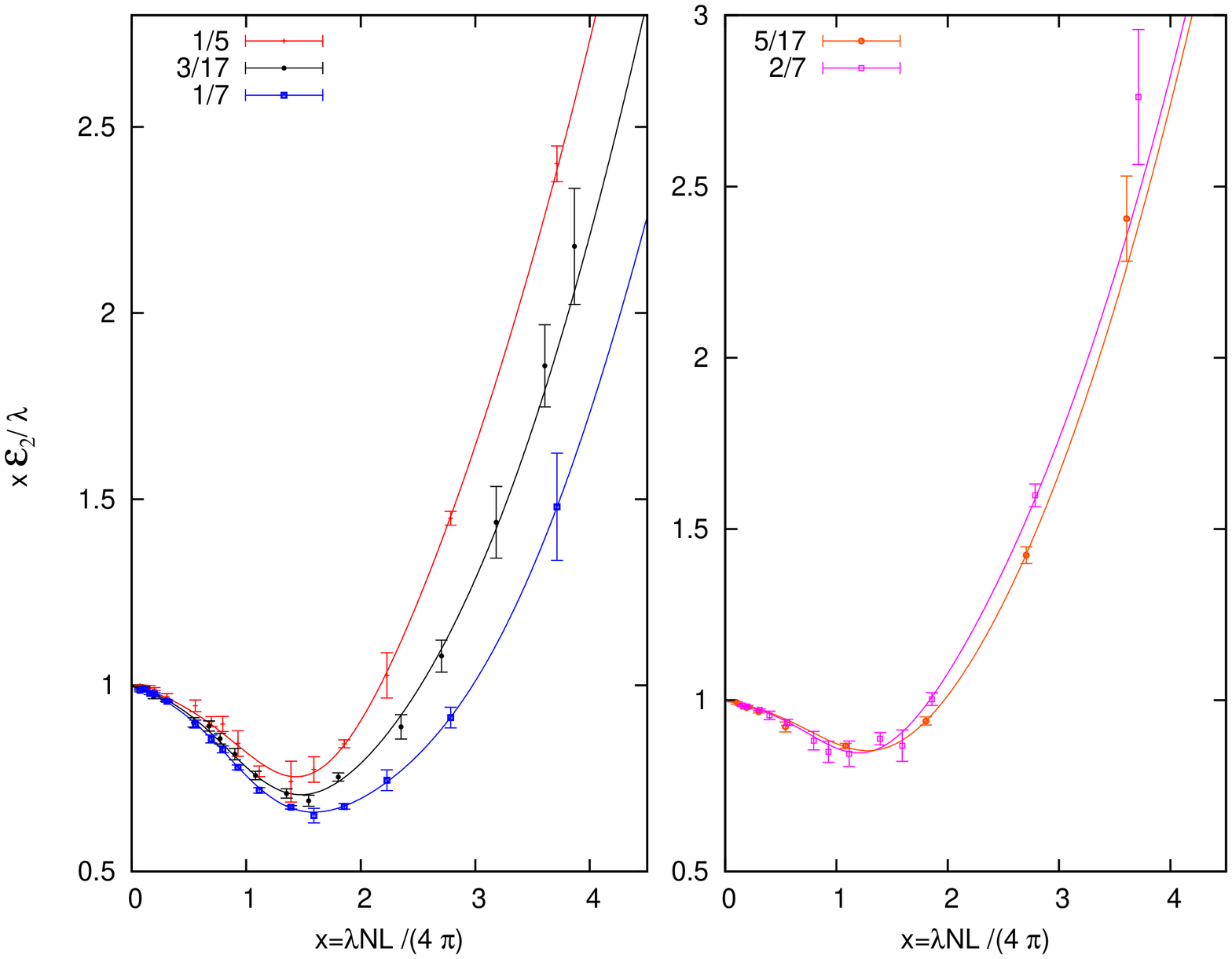,angle=-90,width=15cm}}
\caption{\label{fig.deltaE3}
We display $x \ET_2 /\lambda$ as a function  of  $x$.
The lines are fits derived from Eq.~(\ref{eq.fits}) with parameters determined
as  discussed in section~\ref{s.results}.
The labels in the plot indicate the value of $\kb /N$.
}}

Our results  also allow us to extract relevant information on the values of the string tension for 
these states. Following the discussion around Eq.~\ref{kstrings}, we parameterize $\gamma$ as:
\be
\label{eq.gamma}
\gamma(z) = {4 \pi \sigma'\over \lambda^2} \phi(z) \quad .
\ee
The information on what is the relevant scaling of the $k$-string tension with the electric flux 
is thus encoded in the function $\phi(z)$.
In the right plot of Fig.~\ref{fig.kstring} we display $\gamma(z)$ for our 
data set, which includes all
the results for different gauge gauge groups and magnetic fluxes. The lines in the plot
correspond to two of the most commonly discussed scaling functions for the $k$-string tension:
Casimir scaling: $\phi(z) = z (1-z)$, and the Sine scaling: $\phi(z) = \sin(\pi z)/\pi $. Clearly the latter
is in a much better accordance with the data giving a $\chi^2$ per degree of freedom of order 0.34 
(if the point at largest value of $z$ is excluded). The value obtained from this fit for the
large N fundamental string tension is $\sqrt{\sigma'}/ \lambda =
0.217(1)$. It deviates around  10$\%$ from the 
large $N$ value obtained by Teper and collaborators in Ref.~\cite{teper}: $\sqrt{\sigma}/ \lambda = 
0.19638(9)$. Given the limitations of our numerical analysis, and the absence of a continuum extrapolation,
this  agreement can be considered very satisfactory.

Let us now briefly comment about the results of the $n=2$ momenta
states. The quality of the data is poorer and we have to face
additional complications, as the mixing of states occurring in
pertubation theory. Hence, the results for $x\ET_2/\lambda$ as a
function of $x$,   presented in Fig.~\ref{fig.deltaE3}, do not have 
the quality of the $n=1$ case.  Notice that we have left out of the 
figure the set corresponding to the 
rightmost plot in Fig.~\ref{fig.deltaE}, where perturbation theory indicates the existence
of instabilities at this value of the momentum for small values of $x$.
The corresponding data will be discussed in the next subsection.

The continuum curves in the plot correspond again to the  parameterization of the energy given 
in Eq.~\ref{eq.fits}. Here we had to deal with the complication introduced by the degeneracy at 
zero-order in perturbation theory between one gluon and two collinear gluon states. This degeneracy 
is broken by the self-energy contribution and accordingly we have taken
$-\alpha = {\rm max} \{G(2\kb /N), 4G(\kb /N)\}$. For all the values of $\kb/N$ displayed in the figure 
this consideration favours the second choice, and indeed its value is the one that appropriately 
describes our data. Concerning the parameters $\alpha$ and $\beta$, the precision in this case is not good enough to 
obtain a reliable independent determination of both. Nevertheless, if 
we fix $\gamma(z)$ according to Eq.~\ref{eq.gamma}, 
with $\phi(z)$ and $\sigma'$ given by the Sine scaling formula, the resulting three parameter fit describes 
rather well the data. The corresponding parameters are presented in Table~\ref{t.fits2}.
The analysis of the L\"uscher term based on these results is consistent with an scaling in $z=e/N$
given by the same equation~\ref{eq.rho} that describes the $n=1$ results, but with a coefficient
${\cal C}$ that is about four times larger.

\TABLE{
\begin{tabular}{|l|l|l|l|l|l|l|}
\hline
$2\kb$ & N & $4 G(\kb /N)$& $G(2\kb /N)$&$-\beta$ &
${\cal A}$ & $S_0$ \\
\hline
2&5&0.129046&0.0185816&0.231(23)&282(50)&10.6(6)\\
\hline
2&7&0.1785068&0.0234449&0.295(8)&118(7)&8.2(2)\\
\hline
3&7&0.0937796&0.0446267&0.219(25)&160(30)&8.8(5)\\
\hline
6&17&0.145384&0.0199961&0.294(13)&126(14)&8.5(3)\\
\hline
7&17&0.0915828&0.0183267 &0.187(21)&111(31)&9.0(6)\\
\hline
\end{tabular}
\caption{\label{t.fits2}
We present the set of parameters corresponding to the fits used to describe the
$\ET_2$ data displayed in Fig.~\ref{fig.deltaE3}.
}
}

\subsection{Tachyonic instabilities}
\label{s.tach}

In this subsection we will address the possible occurrence in our results
of tachyonic instabilities,  as suggested by perturbation theory. We
argued   that, if the calculated threshold for instabilities $x_T$ 
does not turn out to be small, their presence has  to be considered
questionable. The data presented in the previous subsection verified
that this is indeed the case: no instabilities occurred. On the contrary,
if $x_T$ is small or very  small, the perturbative indication of an 
instability  should be considered seriously.  Our perturbative discussion
identified  the cases of small  of $\kb /N$ and/or $\m/N$ values as credible
candidates for a tachyonic instability to set in. Within our data set  
there are two  particular points that fall into this category. The
first one corresponds to $n=1$, $N=17$ and $\m=\kb=1$, and the 
second to $n=2$, $N=17$ and $\m=2$. In what follows, we will compare
the expectations with the actual results for these cases. 

In section~\ref{s.remarks} we gave as estimate for the 
threshold for instabilities $x_T = |\vec n|^2/(4G(e/N))$. 
By making use of our fitting function  
Eq.~(\ref{eq.fits}) we are now in condition for improving this estimate.
This will allow us to test, beyond perturbation theory, 
the conditions leading to instabilities.  For the energy to remain non 
tachyonic we should impose: 
\be
\label{eq.tach}
{{\cal A} \over x \sqrt{x}}
\, e^{-{S_0 \over x}} + {|\vec n|^2 \over 4 } + \alpha \, x +  \beta \, x^2 + \gamma^2 \,  x^4 \ge 0
\quad .
\ee
Given that ${\cal A} \ge 0$, a simpler operational expression can be obtained by 
setting ${\cal A} =0$. In that limit, the energy is non-tachyonic whenever 
the quartic equation:
\be
\label{eq.tach2}
 {|\vec n|^2 \over 4 } + \alpha \, x +  \beta \, x^2 + \gamma^2 \,  x^4 = 0
\ee
does not have real roots. This is the case if and only if the discriminant 
is bigger than zero and the following condition, defined in terms of the L\"uscher parameter
$\rho$, is satisfied:  
\be
\label{eq.H}
\HT \equiv {|\vec n|^2\over N^2}- {1 \over 4} \, \rho^2 (e) > 0  \quad .
\ee
It is interesting to point out that $-N^2 \HT$ is the discriminant of the 
equation obtained by setting the parameter $\alpha$ to zero in Eq.~\ref{eq.tach2}.  
In this limit we recover the Nambu-Goto string spectrum from Eq.~\ref{eq.NG}.
Hence, $\HT > 0$ is the bound to avoid tachyonic behaviour coming from the string
component of the equation. 
The condition on the positivity of the discriminant for $\alpha \ne 0$ turns 
out to be a more restrictive one.
After some algebraic manipulation it can be cast in the form:
\be
\label{eq.Disc}
4 |\vec n|^2\, \gamma^2 \HT^2  + 4 \beta \, {\alpha^2\over N^2}\,  (\HT + 8 {|\vec n|^2\over N^2}) -27 \, {\alpha^4\over N^4} > 0 
\quad ,
\ee
Notice that, since $\beta<0$, the bound is violated for $\HT=0$.  
The condition derived from the string
formula is thus replaced by the more stringent one: $\HT > \HT_0$,  
with $\HT_0$ the positive solution of the quadratic equation  that saturates the bound.
For small $\HT_0$ we can still approximate the result by imposing $\HT > 0$.   
Parameterizing now the electric flux dependence of $\rho$ by $N\rho(z)=-{\cal C}/z(1-z)$,
as in the previous subsection, we arrive at the following condition to
prevent the appearance of tachyonic instabilities:
\be
\label{eq.tach3}
|\vec n| \, {|\vec e| \over N }\,  \Big(1-{|\vec e| \over N} \Big ) >  \, 
{{\cal C} \over 2} \quad .
\ee
where the constant ${\cal C}$ is approximately $1/6$.

\FIGURE[ht]{\centerline{
\psfig{file=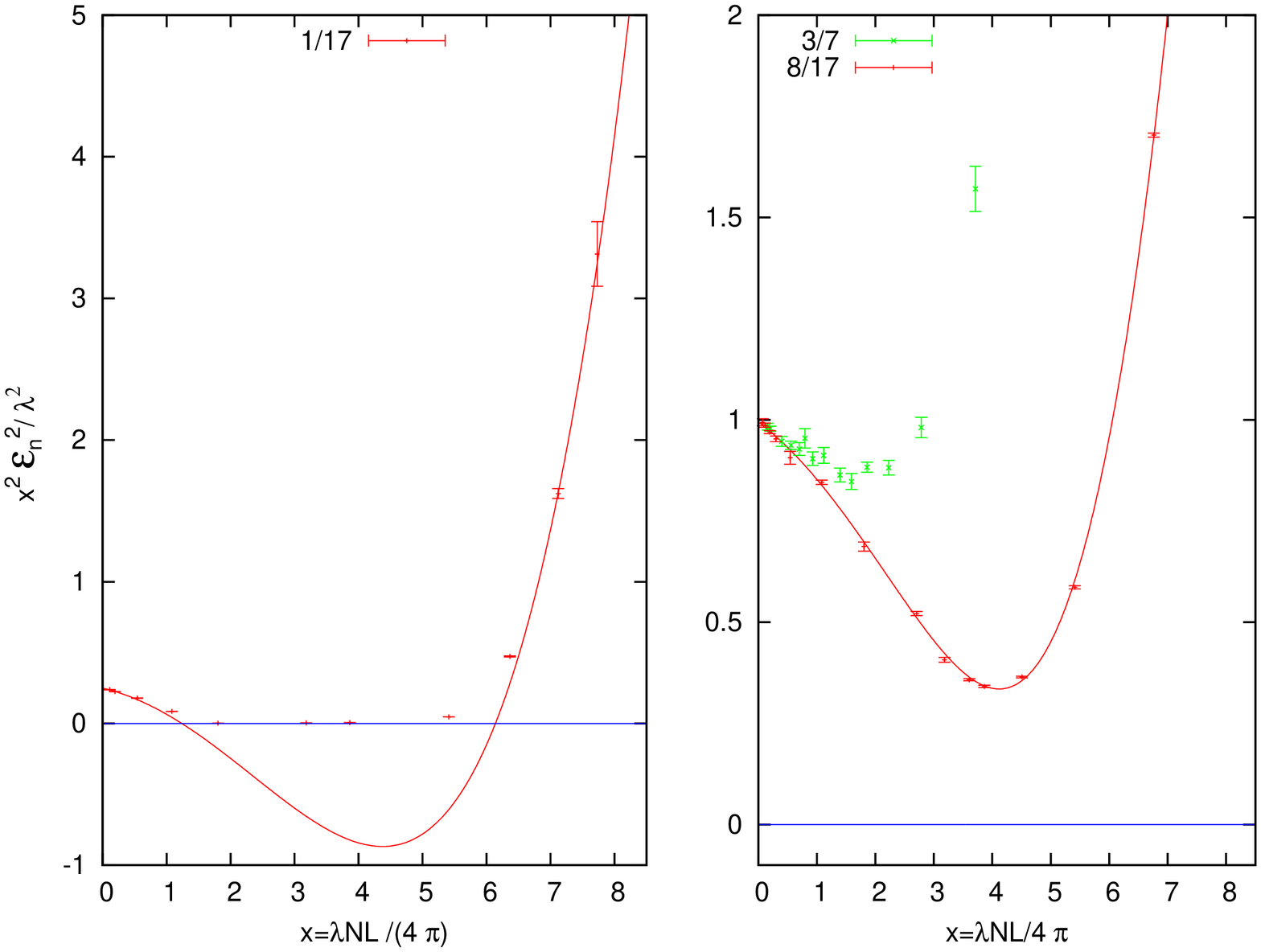,angle=-90,width=15cm}
}
\caption{\label{fig.inst}
We display $x^2 \ET^2_n/\lambda^2$ vs $x$,  in the sector of electric-flux $e=1
$,
for:   Left: $n=1$, $\kb=\m=1$, and  Right: $n=2$, $\m=2$, $\kb=(N-1)/2$.
The continuum lines represent fits to the data derived from Eq.~(\ref{eq.fits}) with parameters
fixed as described in section ~\ref{s.results}. For the fit in the left plot the condition 
${\cal A}=0$ has been imposed. In all cases the choice of fit parameters corresponds
to a single-gluon state. The labels in the figure indicate the values  of $\kb/N$.
}}

Let us apply the previous formulas to the two cases mentioned earlier.
Both correspond to electric flux equal to 1. Hence, for the $n=1$ case 
the bound implies that instabilities are expected to arise   for $N >
12$. If $n=2$, as in our second case, the instability would appear for 
$N>24$. The data in our examples correspond to $N=17$. Thus, according
to these considerations,  tachyonic instabilities should be present in the 
first case and not in the second. 

The numerical results for $N=17$ in the two situations under study 
are presented in Fig.~\ref{fig.inst}, 
where we display $x^2 \ET^2_n/\lambda^2$ as a function of $x$.
The reason to display $\ET^2$ instead of $\ET$ will become clear below.
The $n=1$ case, displayed on the left plot, is very 
different from the ones presented in the previous section. The energy 
decreases  from the tree-level value of $x^2 \ET_1^2/\lambda^2=0.25$
and touches zero at values of $x$ of order one, much 
before the linearly rising contribution is able to reverse this tendency.
The perturbative formula predicts $x_T\sim 2$, but 
non-perturbative effects have anticipated the appearance of the instability.
A fit of the data to Eq.~\ref{eq.fits}, analogous 
to the ones performed in the previous section, is able to capture
rather well the overall behaviour outside the predicted tachyonic region.
Beyond that point, the fitting function becomes negative, while the
data points give a small value of the mass. This might well reflect 
that the correlation function is dominated by the square of the
expectation values of each Polyakov loop, as predicted by the
condensation mechanism~\cite{Guralnik:2002ru,Bietenholz:2006cz}.

The second candidate for instabilities in our data,
corresponding to $n=2$, looks in some respects  more similar to the ones
presented in the previous section. It is represented by  the right
plot of Fig.~\ref{fig.inst}, where we display results for $N=7$ and 17. 
In both cases the linearly rising term compensates the negative decrease and 
the energy does not reach zero, in accordance with our expectations. 
Nevertheless, the results show a  strong decrease of the energy with $N$ at
intermediate values, well accounted  by the continuum red line
in the plot  which represents a fit to Eq.~(\ref{eq.fits}). 
This could anticipate the development of an
instability at even larger values ($N \sim 24$ is our  prediction).
The curve in Fig.~\ref{fig.inst} has been obtained using $G(1/17) =
0.107703$, corresponding to the single gluon self-energy. The initially 
degenerate two-gluon state has a higher energy ($4 G(8/17)=0.0702316$).

\section{Conclusions}
\label{s.conclusions}

In this paper we have studied SU(N) Yang-Mills theory defined on a 
two-dimensional spatial torus of size $L_1\times L_2$ with twisted
boundary conditions. The paper is written in a self-contained form, 
designed to serve  as a useful reference in future papers on the 
subject. There are several new results contained in the paper both
from the analytical as the numerical side. Two alternative choices 
for the twist matrices are given which lead to rather different 
descriptions of the same Physics. One of the formalisms is well suited 
for perturbative calculations, while the other is useful to study
certain type of classical solutions similar to sphalerons, which are
extrema of the potential having a few unstable directions. The
connection between both formalisms is just a gauge transformation which 
is derived and presented in an appendix of the paper. 

The paper then includes a description of the perturbative expansion 
to determine the spectrum of the theory. Calculations for the lowest
energy states in different electric flux sectors are given up to order 
$g^2$. This includes a close analytical expression  for the one loop 
self-energy as a function of the magnetic flux. Inspection of this
formula allows one to investigate the possible emergence of tachyonic
instabilities. It is interesting to notice that to avoid instabilities
at small  values of the coupling constant one is led to a similar
set of conditions as advocated  for the validity of the TEK
model~\cite{TEK3}. Thus, this is an important piece of evidence in
favour of these criteria. A four dimensional extension is most welcome 
and is currently being analysed. 

The perturbative setting indicates that at fixed value of the ratio 
$\kb/N$ ($\kb$ depends on the magnetic flux and is defined by
$\m \kb = 1 ($mod $N)$)
all perturbative expressions depend jointly on the
combination $LN$. This a an important result which extends the
validity of reduction to finite values of $N$ and to the non-zero
electric flux sectors of the theory. As a matter of fact, this merges
with the `t Hooft coupling dependence into the combination
$x= \lambda LN/4\pi $, which expresses the effective coupling for the
energies of the afore-mentioned sectors. 
We warn however, that certain non-perturbative effects might not
follow this pattern. 

This brings us to the third part of the paper in 
which we present certain numerical results obtained in the lattice 
version of the model. Our purpose is mainly focused in testing how the 
previous considerations extend to the non-perturbative large volume 
domain. We track the evolution of the energies of the non-vanishing 
electric flux sectors, restricting ourselves to those having the
minimal momentum and twice this value. Both  the scaling with $x$ and 
the absence of tachyonic instabilities are monitored into the region 
in which the electric flux energies develop into a k-string spectrum,
growing linearly with the linear extent of the box. Our results favour 
the simple sine formula for the k-string string tensions. The approach 
to this regime is also quite interesting since it involves the validity 
of the effective string description of Luscher, Symanzik and Weisz. 
In particular, we have discussed that the relevant description of the
string spectrum on the twisted box corresponds to strings on the 
background of a Kalb-Ramond $B$-field. This is natural if one advocates the connection
with non-commutative field theories~\cite{Seiberg:1999vs}.
The identification $B= - 2 \pi \m /N$, with $\m$ the magnetic flux, gives rise to the correct
field theory non-commutative parameter: $\theta = -L^2/B$, and naturally incorporates in the 
dispersion relation of the effective string the contribution of the tree-level 
field-theory perturbative result. 

It is remarkable that the expectations for the large volume region are
consistent with the $x$-scaling hypothesis. Indeed, our data are fairly 
well described in all regions with a relatively simple parameterization 
which incorporates both the perturbative results as the long distance 
expectations. 

Finally, we have analyzed non-perturbatively the fate of the tachyonic instabilities. Several of our 
numerical results show indeed indications that, for certain values of the magnetic flux, a possible 
condensation of electric fluxes might take place in the intermediate volume regime. Nevertheless, 
this situation is avoided if the set of conditions advocated in Ref.~\cite{TEK3} to preserve the stability of 
the TEK model are imposed. Although our numerical results explore only a limited range of values of $N$, 
we have argued using the conjectured $x$ dependence of the results that this should also hold
in the $N\rightarrow \infty$ limit.

It is clear that the paper could be thoroughly improved on the
numerical aspects, both by an increase in statistics, analysis of 
lattice corrections and extension of the  range of 
observables studied. In that respect, our present study has  left out 
the sector having  vanishing electric flux. It would be very
interesting to see how the energies evolve from the perturbative region 
to those giving rise to the glueball spectrum.  However, 
achieving this goal demands not only a substantial scale up of the 
computer resources used in this paper, but also a more elaborate
multi-operator technique to source the states.  

Another aspect that demands further study is the occurrence of
non-perturbative effects of semiclassical origin. Our numerical 
data seem to suggest their presence, but it is unclear what is their
origin and whether they have any connection to the constant magnetic
field sphaleron-type  configurations studied in our paper. 

Finally, a deeper understanding of the properties of the system 
in the presence of tachyonic instabilities is also welcome. The 
goals set up here for future works seem feasible and should be 
considered together with the obvious extension to 3+1 dimensions. 

\section*{Acknowledgements}
A.~G-A wants to thank the Aspen Center for Physics for the opportunity
to participate in the workshop {\em Strong dynamics beyond the
standard model} that took  place during May-June 2010.
The initial idea to pursue this work emerged from conversations and
discussions  with some other participants as M.~Hanada, R.~Narayanan,
D.~Nogradi, M.~Ogilvie,  E.~Poppitz, M.~Unsal  and very specially H.
Neuberger.

The authors acknowledge interesting conversations about certain
aspects related to  the paper with A.~Armoni and J.L.F.~Barbon.

M.G.P. and A.G-A acknowledge financial support from the MCINN grants FPA2009-08785,
FPA2009-09017, FPA2012-31686, and FPA2012-31880, the Comunidad Aut\'onoma de Madrid 
under the program  HEPHACOS
S2009/ESP-1473, and the European Union under Grant Agreement number
PITN-GA-2009-238353 (ITN STRONGnet). They participate in the Consolider-Ingenio 2010
CPAN (CSD2007-00042). They also thank the Spanish MINECO Centro de
excelencia Severo Ochoa Program under grant SEV-2012-0249.
M. O. is supported by the Japanese MEXT grant No 23540310.
We acknowledge the use of the IFT clusters for our numerical results.

\appendix

\section{Connecting the two different twist matrix formalisms}
\label{appendixa}

To connect the two formalisms one needs to make a gauge transformation
with a matrix $\Omega(x)$ satisfying 
\begin{equation}
\label{omega_cond} 
\Omega(x+L_i e_i)= \exp\{i \half  \bar{B} \epsilon_{i j} L_i x_j \} \Omega(x) \Gamma_i^\dagger 
\end{equation}
Once this matrix is found we can use it to find a connection $A_i$
compatible with the abelian twist matrices and having vanishing
magnetic field and minimal energy. This is given by 
$$ A_i=i \Omega \partial_i \Omega^\dagger $$
Alternatively, its inverse can give a connection with constant field
strength in the formalism with constant twist matrices. These are
local extrema of the action functional.

Let us assume without loss of generality that $\Gamma_1$ is diagonal.
This implies that $\bar{B}$ can be written as a linear combination of
$\Gamma_1$ and its powers. Using this fact we can   parameterize $\Omega(x)$ as
follows:
$$ \Omega(x)= \sum_{n=0}^{N-1} X_n(x)\,   \, \exp\{ i \bar{B}
L_2 n x_1\} \Gamma_2^n
$$
This is completely general if $X_n(x)$ is a diagonal matrix. Now we
will study the conditions on $X_n$ that result from
Eq.~\ref{omega_cond}. Imposing first periodicity in $x_1$, we obtain:
$$ X_n(x+e_1 L_1)=\exp\{ i \half \bar{B}  L_1 x_2/2 \} X_n(x)
\Gamma_1^\dagger $$ 
This constraint can be easily solved writing
$$ X_n(x) = \exp\{ i \bar{B}  x_1 x_2/2 -i I_1 x_1/L_1\} Z_n(x) $$ 
where $Z_n(x)$ is diagonal and periodic in $x_1$ with period $L_1$. 
The diagonal matrix $I_1$ is such that 
$$ e^{i I_1}= \Gamma_1$$ 

The next step is to impose the periodicity condition
Eq.~\ref{omega_cond} for  $i=2$. After some massaging we conclude that
the condition on $Z_n(x)$ is the following: 
$$ Z_{n+1}(x)=  Z_n(x+e_2 L_2) $$
This can be easily solved by writing 
$$ Z_n(x) = Z(x+n e_2 L_2) $$ 
with $Z(x)$ a diagonal matrix which is periodic in $x_1$. There is an
additional condition that follows for $n=N$. This condition reads
$$ \exp\{ i \bar{B} N  L_2 x_1 \}\, Z(x+N L_2 e_2) \Gamma_2^N=Z(x) $$
A simple inspection shows that one particular solution of this
condition is given by 
\be
(Z(x))_{a a }= \left( \exp\{- \frac{\pi x_2^2 }{L_1 L_2 N} \} \,
\theta\Big(\frac{x_1}{L_1}+ i \frac{x_2}{L_1}; i \frac{L_2
N}{L_1}\Big)\right)^{M_a} \nonumber
\ee
where $ 2 \pi M_a= -(\bar{B})_{a a } L_1 L_2 N$. The symbol $\theta$
stands for Jacobi's theta function. 

Instead of using knowledge of theta functions to solve the general
form of $Z(x)$, we might simply use standard algebraic methods. Since 
$Z(x)$ is periodic in $x_1$ with period $L_1$ we can write it as a
Fourier expansion. One can treat all components simultaneously by
writing 
$$ Z(x)= \sum_{Q}\, e^{i 2 \pi Q x_1/L_1} \hat{Z}(x_2,Q) $$
where $Q$ is an integer diagonal matrix. 
The periodicity condition in $x_2$ then gives rise to the following
condition on the Fourier coefficients:
$$\hat{Z}(x_2+L_2N;Q)=\hat{Z}(x_2, Q+\frac{\bar{B}N L_1 L_2}{2 \pi}) $$
Since we can uniquely decompose  $Q=Q'+p\frac{\bar{B}N L_1 L_2}{2
\pi}$, the coefficient, then it is simple to solve the constraint. 
Finally, replacing all the previous results  obtains
\be
\Omega(x)= \exp\{ \frac{i}{2} \bar{B}  x_1 x_2 -i I_1 \frac{x_1}{L_1}\} \sum_{p \in
\mathbf{Z}} \sum_{Q'}\, e^{i 2 \pi Q' \frac{x_1}{L_1}}(x)\,   \, \exp\{ i \bar{B}
L_2 p x_1\} \hat{Z}(x_2+pL_2;Q')  \Gamma_2^p
\ee
where $\hat{Z}(x_2;Q')$ are arbitrary diagonal  matrices.  
One particular class of solutions  is that in which these functions
are exponentials of a polynomial of second degree in $x_2$. It is 
this type which can be written in terms of theta functions. 

The general form of the solution is subject to the additional
condition that $\Omega$ belongs to SU(N).

To understand the structure and properties of the $\Omega(x)$ let us
study the simpler SU(2) case. Here $k=1$ and $M_a=0$. Using the previous
formulas in this specific case, we conclude that:
\be
\label{omegaval}
\Omega(x)= \begin{pmatrix}
\exp \{  - \frac{ i \pi x_1}{2 L_1} (\frac{x_2}{L_2}\!+\!1)\}  
\,C(\frac{x_1}{L_2},\frac{x_2}{L_2}) & \ \  
\exp \{  - \frac{ i \pi x_1}{2 L_1} (\frac{x_2}{L_2}\!+\!1)\}
\,T(\frac{x_1}{L_1},\frac{x_2}{L_2}) \\
\\
-\exp \{  \frac{ i \pi x_1}{2 L_1} (\frac{x_2}{L_2}\!+\!1)\}
\,T^*(\frac{x_1}{L_1},\frac{x_2}{L_2}) & \ \ 
\exp \{ \frac{ i \pi x_1}{2 L_1} (\frac{x_2}{L_2}\!+\!1)\}
\,C^*(\frac{x_1}{L_1},\frac{x_2}{L_2})
\end{pmatrix}
\ee
where the quaternionic structure of $\Omega(x)$ is obvious. To make it
unitary one must demand $|C|^2+|T|^2=1$. The functions $C$ and $T$ are
both periodic of period 1.  
Furthermore periodicity in $x_2$ implies 
\be
 C\Big(\frac{x_1}{L_1},\frac{x_2}{L_2}+2\Big)= -e^{i 2 \pi \frac{x_1}{L_1}} \, C\Big(\frac{x_1}{L_1},\frac{x_2}{L_2}\Big) 
\nonumber
\ee
and a similar equation for $T$. Finally one has:
\be
T\Big(\frac{x_1}{L_1},\frac{x_2}{L_2}\Big)= -i e^{- i \pi  \frac{x_1}{L_1}} \, C\Big(\frac{x_1}{L_1},\frac{x_2}{L_2}+1\Big) 
\nonumber
\ee

By Fourier decomposition in $x_1$ one can give the general solution in
terms of an arbitrary function $\hat{C}$ as follows:
\be
 C(x,y)= \sum_{q {\rm even}} i^q\,  e^{ i\pi x q }\, \hat{C}(y-q) 
\nonumber 
\ee
and 
\be
 T(x,y)= \sum_{q {\rm odd}} i^q\,  e^{ i\pi x q }\, \hat{C}(y-q) 
\nonumber
\ee
Now choosing $\hat{C}(y)=e^{-\pi y^2/2}$, we obtain:
\bea 
C(x,y)&=& e^{-\frac{\pi y^2}{2}}\, \theta_{0,\frac{1}{2}}(x-iy;2i) \nonumber \\
T(x,y)&=& e^{-\frac{\pi y^2}{2}}\, \theta_{\frac{1}{2},\frac{1}{2}}(x-iy;2i) \nonumber
\eea
where $\theta_{a b }$ are theta functions with rational
characteristics, as defined in the book {\em Tata lectures on Theta}
for example. The location of the zeroes of these functions are
well-known and it turns out that $C(x,y)$ and $T(x,y)$ do not vanish
as the same points. This is what we wanted since we can now divide 
the expression both functions by $|C|^2+|T|^2$, to impose that the
matrix $\Omega$ belongs to SU(N). The final form for $\Omega$ is given
by Eq.~\ref{omegaval} with $C$ and $T$ given by
\bea
C(x,y)&=&\frac{\theta_{0,\frac{1}{2}}(x-iy;2i)}{|\theta_{0,\frac{1}{2}}(x-iy;2i)|^2+|\theta_{\frac{1}{2},\frac{1}{2}}(x-iy;2i)|^2}\nonumber \\
T(x,y)&=&\frac{\theta_{\frac{1}{2},\frac{1}{2}}(x-iy;2i)}{|\theta_{0,\frac{1}{2}}(x-iy;2i)|^2+|\theta_{\frac{1}{2},\frac{1}{2}}(x-iy;2i)|^2}\nonumber
\eea

\section{Derivation of the Euclidean self-energy}
\label{appendixb}

In this appendix we will present  the derivation of the formula for the
Euclidean  vacuum polarization at one-loop which is used
in section~\ref{s:euclidean} to compute the gluon self-energy. The
procedure is mostly straightforward, and follows by combining the standard 
continuum  procedure, to be found in many textbooks, and the modified
Fourier expansion associated to twisted boundary conditions. 

The gauge fixed Euclidean Lagrangian density in a generalized covariant gauge 
with gauge parameter $\xi$ reads:
\be
{\cal L} = \half \Tr ( F_{\mu \nu}^2 )+ {1 \over \xi} \Tr (\partial_\mu A_\mu )^2 
- 2 \Tr (\bar c \partial_\mu D^\mu c)\quad ,
\ee 
with  $D_\mu \equiv \partial_\mu - i g A_\mu$, the covariant derivative, and $c$, 
$\bar c$ the ghost fields.

As mentioned earlier, we may now introduce the combined colour/space Fourier 
decomposition presented in section~\ref{method}. The result is
\be
A_\mu(x)=  {\cal N}\int_{-\infty}^{\infty} {d \q_0 \over 2 \pi} \sum'_{\vqt} \hat{A}_\mu(\q)\,  
e^{i \q\dot x}\,  \HG(\vqt)
\ee
with $\vqt= {2 \pi \vec{n} \over N L}$, $n_i \in {\bf Z}$. As
customarily done, the space-time momenta will be labelled by $\q$
without an arrow, to distinguish it from the spatial part.
The prime indicates that we should exclude from the sum the values 
of $\vqt$ corresponding to zero colour momenta, i.e. the values 
of $\vec{n}$ for which $n_i = 0 \ ({\rm mod}\ N),\
 \forall i$. 

This gives rise to the propagators:
\be 
\label{eq:prop}
\tilde P_{\mu \nu} (\p, \q)\equiv P_{\mu \nu} (\q)\  \delta(\q+\p)  = {1 \over \q^2} \Big 
(\delta_{\mu \nu} - (1-\xi) \ {\q_\mu \q_\nu \over \q^2}\Big) \  \delta(q+p) \quad ,
\ee
for the gauge field, and 
\be \tilde P_g(\p,\q) \equiv P_g (\q) \ \delta(\q+\p) =  - {1 \over \q^2}  \delta(\q+\p)\quad ,
\ee
for the ghost fields. 
The vertices contain 3-gluon, ghost-gluon and 4-gluon interactions, which look
very similar to their infinite  volume counterparts, with the structure constants $f_{a,b,c}$ replaced 
by the equivalent $F(\vpt_{(i)},\vpt_{(j)},\vpt_{(k)})$.  
They are given by:
\begin{itemize}
\item
{\bf 3-gluon vertex}
\be
{1 \over 3!}\   V^{(3)}_{\mu_1 \mu_2 \mu_3} (\p_{(1)}, \p_{(2)},\p_{(3)}) 
\   A_{\mu_1}(\p_{(1)})A_{\mu_2}(\p_{(2)})
A_{\mu_3}(\p_{(3)})\, \delta(\p_{(1)}+\p_{(2)}+\p_{(3)})\quad ,
\ee
with
\bea
V^{(3)}_{\mu_1 \mu_2 \mu_3}(\p_{(1)}, \p_{(2)},\p_{(3)})&=& i g 
 {\cal N}  F(\vpt_{(1)},\vpt_{(2)},\vpt_{(3)}) \times \\
\Big (( \p_{(3)}-\p_{(2)})_{\mu_1} \delta_{\mu_2 \mu_3}
   &+&  ( \p_{(1)}-\p_{(3)})_{\mu_2} \delta_{\mu_1 \mu_3}
   +  ( \p_{(2)}-\p_{(1)})_{\mu_3} \delta_{\mu_1 \mu_2}
\Big ) \quad , \nonumber
\eea
where $F(\vpt_{(1)},\vpt_{(2)},\vpt_{(3)})=-\sqrt{2/N}\, \sin ( \theta {\cal
A}_{12})$ (given in Eq.~\ref{eqdeF}),  
${\cal A}_{12} = \half (\vpt_{(1)} \times \vpt_{(2)}) $ and 
$\theta$ is defined in Eq.~\ref{thetadef}.
\item
{\bf Ghost-gluon vertex}
\be
V^{(gh)} =   -i  g  {\cal N}  F(\vpt_{(1)},\vpt_{(2)},\vpt_{(3)}) \  
\p_\mu^{(1)} \  \bar c (\p_{(1)}) A_\mu (\p_{(2)})
c(\p_{(3)}) \, \delta(\p_{(1)}+\p_{(2)}+\p_{(3)}) \quad .
\ee
\item
{\bf 4-gluon vertex}
\bea
&&{1 \over 4!}\   V^{(4)}_{\mu_1 \mu_2 \mu_3 \mu_4} (\p^{(1)},
\p_{(2)},\p_{(3)},\p_{(4)}) \   A_{\mu_1}(\p_{(1)})A_{\mu_2}(\p_{(2)})
A_{\mu_3}(\p_{(3)})
A_{\mu_4}(\p_{(4)}) \times \nonumber \\&& \delta(\p_{(1)}+\p_{(2)}+\p_{(3)}+\p_{(4)})
\eea
with
\bea
 V^{(4)}_{\mu_1 \mu_2 \mu_3 \mu_4}= - g^2   \frac{2{\cal N}^2}{N} \times && \\
\Big( \sin (\theta {\cal A}_{12}) \sin (\theta {\cal A}_{34}) &(&\delta_{\mu_1 \mu_3} \delta_{\mu_2 \mu_4}-\delta_{\mu_2 \mu_3} \delta_{\mu_1 \mu_4})  \nonumber\\
+\sin (\theta {\cal A}_{23}) \sin (\theta {\cal A}_{41}) 
&(&\delta_{\mu_2 \mu_4} \delta_{\mu_3 \mu_1}-\delta_{\mu_3 \mu_4} 
\delta_{\mu_2 \mu_1})  \nonumber\\
+\sin (\theta {\cal A}_{13}) \sin (\theta {\cal A}_{24}) 
&(&\delta_{\mu_1 \mu_2} \delta_{\mu_3 \mu_4}-\delta_{\mu_3 \mu_2} 
\delta_{\mu_1 \mu_4}) \Big )\nonumber \quad .
\eea
\end{itemize}

With the previous rules,  one can easily derive the one-loop correction 
to the self-energy. There are three diagrams contributing at this order:
\begin{itemize}
\item
{\bf The tadpole}
\be 
- g^2 {\cal N}^2 \int_{-\infty}^{\infty} {d \q_0 \over 2 \pi} \sum_{\vqt}' 
F^2(\vpt,\vqt,-\vpt-\vqt) \, 
({d \delta_{\mu \nu}\over \q^2}- P_{\mu \nu} (\q)) \quad , 
\ee
where $P_{\mu \nu} (\q)$ is the gluon propagator, $d$ the space-time dimension, and 
$F(\vpt,\vqt,-\vpt-\vqt)$ can be read from Eq.~\ref{eqdeF}.
\item
The contribution from the {\bf ghost loop}
\be
- g^2 {\cal N}^2 \int_{-\infty}^{\infty} {d \q_0 \over 2 \pi} \sum_{\vqt}' 
F^2(\vpt,\vqt,-\vpt-\vqt)\, 
{(\p+\q)_\nu \q_\mu \over (\p+\q)^2 \q^2} \quad .
\ee
\item
The contribution from two  3-gluon vertices ({\bf eye graph})
\be
 \half g^2 {\cal N}^2 \int_{-\infty}^{\infty} {d \q_0 \over 2 \pi} 
 \sum_{\vqt}' F^2(\vpt,\vqt,-\vpt-\vqt) \, S_{\mu \nu} \quad ,
\ee
where
\bea
S_{\mu \nu} &=& P_{\rho \rho'} (\p+\q) P_{\sigma \sigma'} (\q) \times \\
&&\Big ( (\p+2\q)_\mu \delta_{\rho \sigma} + (\p-\q)_\rho \delta_{\mu \sigma} - (\q+2\p)_\sigma \delta_{\mu \rho}\Big ) \nonumber \\
&&\Big ( (\p+2\q)_\nu \delta_{\rho' \sigma'} + (\p-\q)_{\rho'} \delta_{\nu \sigma'} - (\q+2\p)_{\sigma'} \delta_{\nu \rho'}\Big ) \quad , \nonumber
\eea
\end{itemize}

Adding all contributions together,  we obtain the
one-loop formula for the vacuum polarization in  Feynman gauge
($\xi=1$):
\bea
\Pi_{\mu \nu} &=& 
\half g^2 {\cal N}^2 \int_{-\infty}^{\infty} {d \q_0 \over 2 \pi} \sum_{\vqt}'
F^2(\vpt,\vqt,-\vpt-\vqt) {1 \over \q^2 (\p+\q)^2} \times
\\
&&\Big((2(d-2) \q_\nu - d \p_\nu) (\p_\mu + 2 \q_\mu) 
- 2(d-2)  \delta_{\mu \nu} \q^2 \nonumber 
+ 4(d-1) (\q_\mu \p_\nu - \delta_{\mu \nu} \ \p\cdot \q)  \Big )
\nonumber
\eea
\section{Derivation of the Euclidean self-energy on the lattice}
\label{appendixc}

In this appendix we will, for completeness, reproduce the formula for the 
Euclidean  vacuum polarization at one-loop 
as derived in Refs.~\cite{Snippe:1997ru},~\cite{Snippe:1996bk} 
for the Wilson lattice action with twisted boundary conditions in the x-y plane.
For that we consider a $L^2\times R$ volume discretized both in space and time 
with lattice spacing $a$. The continuum limit is taken by sending $a \rightarrow 0$ 
and the number of lattice points to infinity, while keeping $L=a N_s$ fixed.

We list below the different contributions to the vacuum polarization (we have set the 
lattice spacing $a=1$ in what follows):
\bea
\Pi_{\mu \nu}^{\rm ms } &=&  - {g_L^2 N \over 12 }  \delta_{\mu \nu}\quad ,\\
\Pi_{\mu \nu}^{ \rm gh1} &=&- {1 \over 6} g_L^2 {\cal N}^2 \delta_{\mu \nu} \int_{-{\pi}}^{{\pi }} {d \q_0 \over 2 \pi} \sum_{\vqt}' \FSS\, {\widehat \q_\mu^2 \over \widehat \q^2} \quad ,\\
\Pi_{\mu \nu}^{ \rm gh2} &=&    
 {1 \over 4} g_L^2 {\cal N}^2 \int_{-{\pi }}^{{\pi }} {d \q_0 \over 2 \pi} \sum_{\vqt}' \FSS\, 
{\widehat \p_\mu \widehat \p_\nu - \widehat{(2\q + \p)}_\mu \widehat{(2\q + \p)}_\nu \over \widehat \q^2 \widehat {(\p+\q)}^2}\quad ,\\
\Pi_{\mu \nu}^{ \rm V3} &=& \half  g_L^2 {\cal N}^2 \int_{-{\pi}}^{{\pi}} {d \q_0 \over 2 \pi} 
\sum_{\vqt}' \FSS\, {V_{\mu \lambda \rho}^{(3)} V_{\nu \lambda \rho}^{(3)} \over \widehat \q^2 \widehat {(\p+\q)}^2} \quad ,\\ 
\Pi_{\mu \nu}^{ \rm V4} &=& {1 \over 3} g_L^2 {\cal N}^2 \int_{-{\pi}}^{{\pi}} {d \q_0 \over 2 \pi} \sum_{\vqt}' \FSS
\times \nonumber \\
&&{1 \over \widehat \q^2}   \Big(V_{\lambda \lambda \mu \nu}^{(4)}(\q, -\q, \p,-\p) - V_{\lambda \mu \lambda \nu}^{(4)}(\q, \p,-\q,-\p)\Big) \quad ,\\
\Pi_{\mu \nu}^{ \rm W} &=& {g_L^2 {\cal N}^2 \over 4} 
 \int_{-{\pi}}^{{\pi}} {d \q_0 \over 2 \pi} \sum_{\vqt}' \Big({1\over 2N} - {1 \over 6}
\FSS \Big)   \times \nonumber \\
&&{1 \over \widehat \q^2}  \, \widehat \p_\rho \, (\widehat \p_\rho \delta_{\mu \nu} -\widehat \p_\mu \delta_{\rho \nu}) \,   
(\widehat \q_\mu^{\,2} (1 -2 \delta_{\rho \mu}) + \widehat \q_\rho^{\,2} ) \quad ,
\eea
where $ \widehat \q_\mu = 2 \sin( \q_\mu /2)$, with discretized spatial momenta 
$ \q_i = 2 \pi n_i /N N_s$, for $n_i =0,\cdots, N_s N-1$, and where the sum over 
$\vqt$ excludes momenta with zero colour momenta. As in the continuum:
$\FS = - \sqrt{2/N} \sin (\theta (\vpt \times \vqt) /2)$. 
The three and four gluon vertices read:
\be
V_{\mu \lambda \rho}^{(3)} =  \cos\Big ({ \q_\mu+\p_\mu \over 2 }\Big ) \widehat{(\q-\p)}_\rho 
\delta_{\mu \lambda} - \cos\Big ({  \p_\lambda\over 2}\Big ) \widehat{(2\q+\p)}_\mu 
\delta_{\lambda\rho} +  \cos\Big ({ \q_\rho\over 2}\Big ) \widehat{(2\p+\q)}_\lambda \delta_{\mu \rho}\quad ,
\ee
and
\bea
V_{\mu_1 \mu_2 \mu_3 \mu_4}^{(4)}(k_1,k_2,k_3,k_4) &=& f_{\mu_1} \delta_{\mu_1 \mu_2 \mu_3 \mu_4} + (g_{\mu_1 \mu_4} \delta_{\mu_1 \mu_2 \mu_3 } + {\rm 3 \ cyclic\ perms})\\
&+& h_{\mu_1 \mu_2} \delta_{\mu_1 \mu_3 } \delta_{\mu_2 \mu_4 }
+(h_{\mu_1 \mu_3}^{'} \delta_{\mu_1 \mu_2 } \delta_{\mu_3 \mu_4 } + {\rm 1 \ cyclic\ perm}) \quad , \nonumber
\eea
with:
\bea
f_{\mu_1} &=& {1 \over 6} \sum_{\rho =0}^{2} \Big ( \widehat{(k_1+k_3)}^2_\rho 
- \half \widehat{(k_1+k_2)}^2_\rho -\half \widehat{(k_1+k_4)}^2_\rho + \widehat k_{1\rho} \widehat k_{2\rho}
\widehat k_{3\rho}\widehat k_{4\rho} \Big)\quad , \\
g_{\mu \nu} &=& {1 \over 6} \Big(\cos\Big({k_{3\nu}\over 2}\Big) \widehat{(k_1-k_2)}_\nu 
-\cos\Big({ k_{1\nu}\over 2}\Big) \widehat{(k_2-k_3)}_\nu
- \widehat k_{1\nu } \widehat k_{2\nu }\widehat k_{3\nu }\Big)\widehat k_{4\mu } \ \ \quad ,\\
h_{\mu \nu} &=& 2 \cos\Big({k_{2\mu}-k_{4\mu}\over 2}\Big) \cos\Big({ k_{1\nu}-k_{3\nu}\over 2}\Big) \quad ,\\
h_{\mu \nu}^{'} &=& -\cos\Big({k_{3\mu}-k_{4\mu}\over 2} \Big) \cos\Big({k_{1\nu}-k_{2\nu}\over 2} \Big)
+{1 \over 4} \widehat k_{3\mu } \widehat k_{4\mu }\widehat k_{1\nu}\widehat k_{2\nu} \quad .
\eea

\section{Lattice results for the energy of electric flux}
\label{appendixd}

In this appendix we collect the numerical results 
for the electric-flux spectrum obtained in the lattice simulations 
described in section~\ref{s.numerical}. 

The simulations have been performed for various values of the rank of 
the group $N$, 
the angle $\tilde \theta = 2 \pi \kb/N$, and the inverse 't Hooft
coupling $b=1/(a\lambda)$. The corresponding values for each dataset, 
as well as the lattice size employed ($N_0 \times N_s^2$) in each case, are
indicated on the tables. 
The set of values of $b$ simulated for the different cases is not
always the same, although it covers approximately the same interval in $x$.
The reasons for this are on one hand the limitations of numerical resources,
and on the other the exclusion of points with strong lattice artefacts.
In all these cases a dash replaces the corresponding entry in the table.

The electric-flux energies, given in lattice units, are extracted from the 
exponential decay in 
Euclidean time of correlation functions of Polyakov loops projected over 
gluon-momentum $\vpt= 2\pi \vec n/NL$. The 
results denoted in the tables by $\ET_1$ and $\ET_2$ correspond 
respectively to: $\vec n = (0,1)$ and $\vec n= (0,2)$,  with electric flux 
determined through the relation:
$e_i= \epsilon_{ji} \kb n_j$ (mod $N$). They are the basis for the fits
performed in section~\ref{s.numerical}, collected in Tables~\ref{t.fits} and
~\ref{t.fits2}. 
The few points labelled by an asterisk in the tables 
have been excluded from the fits
in order to attain values of the $\chi^2$ per degree
of freedom of order 1. 

\TABLE
[htbp]{
\begin{tabular}{|l||l|l||l|l|}
\hline
$b$ &  $ a \ET_1\, (\kb=1) $ & $a \ET_2\, (\kb=1) $
&   $a \ET_1\, (\kb=2) $ & $a \ET_2\, (\kb=2) $ \\
\hline
1.5$\ $&    0.2485(17)  &  0.4310(86)&  0.428(11)&0.2963(22)
\\ \hline
02&         0.1438(14)  &  0.2601(33)& 0.2508(33)&0.2057(11)
\\ \hline
2.5 &       0.1067(18)  & 0.184(11)& 0.1714(32)&0.1830(19)
\\ \hline
03  &       0.08668(42) & 0.1513(19)& 0.12924(81)&0.1748(17)
\\ \hline
3.5 &       0.08082(92) & 0.1390(61)& 0.1065(12)&0.1685(38)
\\ \hline
4   &       0.07521(90) & 0.1331(99)& 0.09549(93)&0.1715(30)
\\ \hline
5   &       0.07495(33) & 0.1380(26)& 0.08355(60)&0.1660(26)
\\ \hline
6   &       0.07741(63) & 0.1515(61)& 0.0808(10)&0.1689(18)
\\ \hline
7   &       0.0808(14)  & 0.1609(36)& 0.08238(80)&0.17143(86)
\\ \hline
08  &       $*$ 0.08347(50) & 0.1620(24)&- &-
\\ \hline
10  &       $*$ 0.08522(68) & 0.1697(28)& $*$ 0.08654(35)&0.1730(22)
\\ \hline
18  &       $*$ 0.08734(20) & 0.1740(15)& $*$ 0.08834(26)&0.1768(14)
\\ \hline
28  &       0.08780(29) & 0.17763(81)& 0.08881(26)&0.1786(16)
\\ \hline
35  &       0.08884(27) & 0.1782(12)&- &-
\\ \hline
50  &       0.08859(32) & 0.17839(72)& 0.08923(30)&0.17832(63)
\\ \hline
75  &       0.08878(30) & 0.17928(39)& 0.08951(21)&0.17905(60)
\\ \hline
\end{tabular}
\caption{\label{t1.data}
Electric-flux energies corresponding to gauge group SU(5) and lattice size
$48 \times 14^2$.
}
}

\TABLE
[htbp]{
\begin{tabular}{|l||l|l||l|l|}
\hline
$\tilde b$ &  $ a \ET_1\, (\kb = 1)$ & $a \ET_2\, (\kb = 1)$
&  $a \ET_1\, (\kb = 2)$ & $a \ET_2\, (\kb = 2)$
\\ \hline
1&    0.3679(83) &0.619(12)&  0.536(77)& 0.3736(59)
\\ \hline
1.5$\ $& 0.1486(25) &0.273(11)&  0.2677(66)& 0.1802(31)
\\ \hline
2&    0.08836(70)&0.1601(34)&  0.1459(54)& 0.1238(16)
\\ \hline
3&    0.05460(32)&0.085(10) &  0.0818(11)& 0.1078(10)
\\ \hline
5&    0.04823(33)&0.0888(32) &  0.05193(55)&0.1070(26)
\\ \hline
\end{tabular}
\caption{\label{t2.data}
Electric-flux energies corresponding to gauge group SU(5) and lattice size
$72 \times 22^2$. The bare couplings at which the simulations have been performed
correspond to $b=22 \, \tilde b /14$.
}
}

\TABLE[htbp]{
\begin{tabular}{|l||l|l||l|l||l|l|}
\hline
 $b$ & $a \ET_1\, (\kb =1)$ & $a \ET_2\, (\kb =1)$ &  
$a \ET_1\, (\kb =2)$ & $a \ET_2\, (\kb =2)$ &   $a \ET_1\, (\kb =3)$ & $a \ET_2\, (\kb =3)$
\\ \hline
1.5$\ $&   0.1696(14)& 0.266(26)  & 0.346(16)  &0.496(35)  &0.466(15)& 0.2250(40)
\\ \hline
2  &       0.10190(37)&0.1640(49) & 0.1995(27) &0.2869(60) &0.259(17)&0.1778(23)
\\ \hline
2.5&       0.07879(80)&0.1338(50) & 0.1362(15) &0.1877(20)  &0.1687(44)&0.1686(17)
\\ \hline
3  &       0.07039(17)&0.1212(13) & 0.1100(10) &0.1801(33) &0.1320(15)&0.1687(12)
\\ \hline
3.5&       0.06776(57)&0.1167(34) & 0.09377(65)&0.1557(82) &0.1066(23)&0.1653(19)
\\ \hline
 4&    0.06676(34)& 0.12074(80)& 0.08795(80)&0.1594(33) &0.0969(13)&0.1669(17)
\\ \hline
 5&    $*$ 0.07136(34)& 0.1289(13)  & 0.0795(11) &0.1514(67) &0.0842(14)&0.1715(18)
\\ \hline
 6&    $*$ 0.07495(60)& 0.1400(12) & 0.07950(64)&0.1525(56) &0.08181(76)&0.1707(16)
\\ \hline
 7&    $*$ 0.07885(62)& 0.1487(17) & 0.08064(87)&0.1583(49) &$*$ 0.07978(92)&0.1754(22)
\\ \hline
 8&    $*$ 0.08136(41)&0.1536(19) & 0.08173(35)& $*$ 0.137(36)&$*$0.0799(15)&0.1729(15)
\\ \hline
10&    $*$ 0.08351(28)&0.1608(19) & 0.08365(40)&0.1679(16) &$*$0.08381(38)&0.17379(93)
\\ \hline
14&            -      &     -     & 0.08622(29)&0.1715(23) &0.08677(28)&0.1747(11)
\\ \hline
18 &       0.08648(16)&0.17186(52)& 0.08758(16)&0.17446(53)& 0.08813(34)& -
\\ \hline
28 &       0.08770(16)&0.17547(49)& 0.08858(16)&0.17616(68)&0.08861(20)&0.17755(56)
\\ \hline
35 &       0.08843(14)&0.17562(56)& 0.08858(18)&0.17670(49)&0.08859(19)&0.17801(73)
\\ \hline
50 &       0.08848(15)&0.17780(44)& 0.08877(22)&   -       &0.08937(27)& -
\\ \hline
75 &       0.08925(23)&0.17712(73)& -            &  -          &   -         & -
\\ \hline
\end{tabular}
\caption{\label{t3.data} 
Electric-flux energies corresponding to gauge group SU(7) and lattice size
$32 \times 10^2$.
}
}
\TABLE[htbp]{
\begin{tabular}{|l||l|l||l|l|}
\hline
 $b$ &  $a \ET_1\, (\kb =1)$ & $a \ET_2\, (\kb =1)$ &   $a \ET_1\, (\kb =3)$ & $a \ET_2\, (\kb =3)$ 
\\ \hline
0.6 &  0.5995(73)  &  -         &  -  &  -
\\ \hline
0.7 &  0.336(12)   &  -         &  -  &  -
\\ \hline
0.76&  0.2354(25)  & 0.74(38)   &  -  &  -
\\ \hline
0.85&  0.1272(53)  & 0.300(79)  &  -  &  -
\\ \hline
1   &  0.039862(21)& 0.1827(59) &  -  &  -
\\ \hline
1.4 &  0.01502(11) & 0.0585(35) & 0.2500(41) & 0.403(29)
\\ \hline
1.5 &         -    &  -         & 0.2118(29) & 0.343(20)
\\ \hline
1.7 &  0.011483(71)& 0.0399(11) & 0.1671(11) & 0.266(18)
\\ \hline
2   &         -    &    -       & 0.1245(11) & 0.1993(79)
\\ \hline
2.3 &         -    &    -       & 0.10423(73)& 0.1642(61)
\\ \hline
3   &  0.008588(18)& 0.02217(4) & 0.08276(22)& 0.1393(20)
\\ \hline
3.5 &         -    &    -       & 0.07631(47)& 0.1275(27)
\\ \hline
4   &         -    &    -       & 0.07507(52)& 0.1311(23)
\\ \hline
5   &  0.05383(12) & 0.1556(26)  & 0.07600(41)& 0.1401(21)
\\ \hline
6   &         -    &    -       & 0.07906(66)& 0.1506(28)
\\ \hline
7   &         -    &    -       &  0.08178(53)&0.1583(27)
\\ \hline
8   &         -    &    -       & $*$ 0.08469(43)&0.1646(25)
\\ \hline
10  &  0.07818(22) & 0.1766(19) & $*$ 0.08659(21)&0.1664(23)
\\ \hline
18  &         -    &    -       & $*$ 0.08975(15)&0.17731(53)
\\ \hline
28  &  0.08756(17) & 0.18224(96)& 0.09090(19)&0.17911(96)
\\ \hline
50  &  0.09034(24) & 0.1834(16) & 0.09103(31)&0.18324(91)
\\ \hline
75  &         -    &    -       & 0.09161(21)&0.18321(78)
\\ \hline
\end{tabular}
\caption{\label{t4.data} 
Electric-flux energies corresponding to gauge group SU(17) and lattice size
$32 \times 4^2$.
}
}
\TABLE[htbp]{
\begin{tabular}{|l||l|l||l|l|}
\hline
 $b$ & $a \ET_1\, (\kb =5) $ & $a \ET_2\, (\kb =5) $ &    $a \ET_1\, (\kb =8) $ & $a \ET_2\, (\kb =8) $ 
\\ \hline
0.8$\ $ &         -    &      -     &       -    & 0.2412(35)
\\ \hline
1   &         -    &      -     &       -    & 0.14151(49)
\\ \hline
1.2 &         -    &      -     & 0.765(16)  & 0.11161(32)
\\ \hline
1.4 &         -    &      -     & 0.582(14)  & 0.10802(47)
\\ \hline
1.5 $\ $&  0.3505(96)  & 0.445(23) & 0.495(10)  & 0.11055(39)
\\ \hline
1.7 &         -    &      -     & 0.367(26)  & 0.11792(86)
\\ \hline
2   &  0.1986(18)  & 0.2685(41) & 0.2631(45)  & 0.13347(68)
\\ \hline
3   &  0.10949(63) & 0.1736(22)& 0.13338(25)& 0.1532(12)
\\ \hline
5   &  0.08295(44) & 0.1602(14) & 0.0856(13) & 0.16929(68)
\\ \hline
10  &  0.08697(24) & 0.1724(24) & 0.08701(21)& 0.1759(15)
\\ \hline
18  &  0.08997(36) & 0.17855(55)& 0.09051(33)& 0.18061(59)
\\ \hline
28  &  0.09091(27) & 0.18086(87)& 0.09115(19)& 0.18198(34)
\\ \hline
50  &  0.09199(16)$\ $ & 0.18335(69)& 0.09171(15)$\ $& 0.18357(50)
\\ \hline
75  &         -    &      -     & 0.09214(20)& 0.18425(82)
\\ \hline
\end{tabular}
\caption{\label{t5.data} 
Electric-flux energies corresponding to gauge group SU(17) and lattice size
$32 \times 4^2$.
}
}

\FloatBarrier

\end{document}